\documentclass[prd,floatfix,superscriptaddress,preprintnumbers,nofootinbib]{revtex4}
\usepackage{float}
\usepackage{epsfig}
\usepackage{epstopdf}
\usepackage{amsmath}
\usepackage{hyperref}
\hypersetup{colorlinks=True,citecolor=blue}
\usepackage{amsfonts}
\usepackage{amssymb}
\usepackage{caption}
\usepackage{float}
\usepackage{bm}
\usepackage{graphicx}
\usepackage{color}
\usepackage{verbatim}
\usepackage{subfigure}

\def\beq{\begin{equation}}
\def\eeq{\end{equation}}
\def\ber{\begin{eqnarray}}
\def\eer{\end{eqnarray}}
\def\benu{\begin{enumerate}}
\def\eenu{\end{enumerate}}

\def\l{\left}
\def\r{\right}
\def\d{{\rm d}}
\def\n{\nabla}
\def\pa{\partial}
\def\f{\frac}

\def \lleq {\lower0.9ex\hbox{ $\buildrel < \over \sim$} ~}
\def \ggeq {\lower0.9ex\hbox{ $\buildrel > \over \sim$} ~}

\def\prl{{Phys.\@ Rev.\@ Lett.\ }}
\def\prd{{Phys.\@ Rev.\@ D\ }}

\def\plb {{Phys.\@ Lett.\@ B\ }}
\def\pla {{Phys.\@ Lett.\@ A\ }}
\def \jetpl {JETP Lett.\ }

\def\ie {{\it i.e. \,}}
\def\n {\noindent}

\begin{document}

%\title{How general is Inflation ?}
% A study of different models}
\title{Initial conditions for Inflation in an FRW Universe}
\author{Swagat S. Mishra}
\email{swagat@iucaa.in}
\affiliation{Inter-University Centre for Astronomy and Astrophysics,
Post Bag 4, Ganeshkhind, Pune 411~007, India}

\author{Varun Sahni}
\email{varun@iucaa.in}
\affiliation{Inter-University Centre for Astronomy and Astrophysics,
Post Bag 4, Ganeshkhind, Pune 411~007, India}

\author{Alexey V. Toporensky}
\email{atopor@rambler.ru}
\affiliation{Sternberg Astronomical Institute, Moscow State University,
Universitetsky Prospekt, 13, Moscow 119992, Russia}
\affiliation{Kazan Federal University, Kremlevskaya 18, Kazan, 420008, Russia}

\date{\today}

\begin{abstract}
We examine the class of initial conditions which give rise to 
inflation. Our analysis is carried out for several popular models including: Higgs inflation, Starobinsky inflation, chaotic inflation, axion monodromy inflation
and non-canonical inflation. In each case we determine the set of initial conditions which give rise to sufficient inflation, with at least $60$ e-foldings.
A phase-space analysis has been performed for each of these models and the effect of the initial inflationary energy scale on inflation has been studied numerically.
This paper discusses two scenarios of Higgs inflation:
(i) the Higgs is coupled to the scalar curvature, (ii) the Higgs Lagrangian contains a non-canonical
kinetic term.
In both cases we find Higgs inflation to be very robust since it can arise for a large class of
initial conditions.
One of the central results of our analysis is that, for plateau-like potentials associated with
the Higgs and Starobinsky models, inflation can
be realized even for  initial scalar field values 
which lie close to the {\em minimum} of the potential.
This dispels a misconception relating to plateau potentials prevailing in the literature.
We also find that inflation in all models is more robust for larger values of the initial energy scale.
\end{abstract}

%\arxivnumber{....}

\maketitle

\section{Introduction}
\label{sec:intro}

Since its inception in the early 1980's, the inflationary scenario has emerged
as a popular paradigm for describing the physics of the very early
universe \cite{star80,guth81,linde82,alstein82,baumann07}. A major reason for the success of the inflationary scenario is that,
in tandem with explaining many observational features of our universe --
including its homogeneity, isotropy and spatial flatness, it can also account
for the existence of galaxies, via the mechanism of
tiny initial (quantum) fluctuations which are subsequently amplified through
gravitational instability \cite{mukhanov81,hawking82,star82,guth-pi82}.

An important issue that needs to be addressed by a successful model of inflation
is whether the universe can inflate starting from a sufficiently large class of 
initial conditions. This issue was affirmatively answered for chaotic
inflation in the early papers \cite{belinsky85,belinsky88}. Since then
the inventory of inflationary models has rapidly increased. In this paper
we attempt to generalize the analysis of \cite{belinsky85,belinsky88}
to other popular inflationary models including Higgs inflation, 
Starobinsky inflation etc., emphasising the distinction between power law potentials and 
asymptotically flat `plateau-like' potentials. As we shall show,
our results for asymptotically flat potentials do not provide support to
 the  `$unlikeliness ~problem$' raised in \cite{stein13}\footnote{See \cite{linde2017} for an analysis of other problems with plateau-like potentials raised in \cite{stein13}.}. 

Our paper is organized as follows. We introduce our method of analysis
in section \ref{sec:methodology}. Section \ref{sec:chaotic} discusses power law potentials and
includes chaotic inflation and monodromy inflation. 
Section \ref{sec:higgs} discusses Higgs inflation in the context of both
the non-minimal as well as the non-canonical framework\footnote{As pointed out in \cite{sanil_varun} non-canonical scalars permit the Higgs field to play the role of the inflaton.}.
Section \ref{sec:starobinsky} is
devoted to Starobinsky inflation. 
%Section \ref{sec:power_vs_flat} compares
% power-law and asymptotically flat potentials. 
Our results are presented in 
section \ref{sec:discussion}.
  
We work in the units $c,\hbar =1$ and the reduced Planck mass 
is assumed to be
$m_{p}=\frac{1}{\sqrt{8\pi G}}$. The metric signature is 
 $(-,+,+,+)$. For simplicity
we assume that the pre-inflationary patch which resulted
in inflation was homogeneous, isotropic and spatially flat.
An examination of inflation within a more general cosmological setting can be found in 
\cite{branden16}.

\section{Methodology}
\label{sec:methodology}

The action for a scalar field which couples minimally to gravity has
 the following general form
\begin{equation}
%\beq
S[\phi]=\int\d^{4}x\, \sqrt{-g}\; {\cal L}(F,\phi),\label{eqn: action}
%\eeq
\end{equation}
where
 the Lagrangian density ${\cal L}(\phi , F)$
is a function of the field $\phi$ and the
kinetic term
\begin{equation}
F=\frac{1}{2}\pa_{\mu}\phi\; \pa^{\mu}\phi.\label{eqn: X-phi}
\end{equation}
Varying (\ref{eqn: action}) with respect to $\phi$ results in the
equation of motion
\begin{equation}
\frac{\pa {\cal L}}{\pa \phi} - \l(\frac{1}{\sqrt{-g}}\r)\pa_{\mu}\l(\sqrt{-g}\frac{\pa {\cal L}}{\pa \l(\pa_{\mu}\phi\r)}\r) = 0.\label{eqn: EOM1}
\end{equation}

The energy-momentum tensor
associated with the scalar field is
\begin{equation}
T^{\mu\nu}
=\l(\f{\pa{\cal L}}{\pa F}\r)\, \l(\pa^{\mu}\phi\; \pa^{\nu}\phi\r)
- g^{\mu\nu}\, {\cal L}~.\label{eqn: SET}
\end{equation}
Specializing to a spatially flat FRW universe
and a homogeneous scalar field, one gets
\begin{equation} 
\d s^2 = -\d t^2+a^{2}(t)\; \l[\d x^2 + \d y^2 + \d z^2\r],
\label{eqn: FRW}
\end{equation}
\begin{equation}
T^{\mu}_{\;\:\;\nu} = \mathrm{diag}\l(-\rho_{_{\phi}}, p_{_{\phi}},  p_{_{\phi}},  p_{_{\phi}}\r),
\end{equation}
where the energy density, $\rho_{_{\phi}}$, and pressure, $p_{_{\phi}}$, are given by
\begin{eqnarray}
\rho_{_{\phi}} &=& \l(\f{\pa {\cal L}}{\pa F}\r)\, (2\, F)- {\cal L}\label{eqn: rho-phi},\\
p_{_{\phi}} &=& {\cal L}\label{eqn: p-phi},
\end{eqnarray}
and  $F = -({\dot \phi}^{2}/2)$.
The evolution of the scale factor $a(t)$ is governed by the Friedmann equations:
\begin{eqnarray}
\l(\frac{\dot{a}}{a}\r)^{2} &=& \l(\frac{8 \pi G}{3}\r)\rho_{_{\phi}},\label{eqn: Friedmann eqn1}\\
\frac{\ddot{a}}{a} &=& -\l(\frac{4 \pi G}{3}\r)\l(\rho_{_{\phi}} + 3\,p_{_{\phi}}\r),\label{eqn: Friedmann eqn2}
\end{eqnarray}
where $\rho_{_{\phi}}$ satisfies the conservation equation
\begin{equation}
{\dot \rho_{_{\phi}}} = -3\, H \l(\rho_{_{\phi}} + p_{_{\phi}}\r), ~~ H \equiv \frac{\dot a}{a}~.
\label{eqn: conservation eqn}
\end{equation}
For a canonical scalar field 
\begin{equation}
{\cal L}(F,\phi) = - F - V(\phi),
\label{eqn: Lagrangian}
\end{equation}

Substituting (\ref{eqn: Lagrangian}) into
(\ref{eqn: rho-phi}) and (\ref{eqn: p-phi}), we find
\begin{eqnarray}
\rho_{_{\phi}} &=& \frac{1}{2}{\dot\phi}^2 +\;  V(\phi),\nonumber\\
p_{_{\phi}} &=& \frac{1}{2}{\dot\phi}^2 -\; V(\phi), ~~
\label{eqn: p-model}
\end{eqnarray}
consequently
the two Friedmann equations (\ref{eqn: Friedmann eqn1}) and (\ref{eqn: Friedmann eqn2}) become
\begin{eqnarray}
H^{2} &=& \frac{8 \pi G}{3}\l[\frac{1}{2}{\dot\phi}^2 +\;
V(\phi)\r]~,\label{eqn: FR-eqn1 model}\\
\frac{\ddot{a}}{a} &=& -\frac{8 \pi G}{3}\l[{\dot\phi}^2 -\;  V(\phi)\r]~.
\label{eqn: FR-eqn2 model}
\end{eqnarray}
Noting that ${\dot H} + H^2 = {\ddot a}/a$ one finds
${\dot H} = -4\pi G{\dot\phi}^2 < 0$, which informs us that the expansion rate
is a monotonically decreasing function of time for canonical scalar fields which
couple minimally to gravity.
The scalar field
 equation of motion follows from (\ref{eqn: EOM1})
\begin{equation}
{\ddot \phi}+ 3\, H {\dot \phi} + V'(\phi) = 0.
\label{eqn:motion}
\end{equation}
Within the context of inflation, a scalar field rolling down its potential is usually
associated with the Hubble slow roll parameters \cite{baumann07}
\begin{equation}
\epsilon_{H} = 2m_{p}^{2}\left(\frac{H'(\phi)}{H(\phi)}\right)^2,~~
 \eta_{H} = 2m_{p}^{2}\frac{H''(\phi)}{H(\phi)}
\label{eqn:slow_roll}
\end{equation}
and the potential slow-roll parameters \cite{baumann07}
\begin{equation}
\epsilon = \frac{m_{p}^2}{2}\left (\frac{V'}{V}\right )^2, ~~
\eta = m_{p}^2\frac{V''}{V}.
\label{eqn:slow_roll1}
\end{equation}
For small values of these parameters  $\epsilon_{H} \ll 1, \eta_{H} \ll 1$,
 one finds $\epsilon_{H} \simeq \epsilon$
and $\eta_{H} \simeq \eta - \epsilon$.
The expression for $\epsilon_{H}$ in (\ref{eqn:slow_roll}) can be rewritten as
$\epsilon_{H} = -\frac{\dot H}{H^2}$
which implies that the universe accelerates, ${\ddot a} > 0$, when $\epsilon_{H} < 1$.
For the scalar field models discussed in this paper 
${\dot H} = -4\pi G{\dot\phi}^2$ so that $\epsilon_{H} = 4\pi G{\dot\phi}^2/{H^2}$,
which reduces to $\epsilon_{H} \simeq \frac{3}{2}{\dot\phi}^2/V$ when
${\dot\phi}^2 \ll V$.

The slow-roll parameters play an important role in determining the spectral index
of scalar perturbations, since\footnote{Here $n_{_{S}} - 1 \equiv  \frac{d\, \mathrm{ln} \mathcal{P}_{_{S}}}{d\, \mathrm{ln} k}$,
where $\mathcal{P}_{_{S}}$ is the power spectrum of scalar curvature perturbations.}, $n_{_{S}} - 1 = -6\epsilon + 2\eta$.
Observations
indicate \cite{CMB} $n_{_{S}} \simeq 0.97$ which suggests that $\epsilon, \eta \ll 1$ on scales
associated with the present cosmological horizon.
The fact that $\epsilon, \eta$ are required to be rather small 
might appear to imply that successful inflation can only arise under a very restricted
set of initial conditions, namely those for which ${\dot\phi}^2/V(\phi) \ll 1$.
%that guarantee that $\epsilon, \eta \ll 1$
%in (\ref{eq:slow_roll1}). 
This need not necessarily be the case.
As originally demonstrated in the context of chaotic inflation
\cite{belinsky85,belinsky88}, a scalar field rolling down a 
power law potential can 
arrive at the attractor
trajectory $\epsilon, \eta \ll 1$ from a very wide range of initial conditions. 
In this paper we shall apply the methods developed in \cite{belinsky85,belinsky88,felder02}
to several inflationary models with power law and
 plateau-like potentials in order
to assess the impact of initial conditions on these models.

In addition to the field equations developed earlier, we shall find it convenient to
work with the parameter
\begin{equation}
N_e= \log{\frac{a(t_{\rm end})}{a(t_{\rm initial})}} = {\int_{t_i}^{t_e} H dt}
\equiv -\int^{\phi}_{\phi_{_{e}}}\l(\frac{H}{\dot{\phi}}\r)\d\,\phi
\label{eqn:efolds}
\end{equation}
which describes the number of inflationary e-foldings since the onset of inflation.
For our purpose it will also be 
instructive to rewrite the Friedman equation (\ref{eqn: FR-eqn1 model}) as
\begin{equation}
R^{2}=X^{2}+Y^{2}
\label{eqn:circle}
\end{equation}
where 
\begin{equation}
R=\sqrt{6}\frac{H}{m_{p}}, ~~
X=\hat{\phi}\frac{\sqrt{2V(\phi)}}{m_{p}^{2}},~~
Y=\frac{1}{m_{p}^{2}} \frac{d\phi}{dt},
\label{eqn:circle1}
\end{equation}
where $\hat{\phi}=\frac{\phi}{|\phi|}$ is the sign of $\phi$ (this definition ensures that $X$ and $\phi$ have the same sign).
Clearly, holding $R$ fixed and
varying $X$ and $Y$, one arrives at a set of initial conditions 
which satisfy the constraint
equation (\ref{eqn:circle}) 
 defining the boundary  of
 a circle of radius $R$. Adequate inflation is then 
qualified by the range of initial values of $X$ and $Y$ for which the universe
inflates by at least 60 e-foldings, \ie $N_e \geq 60$.

We commence our discussion of inflationary models by an analysis of power law potentials
which are usually associated with Chaotic inflation \cite{linde83,belinsky88}.

\section{Inflation with Power-law Potentials}  
\label{sec:chaotic}

\subsection{Chaotic Inflation}
\begin{figure}[h]
\centering
{\includegraphics[width=0.6\textwidth]{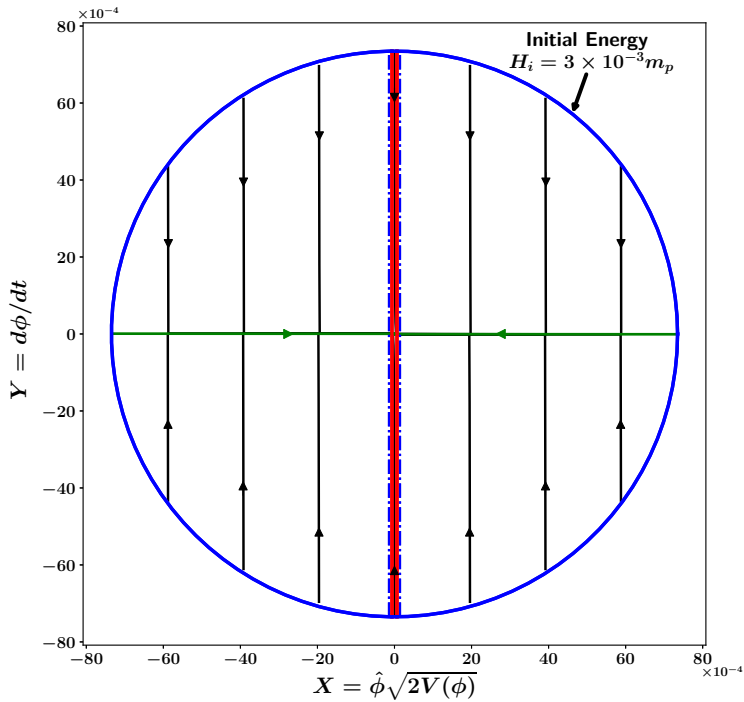}}
\captionsetup{
	justification=raggedright,
	singlelinecheck=false
}
\caption{This figure illustrates the phase-space  of chaotic inflation described by 
the potential (\ref{eqn:chaotic}). $Y$ ($=\dot{\phi}$) is plotted against 
$X$ ($=\hat{\phi}\sqrt{2V(\phi)}$) for 
different initial conditions all of which commence on the
circumference of a circle (blue) with radius $R=\sqrt{6}H_i/m_p$
corresponding to the initial energy scale $H_{i}=3\times 10^{-3}m_{p}$. 
($\hat{\phi}=\frac{\phi}{|\phi|}$ is the sign of field $\phi$.)
One finds that commencing
from the circle, the different inflationary trajectories rapidly
 converge towards one of the two inflationary separatrices 
(green horizontal lines).
After this, the scalar field moves towards the minimum of the potential $V(\phi)$
at $X=0, Y=0$. 
The thin vertical central band (red)
corresponds to the region in phase-space that {\em does not}
 lead to adequate inflation ($N_e < 60$). This central region
is shown greatly magnified in figure \ref{fig:chaotic1}.}	
\label{fig:chaotic}
\end{figure}

We first consider the potential \cite{linde83}
\begin{equation}
V(\phi)=\frac{1}{2}m^{2}\phi^{2}
\label{eqn:chaotic}
\end{equation}
where $m\simeq 5.97\times 10^{-6}m_{p}$ is assumed, in agreement with observations of the cosmic microwave background \cite{CMB,linde06} (see Appendix \ref{App:AppendixA}) . The 
generality of this model is studied by plotting the phase-space diagram ($Y$ vs $X$) and 
determining  the region of initial conditions which gives rise to $N_{e}\geq 60$. 
Equations \eqref{eqn: FR-eqn2 model},
\eqref{eqn:motion}, \eqref{eqn:efolds} have been solved numerically for different initial energy scales $H_{i}$.  The phase-space diagram corresponding to $H_{i}=3\times 10^{-3} m_{p}$ is shown in
figure \ref{fig:chaotic}.

\begin{figure}[htb]
\begin{center}
\includegraphics[width=0.55\textwidth]{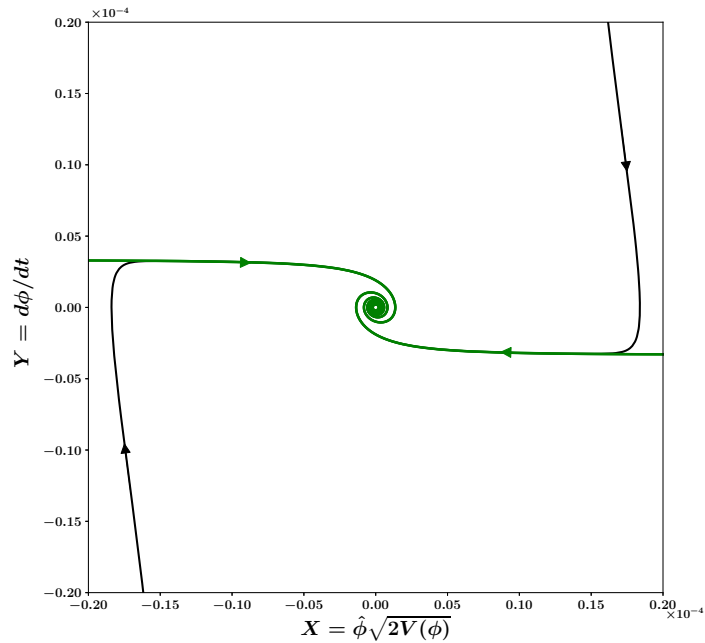}
\captionsetup{
	justification=raggedright,
	singlelinecheck=false
}
\caption{A zoomed-in view of the central region of figure \ref{fig:chaotic}.
Note that $\hat{\phi}=\frac{\phi}{|\phi|}$ gives the sign of $\phi$.
Inflationary
trajectories  (black) corresponding to different initial values of $\phi$ and $\dot{\phi}$, first converge onto the
slow-roll inflationary separatrices (green horizontal lines) 
before winding up to spiral towards the center.}
\label{fig:chaotic1}
\end{center}
\end{figure}

\begin{figure}[htb]
\begin{center}
\includegraphics[width=0.6\textwidth]{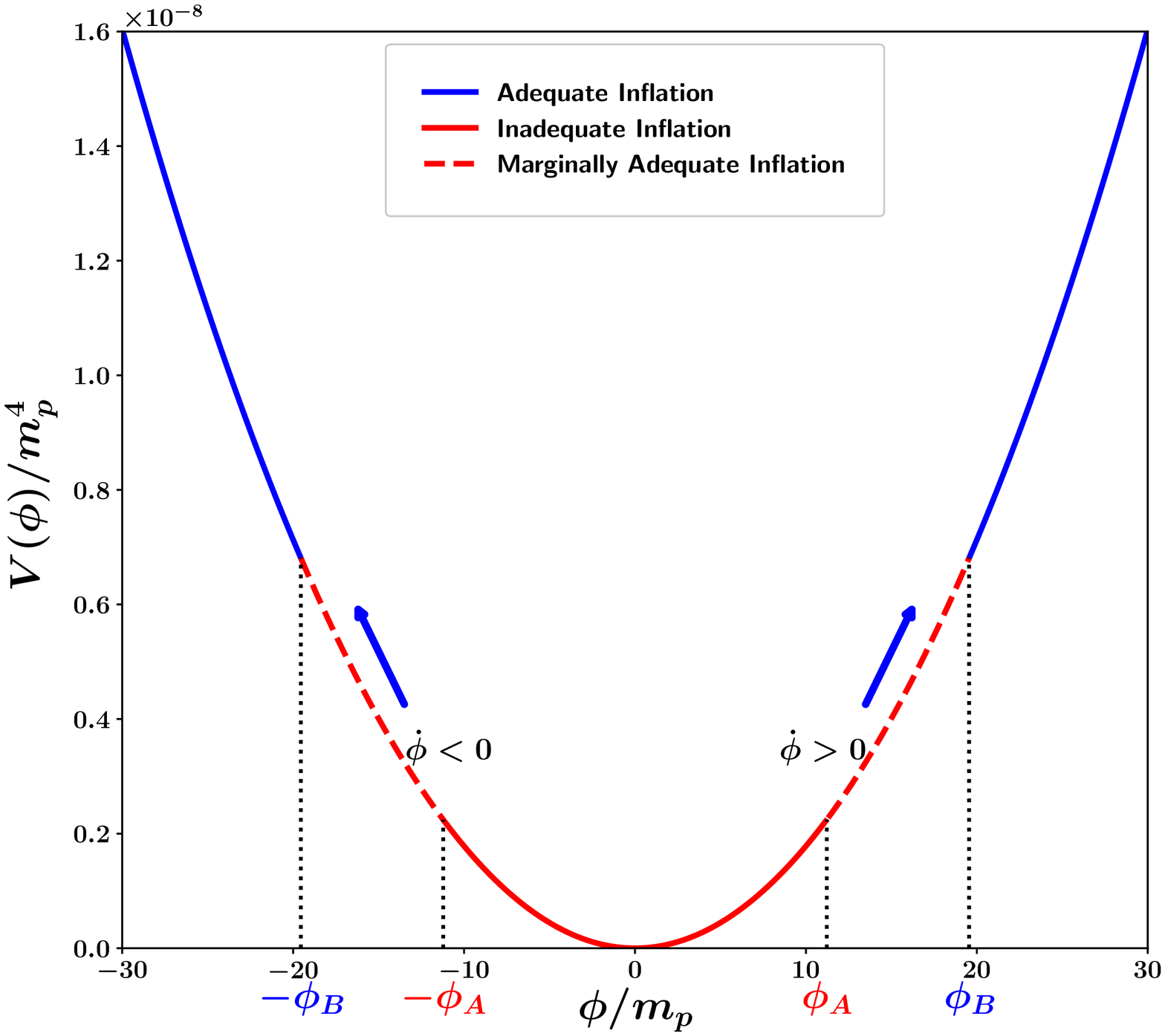}
\captionsetup{
	justification=raggedright,
	singlelinecheck=false
}
\caption{Initial field values, $\phi_i$,
which lead to adequate inflation with $N_e \geq 60$ (blue), marginally adequate (dashed red)  and 
inadequate (red) inflation are schematically shown
 for chaotic inflation (\ref{eqn:chaotic}). 
The blue 
lines represent regions of adequate inflation.
Initial values of $\phi_i$ lying in the blue region result in
adequate inflaion {\em irrespective} of the sign of $\dot\phi_i$.
 The red lines come in two styles: dashed/solid
and correspond to the following two possibilities:
(i) The solid red line represents initial values of $\phi_i$ for which inflation is never adequate 
irrespective of the direction of ${\dot\phi_i}$.
(ii) In the region shown by the dashed red line one gets 
 adequate inflation only when ${\dot\phi_i}$ is directed towards  increasing 
values of $V(\phi)$ (shown by blue arrows).
Note that only a small portion of the full potential is shown in this figure
which corresponds to the initial energy scale $H_{i}=3\times 10^{-3}m_{p}$.
}
%While for initial $\phi\in[-\phi_{B},-\phi_{A}]\cup [\phi_{A},\phi_{B}]$, represented by the red color dashed lines,  we get adequate inflation only when the initial velocity $y$ is upwards.}
\label{fig:chaotic2}
\end{center}
\end{figure}

To study the effect of different energy scales on inflation, 
we take different values of $R$ ($\equiv \sqrt{6}H_{i}/m_p$) and determine the 
range of initial values of 
$\phi$ that lead to adequate inflation with $N_e \geq 60$. 
(The initial value of ${\dot\phi}$ is
conveniently determined from the consistency relation (\ref{eqn:circle}).)
Our results are summarized in figure \ref{fig:chaotic2}.
The solid blue lines correspond to initial values, $\phi_i$, which 
always result in adequate inflation 
(irrespective of the sign of ${\dot\phi_i}$).
The dashed red lines corresponding to
$\phi_i\in \big[-\phi_{B},-\phi_{A}\big]\cup\big[\phi_{A},\phi_{B}\big]$, result in
 adequate inflation
only when ${\dot\phi_i}$ points in the direction of increasing $V(\phi)$ (represented by blue arrows).
Inadequate inflation is associated with the region
$\phi_{i}\in [-\phi_{A},\phi_{A}]$. If the initial scalar field value falls within
this region then
one does not get adequate inflation {\em irrespective of the sign} of 
${\dot\phi_i}$. This region is shown in  figure \ref{fig:chaotic2} by 
the solid red line.
The dependence of $\phi_{A}$ and $\phi_{B}$ on the initial energy scale 
$H_{i}$ is given in table  \ref{table:1}.

\par
\begin{figure}[htb]
\begin{center}
\includegraphics[width=0.55\textwidth]{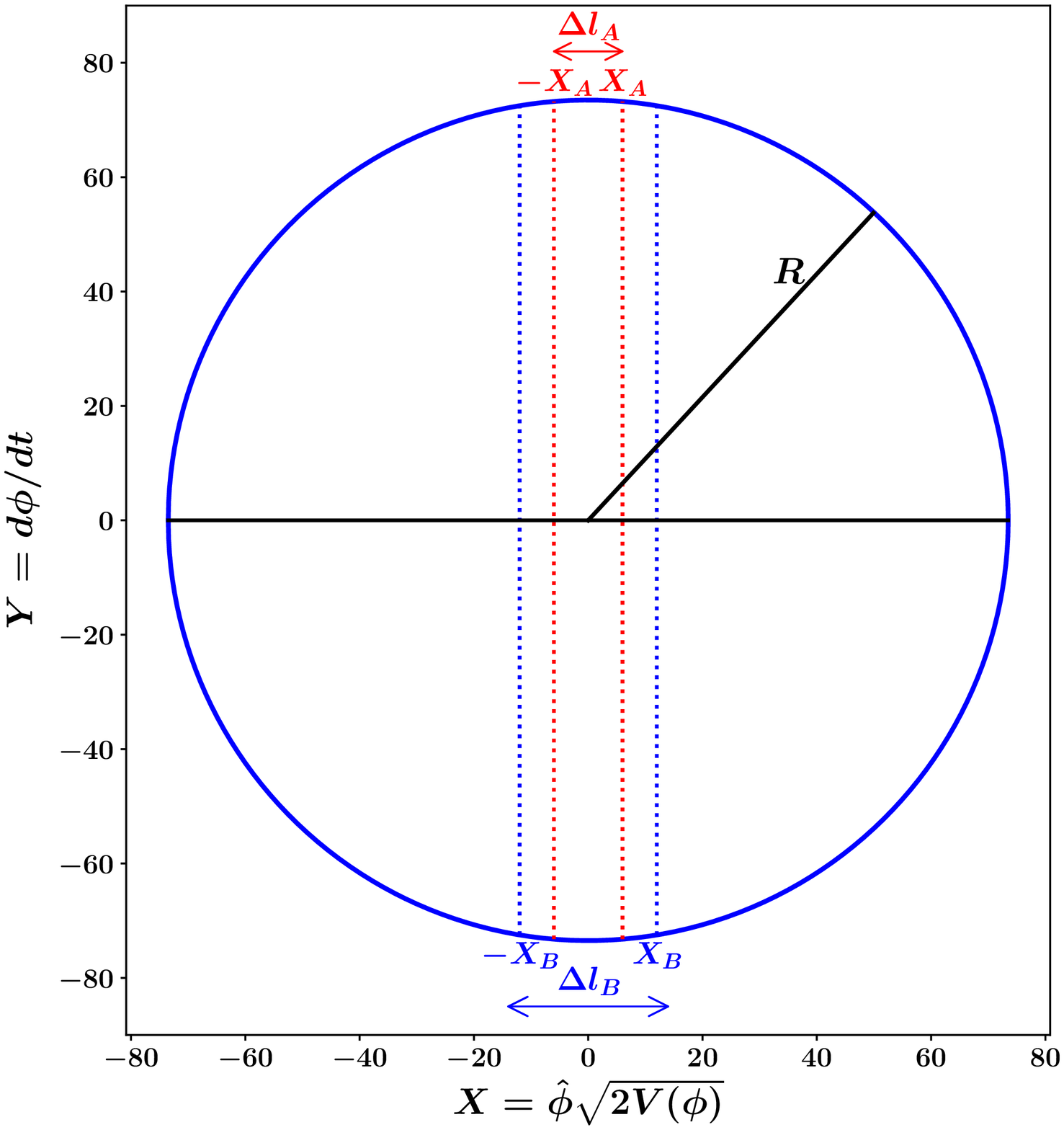}
\captionsetup{
	justification=raggedright,
	singlelinecheck=false
}
\caption{This figure illustrates how one can determine the degree of
adequate/inadequate inflation for power law potentials characterizing
 chaotic inflation and monodromy inflation. The fraction of initial conditions 
(corresponding to $\phi_{A}$ and $\phi_{B}$ in figure \ref{fig:chaotic2})
that leads either to inadequate inflation 
or marginally adequate inflation, is given by $2\frac{\Delta l_{A}}{l}$ and 
$2\frac{\Delta l_{B}}{l}$ respectively, where $l=2\pi R$. Adequate inflation 
with $N_e \geq 60$ is described by the fraction
$1 - 2\frac{\Delta l_{B}}{l}$. ($\hat{\phi}=\frac{\phi}{|\phi|}$ is the sign of field $\phi$.)}
\label{fig:chaotic3}
\end{center}
\end{figure}

 
 To determine the fraction of initial conditions that do not
 lead to adequate inflation (we call this `the degree of inadequate inflation'), 
we consider a uniform measure on the distribution of initial conditions for $Y_i (\equiv\dot{\phi}_i)$ 
and $X_i(\equiv \hat{\phi_i}\sqrt{2V(\phi_i)})$. These initial conditions are
described by a circle of
circumference $l=2\pi R$ with $R=\sqrt{6} H_{i}$ (in Planck units) which
is illustrated in figure \ref{fig:chaotic3}. 
The degree of inadequate inflation and marginally adequate inflation
(corresponding respectively
 to  $\phi_{A}$ and $\phi_{B}$ 
in
figure \ref{fig:chaotic2})
is
$2\frac{\Delta l_{A}}{l}$ and $2\frac{\Delta l_{B}}{l}$, where
$\Delta l_{A}$ and $\Delta l_{B}$ are illustrated in figure \ref{fig:chaotic3}.

 The dependence of  $\phi_{A}$, $\phi_{B}$ and $\frac{\Delta l_{A}}{l}$,
 $\frac{\Delta l_{B}}{l}$ on the commencement scale of inflation is shown in table \ref{table:1}. 
We see that the fraction of initial conditions that leads to inadequate inflation, $2\frac{\Delta l_{A}}{l}$,
 {\em decreases}
 with 
an increase in the initial energy scale $H_{i}$. This result is also 
illustrated in figures \ref{fig:scaling_mono1} and \ref{fig:scaling_mono2} where we compare chaotic inflation with monodromy inflation.

\begin{table}[htb]
\begin{center}
 \begin{tabular}{||c|c|c|c|c|c|c||} 
 \hline
 $H_{i}$ (in $m_{p}$) & $\phi_{A}$ (in $m_{p}$) & $\phi_{B}$ (in $m_{p}$) &  $2\frac{\Delta l_{A}}{l}$   & $2\frac{\Delta l_{B}}{l}$\\ [1ex] 
 \hline\hline
  $3\times 10^{-3}$ & $11.22$ & $19.55$ & $5.80 \times 10^{-3}$ & $1.01 \times 10^{-2}$ \\ [1.2ex] 
 \hline
 $3\times 10^{-2}$ &  $9.33$ & $21.38$ & $4.83 \times 10^{-4}$ & $1.11 \times 10^{-3}$ \\ [1.2ex] 
 \hline
 $3\times 10^{-1}$ &  $7.47$ & $23.27$ & $3.86 \times 10^{-5}$ & $1.20 \times 10^{-4}$ \\ [1.2ex] 
 \hline
\end{tabular}
\captionsetup{
	justification=raggedright,
	singlelinecheck=false
}
\caption{Dependence of $\phi_{A}$, $\phi_{B}$, $\frac{\Delta l_{A}}{l}$ and $\frac{\Delta l_{B}}{l}$  on the initial energy scale $H_{i}$ for quadratic chaotic inflation; see figure \ref{fig:chaotic3}. 
Here $l = 2\pi R \equiv 2\pi\sqrt{6} H_i/m_p$. 
Note that the fraction of initial conditions which leads to inadequate inflation, 
$2\frac{\Delta l_{A}}{l}$,
{\em decreases} as $H_{i}$ is increased. The same is true for the fraction of initial conditions 
giving rise to marginally adequate inflation, $2\frac{\Delta l_{B}}{l}$.
The fraction of initial conditions leading to adequate inflation, with $N_e \geq 60$, is given by
$1-2\frac{\Delta l_{B}}{l}$.
Thus inflation proves to be more general for larger values of the initial energy scale $H_{i}$,
since a larger initial region in phase space gives rise to adequate inflation with $N_e \geq 60$.}
\label{table:1}
\end{center} 
\end{table}

\subsection{Monodromy Inflation}
A straightforward extension of chaotic inflation, called
 Axion Monodromy, was discussed in \cite{monodromy1,monodromy2} in the context of String Theory\footnote{See \cite{kai1,kai2,kai3} for a field theory analogue of  monodromy inflation.} and tested against the CMB in \cite{Flauger:2009ab,CMB}. 
The potential for monodromy inflation, which contains a  monomial term along with axionic sinusoidal modulations, is given by 
\begin{equation}
V(\phi)=V_{0}\left|\frac{\phi}{m_{p}}\right|^p+\Lambda^4\left(\cos{\frac{\phi}{f}}-1\right)
\label{eqn:monodromy_axion}
\end{equation}
for $0 < p \leq 1$, where $f$ is the axion decay constant while $\Lambda$ is the scale corresponding to non-perturbative effects. In this paper our focus will be
on two values of $p$, namely $p=1,\frac{2}{3}$. (However our methods are very general and 
easily carry over to other values of $p$.)

Demanding the monotonicity of the potential (\ref{eqn:monodromy_axion}) one gets
\begin{equation}
b\left|\frac{\phi}{m_p}\right|^{1-p}\sin{\frac{\phi}{f}}<1~,
\label{eqn:monotonic}
\end{equation}
where $b=\frac{1}{p}\frac{\Lambda^4}{V_0}\frac{m_p}{f}$. Since $p\leq 1$ and
the observable period of inflation corresponds to $\phi>m_p$, the monotonicity 
condition \eqref{eqn:monotonic} implies $b<1$. Furthermore
for $b<1$, observational 
constraints\cite{Flauger:2009ab} from the CMB (combined with microphysical constraints from String Theory)  require $b\ll 1$ and $f\ll m_p$.
This implies that the amplitude of modulation 
$\Lambda^4 = V_0 \frac{f}{m_p}bp$ is much smaller than 
the monomial term, i.e $\Lambda^4 \ll V_0$. In other words,
the sinusoidal axionic term has a negligible effect on the background dynamics
so that, in an analysis of inflation, one can safely approximate 
the potential by its monomial term, namely\footnote{Note that for $b\geq 1$, the monodromy potential \eqref{eqn:monodromy_axion}
can have quite complicated but interesting features. However in this 
work we shall confine ourselves to the case $b<1$ as discussed in \cite{Flauger:2009ab,CMB}.}
 
\begin{equation}
V(\phi)=V_{0}\left|\frac{\phi}{m_{p}}\right|^p~.
\label{eqn:monodromy}
\end{equation} 
It is important to mention that for $p\leq 1$ the potentials (\ref{eqn:monodromy_axion}) as well as  (\ref{eqn:monodromy})  are
not differentiable at the origin. This might lead to  problems when $\phi$ 
rapidly oscillates around $\phi = 0$ after the end of inflation at $\phi=\phi_{end}$. We circumvent this problem by the following useful generalization\footnote{See \cite{mono_modify} for a similar modification of (\ref{eqn:monodromy}).} of (\ref{eqn:monodromy})
%\begin{equation}
%V(\phi)=V_1\left|\frac{\phi}{\phi_{0}}\right|^p\frac{1}{\big[1+\big(\frac{\phi_{0}}{\phi}\big)^{n}\big]^{\frac{2-p}{n}}}~,
%\label{eqn:monodromy1}
%\end{equation}
\begin{equation}
V(\phi) = V_1\left|\frac{\phi}{\phi_{c}}\right|^p W(\phi)
\label{eqn:monodromy1}
\end{equation}
where $W(\phi) = \left\lbrack 1 + \left (\frac{\phi_c}{\phi}\right )^n\right\rbrack^\frac{p-2}{n}$,
$V_1 = V_0 \left (\phi_c/m_p \right )^p$ and $n>1$ is an integer (we assume $n=4$
 in the ensuing analysis).
In this expression the value of $\phi_{c}$ is chosen  so that 
$V(\phi) \sim |\phi|^{p}$ for $|\phi|\gg|\phi_{c}|$ whereas
$V(\phi)\sim \phi^{2}$ for $|\phi|\ll|\phi_{c}|$. It is well known that inflation ends when the slow-roll parameter
$\epsilon$ in (\ref{eqn:slow_roll1}) grows to unity. 
Substituting (\ref{eqn:monodromy}) in (\ref{eqn:slow_roll1}) and setting $\epsilon \simeq 1$ 
one finds $\phi_{end}=\frac{p}{\sqrt{2}}m_{p}$ which can be used as a reference to set a value to $\phi_{c}$, namely $\phi_{c} \ll \phi_{end}$. One should note that the monomial part of the actual potentials of Axion Monodromy inflation  for $p=1,~\frac{2}{3}$ do not have cusps at the origin. For example 
for $p=1$, the monomial term has the form \cite{monodromy2} 
$V(\phi)=V_0\left(\sqrt{(\phi/m_p)^2+(\phi_c/m_p)^2}-(\phi_c/m_p)\right)$ which displays smooth 
quadratic behaviour near $\phi=0$. 
% where $\phi_c$ is equivalent to $\phi_0$ in (\ref{eqn:monodromy1}). 
Likewise, for a general value of the monodromy parameter
 p, it is convenient to modify the potential around 
$\phi=0$ without changing any of the results for the background dynamics 
as done in \cite{mono_modify}. Our introduction of a smoothing kernel $W$ in \eqref{eqn:monodromy1}
 follows a 
similar line of reasoning. 
It is important to emphasize that our results are quite insensitive to the values of $n$ and
$\phi_c$ in \eqref{eqn:monodromy1} provided $\phi_c\ll \phi_{end}$ and $n>1$.
%One should note that for $p\leq 1$ the potential (\ref{eqn:monodromy}) is
%not differentiable at the origin. This might lead to  problems with reheating
%since the latter usually occurs during rapid oscillations of $\phi$ around $\phi = 0$.
%We circumvent this problem by the following useful generalization of (\ref{eqn:monodromy})
%\begin{equation}
%V(\phi)=V_1\left|\frac{\phi}{\phi_{0}}\right|^p\frac{1}{\big[1+\big(\frac{\phi_{0}}{\phi}\big)^{n}\big]^{\frac{2-p}{n}}}~,
%\label{eqn:monodromy1}
%\end{equation}
%where $n>1$ is an integer (we have taken n=4).
  % In this expression the value of $\phi_{0}$ is chosen so that 
%$V(\phi) \sim |\phi|^{p}$ for $|\phi|\gg|\phi_{0}|$ whereas
%$V(\phi)\sim \phi^{2}$ for $|\phi|\ll|\phi_{0}|$. It is well known that inflation ends when the slow-roll parameter
%$\epsilon$ in (\ref{eqn:slow_roll1}) grows to unity. 
%Substituting (\ref{eqn:monodromy}) in (\ref{eqn:slow_roll1}) and setting $\epsilon \simeq 1$ 
%one finds $\phi_{end}=\frac{p}{\sqrt{2}}m_{p}$ which can be used to set a value
%to $\phi_{0}$, namely $\phi_{0} = \phi_{end}$. With this value of $\phi_{0}$

Next we proceed with a generality analysis for $p=1$ which will be followed by a similar analysis for $p=2/3$.
%(Note that our main results are quite insensitive to the value of $\phi_{0}$.)

\subsection*{Linear Monodromy Inflation}
\label{sec:mono_lin}
\begin{figure}[htb]
\begin{center}
\includegraphics[width=0.6\textwidth]{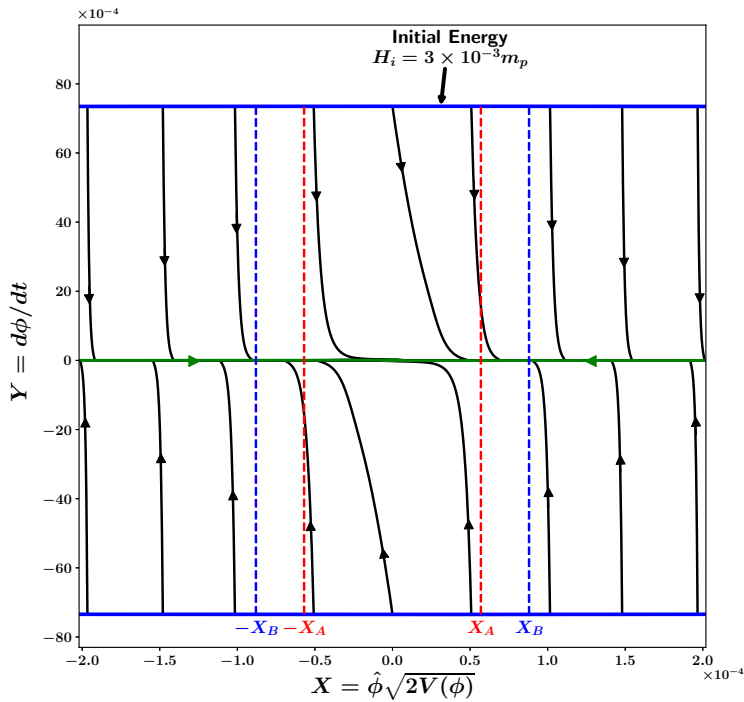}
\captionsetup{
	justification=raggedright,
	singlelinecheck=false
}
\caption{This figure shows a portion of the phase-space  of monodromy inflation 
$V \propto |\phi|$. 
The variable $Y$ ($=\dot{\phi}$) is plotted against $X$ ($=\hat{\phi}\sqrt{2V(\phi)}$).
($\hat{\phi}=\frac{\phi}{|\phi|}$ is the sign of field $\phi$.)
The initial conditions are specified 
 on arcs which form the blue colored boundary. 
Note that these arcs correspond to 
 a {\em very small portion} of the full `initial conditions' circle $R$, and therefore appear to be
straight lines. In this analysis we assume 
$R =\sqrt{6} H_{i}/m_{p}$, with $H_i = 3\times 10^{-3} m_p$. We find that, commencing at the boundary, most solutions quickly converge to
the two slow-roll inflationary separatrices (green horizontal lines) before travelling to the 
origin where $\lbrace \dot\phi, \phi\rbrace = \lbrace 0,0\rbrace$. A blow up 
of the central portion of this figure is shown in figure \ref{fig:monodromy6}.}
\label{fig:monodromy5}
\end{center}
\end{figure}

Consider first the linear potential
\begin{equation}
V(\phi)=V_{0}\left|\frac{\phi}{m_{p}}\right|
\label{eqn:mono}
\end{equation}
where $V_{0}\simeq 1.97\times 10^{-10}m_{p}^{4}$ is in agreement with the CMB \cite{CMB} (see Appendix \ref{App:AppendixA}).
The phase-space diagram for this potential, shown in figure \ref{fig:monodromy5}, was obtained by solving
the equations \eqref{eqn: FR-eqn2 model},
\eqref{eqn:motion}, \eqref{eqn:efolds}, \eqref{eqn:mono}
numerically, for the initial energy scale $H_{i}=3\times 10^{-3}m_{p}$. 

\begin{figure}[htb]
\begin{center}
\includegraphics[width=0.55\textwidth]{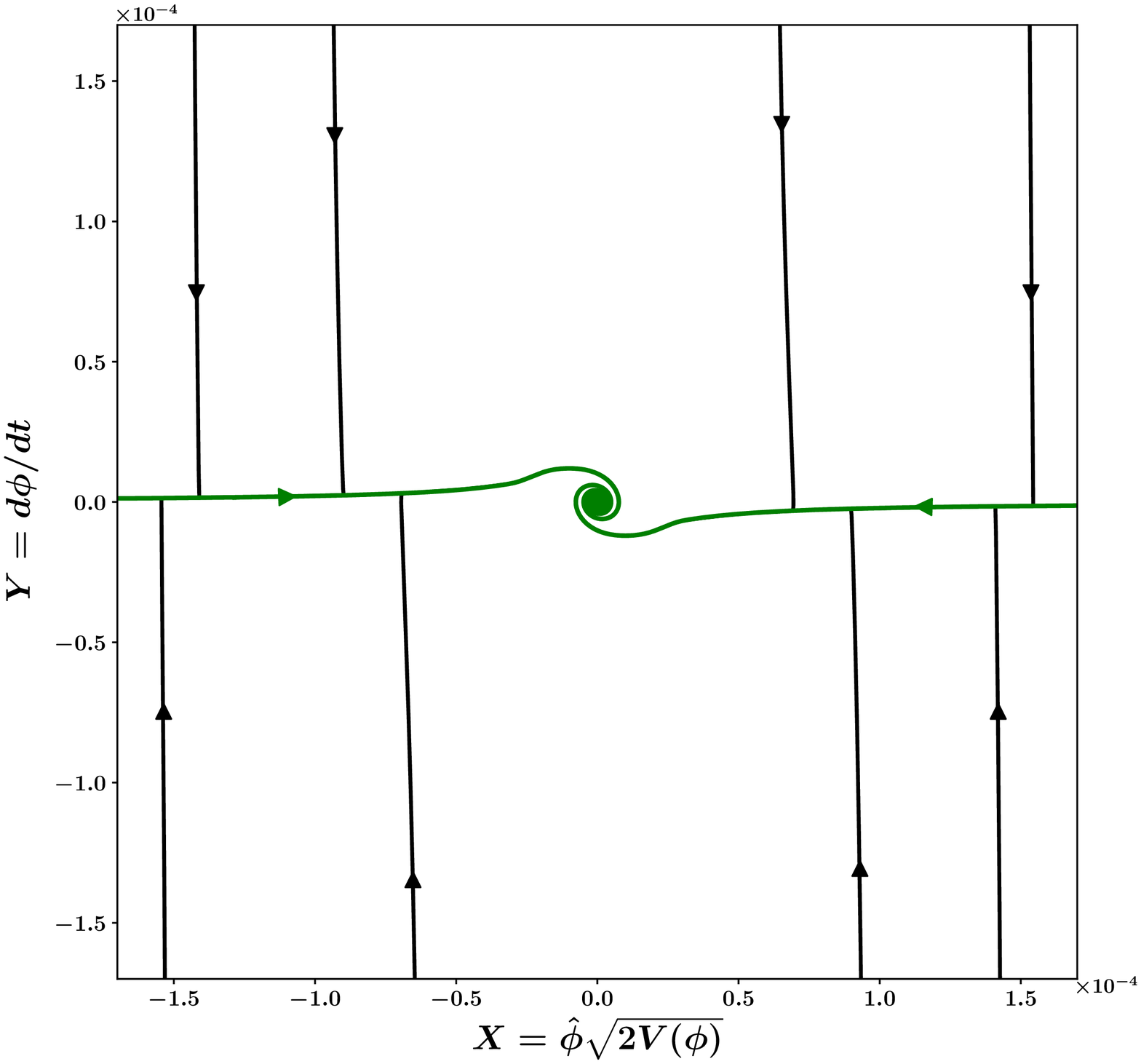}
\captionsetup{
	justification=raggedright,
	singlelinecheck=false
}
\caption{A zoomed-in view of the phase-space of monodromy inflation with $V \propto |\phi|$.
Note that scalar field trajectories initially converge towards the
slow-roll inflationary separatrices (horizontal green lines), moving from there towards $\phi = 0$, where the
field oscillates.}
\label{fig:monodromy6}
\end{center}
\end{figure}

\begin{figure}[htb]
\begin{center}
\includegraphics[width=0.6\textwidth]{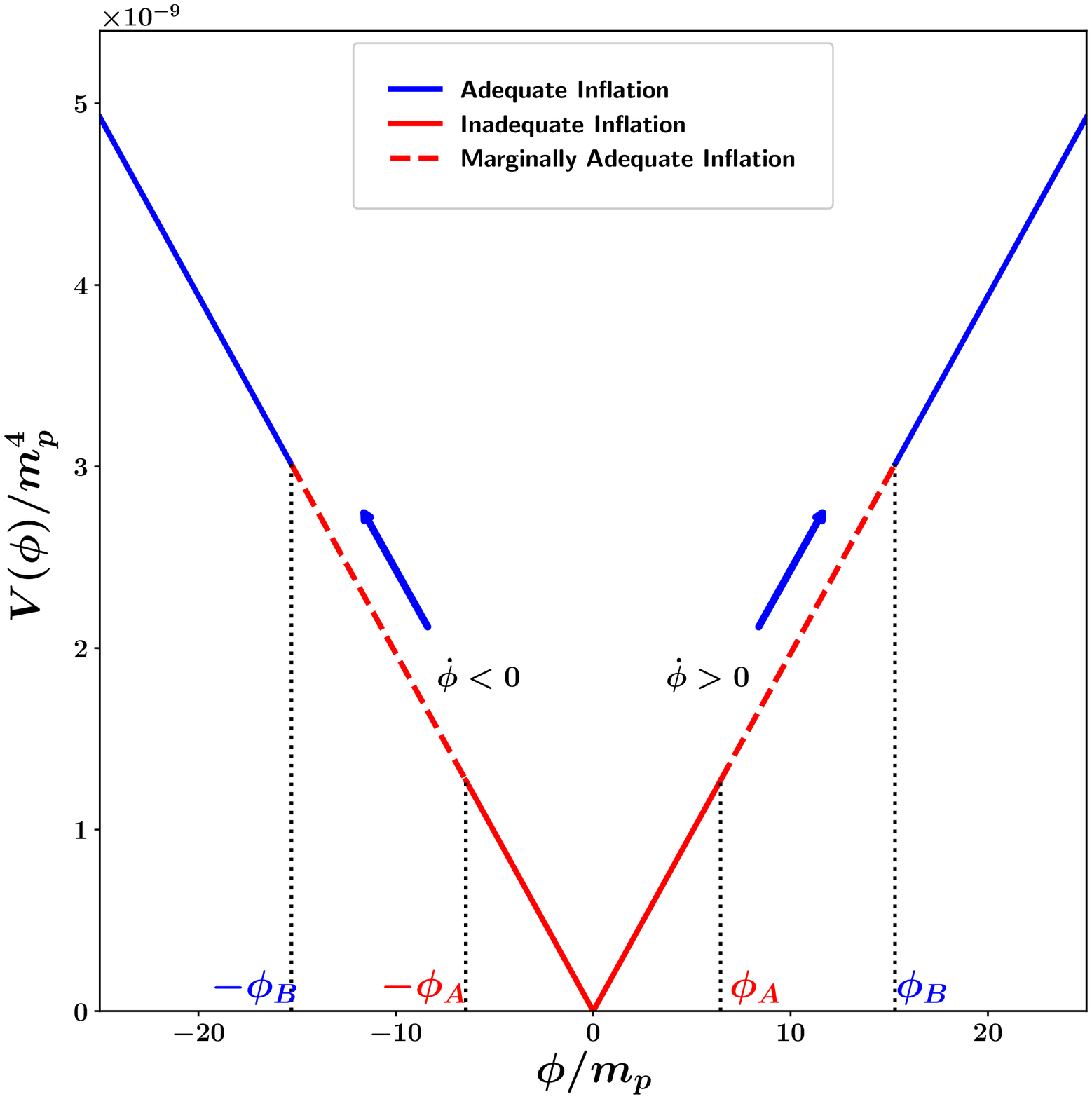}
\captionsetup{
	justification=raggedright,
	singlelinecheck=false
}
\caption{
This figure schematically shows initial field values which result in adequate
inflation with $N_e \geq 60$ (blue), marginally adequate (dashed red)  and inadequate
inflation (red) for the monodromy potential $V \propto |\phi|$.
 The initial energy scale is $H_{i}=3\times 10^{-3}m_{p}$.
As earlier, blue
lines represent regions of adequate inflation. The red lines come in two styles: dashed/solid
and correspond to the two possible initial directions of ${\dot\phi}$.
The solid red line represents initial values of $\phi$ for which inflation is never adequate
irrespective of the direction of ${\dot\phi_i}$.
In the region shown by the dashed line one gets
 adequate inflation only when ${\dot\phi_i}$ points in the direction (shown by blue arrows) of increasing $V(\phi)$.
Note that only a small portion of the full potential is shown in this figure.
}
\label{fig:mono3}
\end{center}
\end{figure}

Initial values of $\phi$ that  lead to adequate or inadequate inflation  are schematically shown in 
figure \ref{fig:mono3}. Inadequate inflation arises when the scalar field originates in the region
$\phi_{i}\in [-\phi_{A},\phi_{A}]$,  shown by solid red line.
Blue lines represent initial field values  $\phi_i\in \big(-\phi_{m},-\phi_{B}\big)\cup\big(\phi_{B},\phi_{m}\big)$,
which always result in adequate inflation. Note that $\phi_{m}$ is the 
maximum allowed value of $\phi_i$ for a given
initial energy scale, as determined from the consistency equations (\ref{eqn: FR-eqn1 model}), (\ref{eqn:circle}).
Initial conditions $\phi_i\in \big[-\phi_{B},-\phi_{A}\big]\cup\big[\phi_{A},\phi_{B}\big]$, shown by dashed red lines, lead to
 adequate inflation only when ${\dot\phi_i}$ points  in the direction (shown by blue arrows) of increasing $V(\phi)$ .
The dependence of $\phi_{A}$ and $\phi_{B}$ on the initial energy scale $H_{i}$ is shown in table \ref{table:2}.

\par
The values of $2\frac{\Delta l_{A}}{l}$ and $2\frac{\Delta l_{B}}{l}$ in table \ref{table:2}
have been determined by
assuming a uniform distribution of initial values of 
$Y=\dot{\phi}$ and $X=\hat{\phi}\sqrt{2V(\phi)}$ on the circular boundary (\ref{eqn:circle}).  
We find that $2\frac{\Delta l_{A}}{l}$ and $2\frac{\Delta l_{B}}{l}$ decrease 
with an increase in $H_{i}$, as expected.
\begin{table}[htb]
\begin{center}
 \begin{tabular}{||c|c|c|c|c|c|c||} 
 \hline
 $H_{i}$ (in $m_{p}$) & $\phi_{A}$ (in $m_{p}$) & $\phi_{B}$ (in $m_{p}$) &  $2\frac{\Delta l_{A}}{l}$   & $2\frac{\Delta l_{B}}{l}$\\ [1ex] 
 \hline\hline
  $3\times 10^{-3}$ & $6.45$ & $15.29$ & $4.37 \times 10^{-3}$ & $6.73 \times 10^{-3}$ \\ [1.2ex] 
 \hline
 $3\times 10^{-2}$ &  $4.58$ & $17.18$ & $3.68 \times 10^{-4}$ & $7.13 \times 10^{-4}$ \\ [1.2ex] 
 \hline
 $3\times 10^{-1}$ &  $2.69$ & $19.06$ & $2.84 \times 10^{-5}$ & $7.51 \times 10^{-5}$ \\ [1.2ex] 
 \hline
\end{tabular}
\captionsetup{
	justification=raggedright,
	singlelinecheck=false
}
\caption{Dependence of $\phi_{A}$, $\phi_{B}$, $\frac{\Delta l_{A}}{l}$ and $\frac{\Delta l_{B}}{l}$  on the 
initial energy scale $H_{i}$ for monodromy inflation $V \propto |\phi|$. Here $l = 2\pi R \equiv 2\pi\sqrt{6} H_i/m_p$
and $\frac{\Delta l_{A}}{l}$, $\frac{\Delta l_{B}}{l}$ were defined in figure \ref{fig:chaotic3}.
Note that the fraction of initial conditions which leads to inadequate inflation,
$2\frac{\Delta l_{A}}{l}$,
{\em decreases} as $H_{i}$ is increased. The same is true for the fraction of initial conditions
giving rise to marginally adequate inflation, $2\frac{\Delta l_{B}}{l}$.
The fraction of initial conditions leading to adequate inflation, with $N_e \geq 60$, is given by
$1-2\frac{\Delta l_{B}}{l}$.
Thus inflation proves to be more general for larger values of the initial energy scale $H_{i}$,
since a larger initial region in phase space gives rise to adequate inflation with $N_e \geq 60$.}
\label{table:2}
\end{center} 
\end{table}

\subsection*{Fractional Monodromy Inflation}
\label{sec:mono_frac}
\begin{figure}[htb]
\begin{center}
\includegraphics[width=0.6\textwidth]{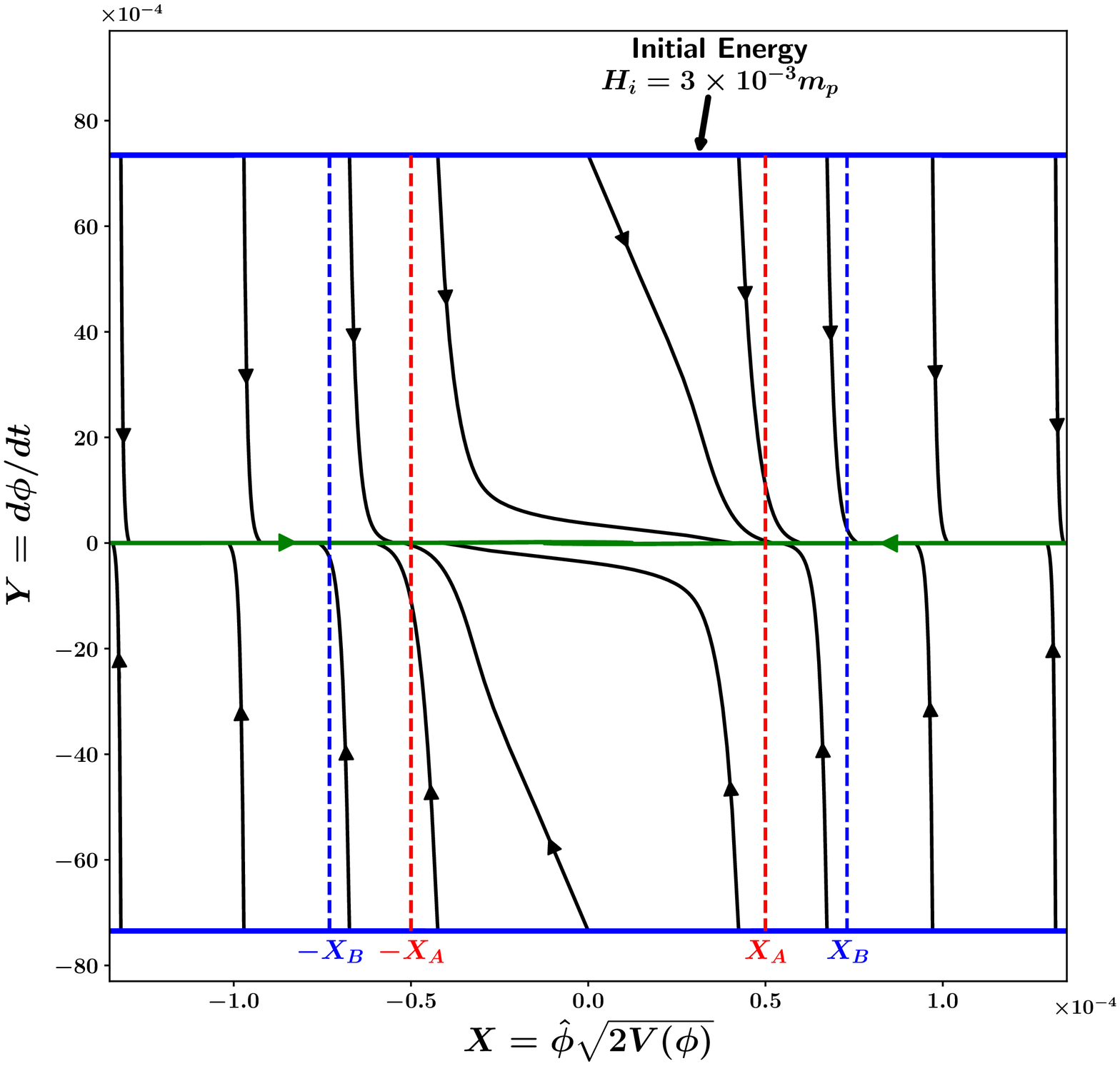}
\captionsetup{
	justification=raggedright,
	singlelinecheck=false
}
\caption{This figure shows a portion of the phase-space  of monodromy inflation with $V \propto |\phi|^{2/3}$.
The variable $Y$ ($=\dot{\phi}$) is plotted against $X$ ($=\hat{\phi}\sqrt{2V(\phi)}$).
($\hat{\phi}=\frac{\phi}{|\phi|}$ is the sign of field $\phi$.)
Initial conditions are specified 
 on arcs which form the blue colored boundary. Note that since these arcs correspond to 
 a {\em very small portion} of the full `initial conditions' circle $R$, they appear to be
straight lines. As in the previous analysis for chaotic inflation we again assume
$R =\sqrt{6} H_{i}/m_{p}$, with $H_i= 3\times 10^{-3} m_p$. One finds that, 
commencing at the boundary, most solutions quickly converge to 
the two slow-roll inflationary separatrices (green horizontal lines) before travelling to the 
origin where $\lbrace \dot\phi, \phi\rbrace = \lbrace 0,0\rbrace$. A blow up 
of the central portion of this figure is shown in figure \ref{fig:monodromy2}.}
\label{fig:monodromy1}
\end{center}
\end{figure}

Next we consider 
\begin{equation}
V(\phi)=V_{0} \left | \frac{\phi}{m_{p}} \right | ^{\frac{2}{3}}
\label{eqn:monodromy2}
\end{equation}
 where CMB constraints imply $V_{0}=3.34\times 10^{-10}m_p^{4}$ \cite{CMB} (see Appendix \ref{App:AppendixA}). 
The phase-space diagram for this potential, shown in figure \ref{fig:monodromy1}, was obtained by solving 
the equations \eqref{eqn: FR-eqn2 model},
\eqref{eqn:motion}, \eqref{eqn:efolds}
numerically for the initial energy scale $H_{i}=3\times 10^{-3}m_{p}$. 

\begin{figure}[htb]
\begin{center}
\includegraphics[width=0.48\textwidth]{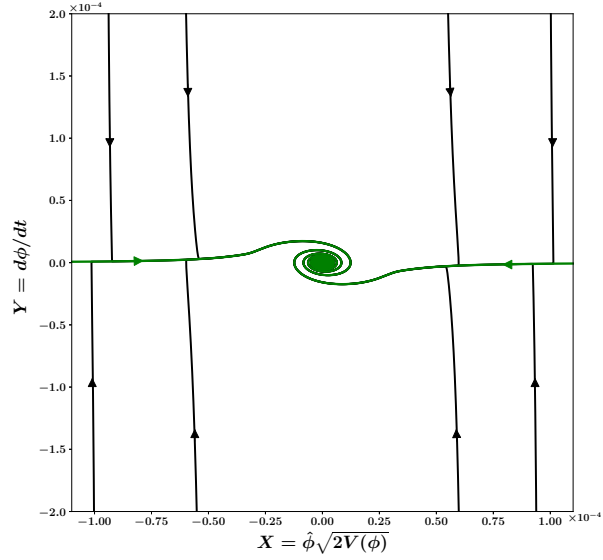}
\captionsetup{
	justification=raggedright,
	singlelinecheck=false
}
\caption{A zoomed-in view of the phase-space of monodromy inflation with $V \propto |\phi|^{2/3}$.
One notes that the motion of the scalar field is initially towards the
slow-roll inflationary separatrices (horizontal green lines) and from there towards $\phi = 0$, where the
field oscillates.}
\label{fig:monodromy2}
\end{center}
\end{figure}

\begin{figure}[htb]
\begin{center}
\includegraphics[width=0.6\textwidth]{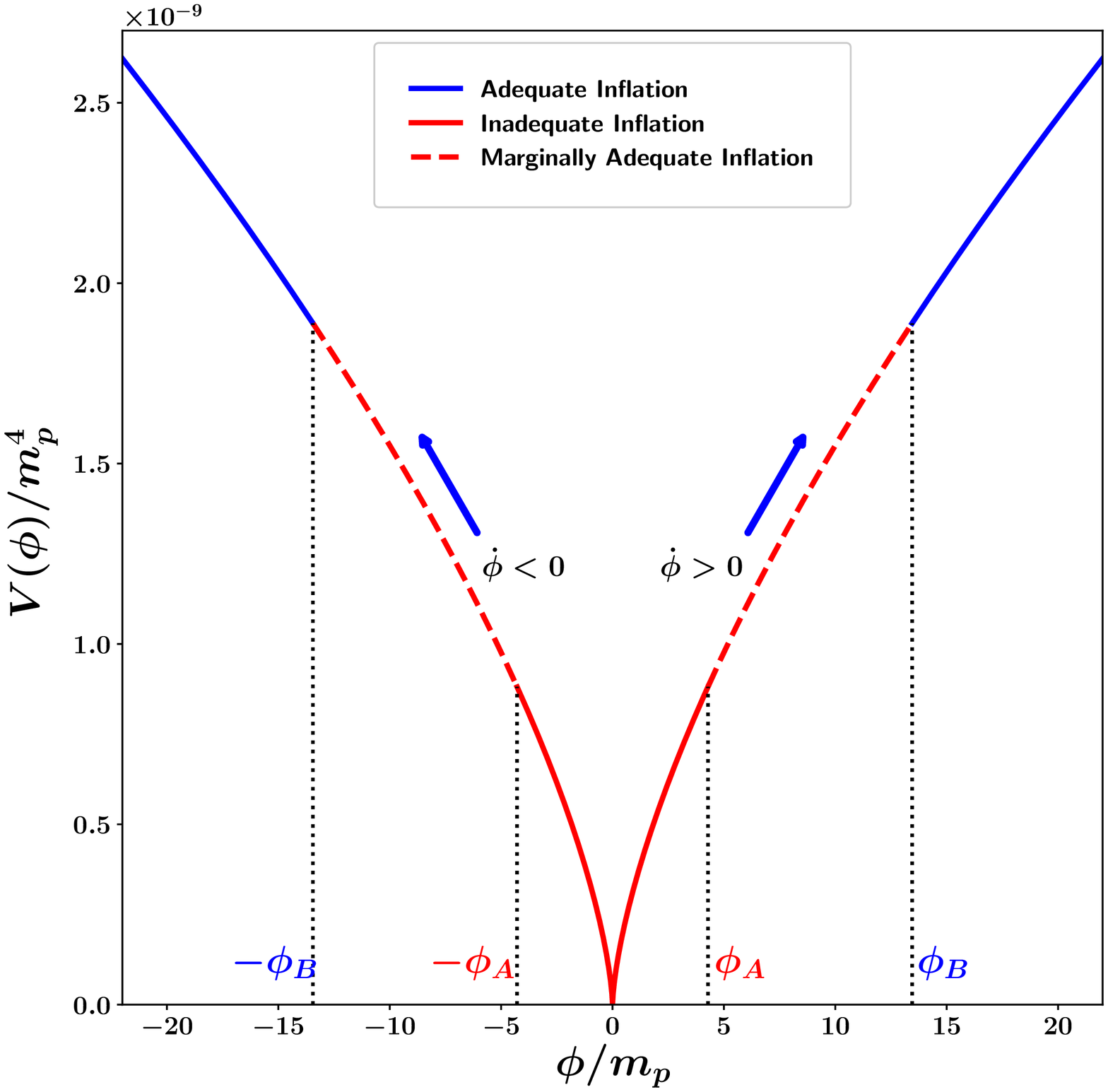}
\captionsetup{
	justification=raggedright,
	singlelinecheck=false
}
\caption{This figure schematically shows initial field values which result in adequate
inflation with $N_e \geq 60$ (blue), marginally adequate (dashed red)  and inadequate 
inflation (red) for the monodromy potential (\ref{eqn:monodromy2}).
 The initial energy scale is $H_{i}=3\times 10^{-3}m_{p}$. 
As earlier, blue
lines represent regions of adequate inflation. The red lines come in two styles: dashed/solid
and correspond to the two possible initial directions of ${\dot\phi}$.
The solid red line represents initial values of $\phi$ for which inflation is never adequate 
irrespective of the direction of ${\dot\phi_i}$.
In the region shown by the dashed red line one gets  
 adequate inflation only when ${\dot\phi_i}$ points in the direction (shown by blue arrows) of increasing $V(\phi)$.
Note that only a small portion of the full potential is shown in this figure.
}
\label{fig:monodromy3}
\end{center}
\end{figure}

Initial values of $\phi$ that  lead to adequate or inadequate inflation  are schematically shown in figure \ref{fig:monodromy3}. Inadequate inflation arises when the scalar field originates in the region 
$\phi_{i}\in [-\phi_{A},\phi_{A}]$,  shown by solid red lines. 
Blue lines represent initial field values  $\phi_i \in \big(-\phi_{m},-\phi_{B}\big)\cup\big(\phi_{B},\phi_{m}\big)$, 
which always result in adequate inflation. Note that $\phi_{m}$ is the maximum allowed value of $\phi$ for a given 
initial energy scale, as determined from the consistency equations (\ref{eqn: FR-eqn1 model}), (\ref{eqn:circle}). 
The initial conditions $\phi_i \in \big[-\phi_{B},-\phi_{A}\big]\cup\big[\phi_{A},\phi_{B}\big]$, shown by dashed red lines, lead to
 adequate inflation only when ${\dot\phi}_i$ points in the direction (shown by blue arrows) of increasing $V(\phi)$.
We refer to this as marginally adequate inflation. 
The dependence of $\phi_{A}$ and $\phi_{B}$ on the initial energy scale $H_{i}$ is shown in Table \ref{table:3}.

As in the case of chaotic inflation, we determine the fraction of initial conditions that do not
 lead to adequate inflation (the degree of inadequate inflation), by assuming a uniform distribution of
initial values of $Y=\dot{\phi}$ and $X=\hat{\phi}\sqrt{2V(\phi)}$ on the circular boundary (\ref{eqn: FR-eqn1 model}),
(\ref{eqn:circle}) with $V(\phi)$ given by (\ref{eqn:monodromy2}).
Our results are given in table \ref{table:3}. As was the case for quadratic chaotic inflation, we once more
 find that $\frac{\Delta l_{A}}{l}$ 
and $\frac{\Delta l_{B}}{l}$ decrease with an increase in $H_{i}$; see table \ref{table:3}, figures \ref{fig:scaling_mono1}
 and \ref{fig:scaling_mono2}.

\begin{table}[htb]
\begin{center}
 \begin{tabular}{||c|c|c|c|c|c|c||} 
 \hline
 $H_{i}$ (in $m_{p}$) & $\phi_{A}$ (in $m_{p}$) & $\phi_{B}$ (in $m_{p}$) &  $2\frac{\Delta l_{A}}{l}$   & $2\frac{\Delta l_{B}}{l}$\\ [1ex] 
 \hline\hline
  $3\times 10^{-3}$ & $4.29$ & $13.45$ & $3.64 \times 10^{-3}$ & $5.33 \times 10^{-3}$ \\ [1.2ex] 
 \hline
 $3\times 10^{-2}$ &  $2.41$ & $15.33$ & $3.0 \times 10^{-4}$ & $5.56 \times 10^{-4}$ \\ [1.2ex] 
 \hline
 $3\times 10^{-1}$ &  $0.61$ & $17.22$ & $2.08 \times 10^{-5}$ & $5.78 \times 10^{-5}$ \\ [1.2ex] 
 \hline
\end{tabular}
\captionsetup{
	justification=raggedright,
	singlelinecheck=false
}
\caption{Dependence of $\phi_{A}$, $\phi_{B}$, $\frac{\Delta l_{A}}{l}$ and $\frac{\Delta l_{B}}{l}$  on the initial energy scale $H_{i}$ for monodromy inflation with $p=\frac{2}{3}$. Here $l = 2\pi R \equiv 2\pi\sqrt{6} H_i/m_p$
and $\frac{\Delta l_{A}}{l}$, $\frac{\Delta l_{B}}{l}$ were defined in figure \ref{fig:chaotic3}. Note that the fraction of initial conditions which leads to inadequate inflation,
$2\frac{\Delta l_{A}}{l}$,
{\em decreases} as $H_{i}$ is increased. The same is true for the fraction of initial conditions
giving rise to marginally adequate inflation, $2\frac{\Delta l_{B}}{l}$.
The fraction of initial conditions leading to adequate inflation, with $N_e \geq 60$, is given by
$1-2\frac{\Delta l_{B}}{l}$.
Thus inflation proves to be more general for larger values of the initial energy scale $H_{i}$,
since a larger initial region in phase space gives rise to adequate inflation with $N_e \geq 60$.}
\label{table:3}
\end{center} 
\end{table}

\subsection{Comparison of power law potentials}

\begin{figure}[htb]
\centering
\subfigure[]{
\includegraphics[width=0.49\textwidth]{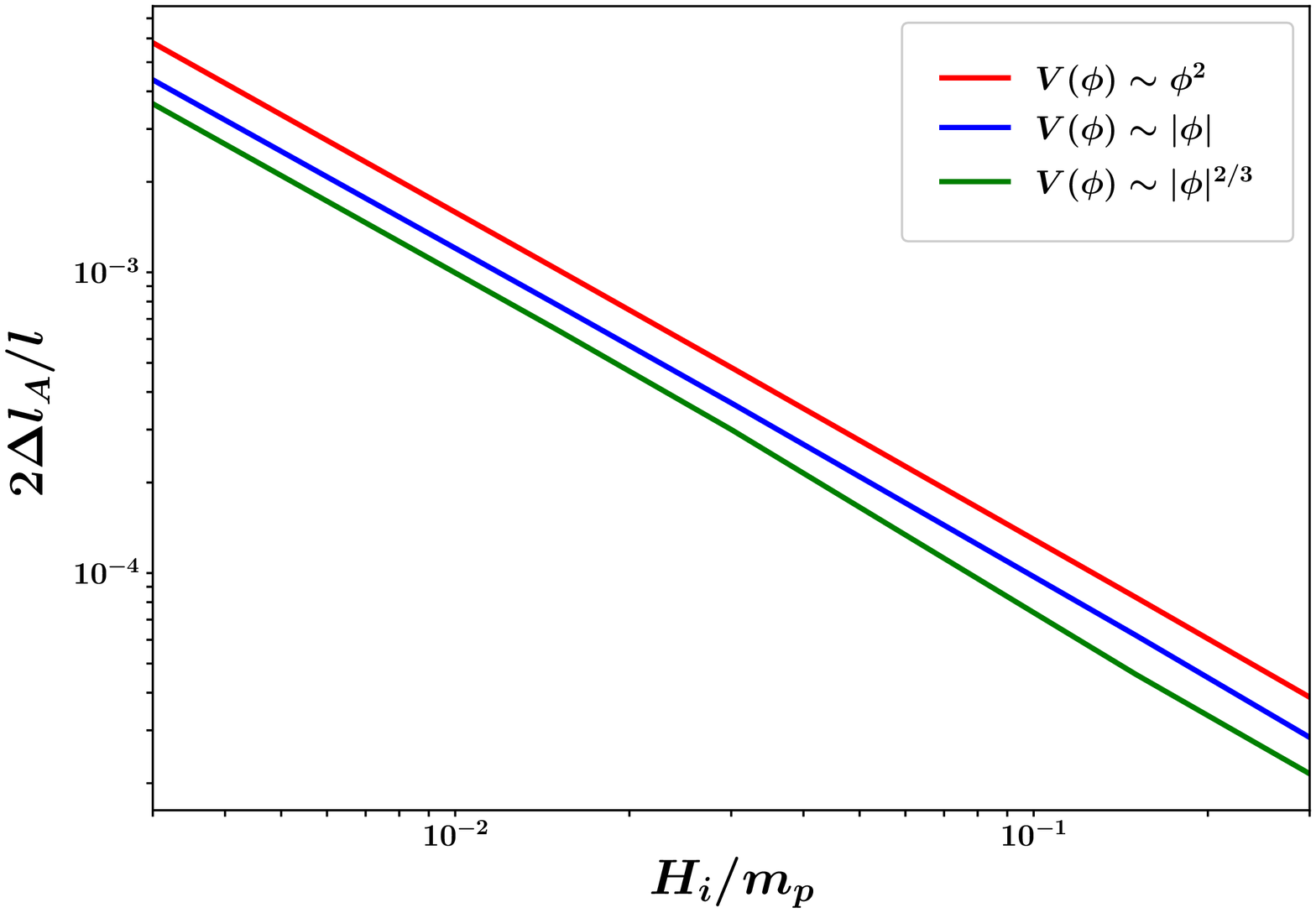}\label{fig:scaling_mono1}}
\subfigure[]{
\includegraphics[width=0.49\textwidth]{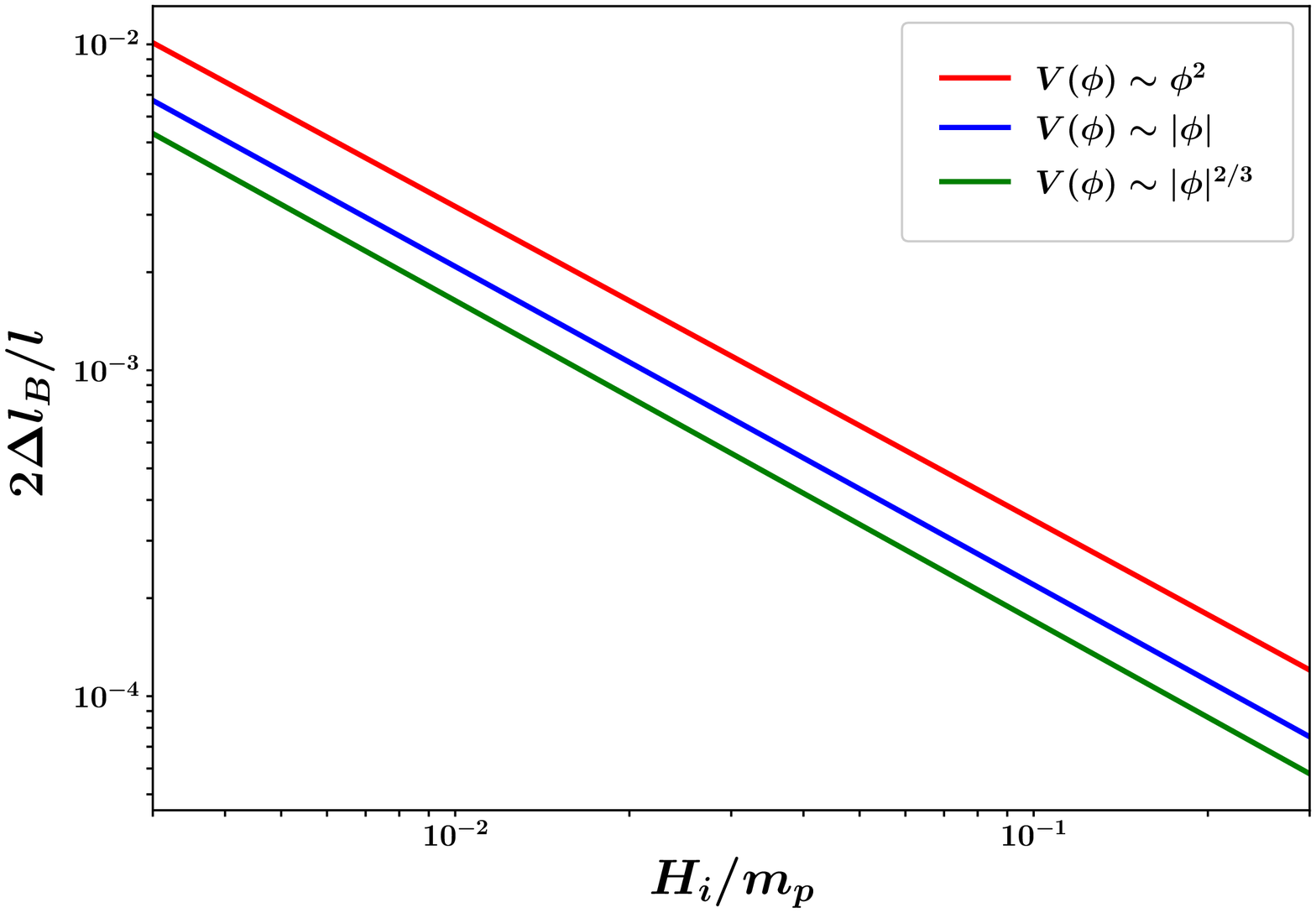}\label{fig:scaling_mono2}}
\captionsetup{
	justification=raggedright,
	singlelinecheck=false
}
\caption{This figure shows the fraction of initial conditions that  leads to \textbf{(a)} inadequate inflation, 
 $2\frac{\Delta l_{A}}{l}$ and \textbf{(b)} marginally adequate inflation,
 $2\frac{\Delta l_{B}}{l}$, plotted against the initial energy scale of inflation, 
$H_{i}$. For the definition of $\frac{\Delta l_{A}}{l}$ and $\frac{\Delta l_{B}}{l}$, see figure \ref{fig:chaotic3}.
 The red curve shows results for $V \propto \phi^2$ while the blue and green curves represent 
monodromy potentials with  $V \propto |\phi|, |\phi|^{2/3}$ respectively.
The decrease in $2\frac{\Delta l_{A}}{l}$ and $2\frac{\Delta l_{B}}{l}$,
which accompanies an increase in $H_{i}$ is indicative of the fact that the set of initial conditions which
give rise to adequate inflation (with $N_e \geq 60$) increases with the energy scale of inflation, $H_{i}$.
This figure also demonstrates that 
inflation is sourced by a larger set of initial conditions for the monodromy potential
$V \propto |\phi|^\frac{2}{3}$,
which is followed by $V \propto |\phi|$ and finally $V \propto \phi^2$.} 
\label{fig:scaling_mono}
\end{figure}

 In this subsection we compare the generality of inflation for the power law family of potentials,
$V \propto |\phi|^p$, 
by plotting the fraction of initial conditions that {\em do not lead to} adequate inflation 
($2\frac{\Delta l_{A}}{l}$ and $2\frac{\Delta l_{B}}{l}$) in figures \ref{fig:scaling_mono1} and
 \ref{fig:scaling_mono2};
 also see
tables 1-3. % table 2 and table 3. 
These figures demonstrate that
the set of initial conditions which
give rise to adequate inflation (with $N_e \geq 60$) increases with the energy scale of inflation, $H_{i}$.
We also find that inflation is sourced by a larger set of initial conditions for the monodromy potential 
$V \propto |\phi|^\frac{2}{3}$,
which is followed by $V \propto |\phi|$ and $V \propto \phi^2$ respectively.
Finally we draw attention to the fact that our conclusions remain
unchanged if we determine the degree of inflation by a different measure such as
$\frac{\Delta\phi_{A}}{\phi_{max}}$ and $\frac{\Delta\phi_{B}}{\phi_{max}}$, where $\phi_{max}$ is the maximum allowed value of $\phi$ for a given inflationary energy scale.

\section{ Higgs Inflation} 
\label{sec:higgs}

It would undoubtedly be interesting if 
inflation could be realized within the context of the Standard Model ($SM$) 
of particle physics. Since the $SM$ has only a single scalar degree of freedom, namely the Higgs field, one can ask whether the Higgs field 
(\ref{eqn:higgs_pot1}) can source inflation. 
Unfortunately the self-interaction coupling of the Higgs field, 
$\lambda$ in (\ref{eqn:higgs_pot1}), is far too large to be
consistent with the small amplitude of scalar fluctuations observed
by the cosmic microwave background 
\cite{CMB}. %Hence the $SM$  Higgs field, with canonical Lagrangian and minimally coupled to gravity, has to be a sub-dominant component in the universe during inflation in order to explain observation.

%However, a remarkable feature of the Higgs field is that it can source inflation provided we relax one of the following properties : minimal coupling to gravity and canonical Lagrangian. As  first demonstrated in \cite{higgs1}, inflation can be sourced by the $SM$ Higgs potential if  the Higgs field is assumed to couple non-minimally to
This situation can however be remedied if either of the following possibilties
is realized:
(i) the Higgs couples non-minimally to gravity, or
(ii) the Higgs field is described by a non-canonical Lagrangian\footnote{Another means of
reconciling the $\frac{1}{4}\lambda\phi^4$ ($\lambda \sim 0.1$) potential with observations is
through a field derivative coupling with the Einstein tensor of the form
$G^{\mu\nu}\partial_\mu\partial_\nu\phi/M^2$. This approach has been discussed in
\cite{field_derivative}.}

Indeed, as first demonstrated in \cite{higgs1}, inflation can be sourced by the $SM$ Higgs potential if  the Higgs field is assumed to couple non-minimally to
the Ricci scalar.
The resultant inflationary model provides a good fit to observations and has been extensively developed and examined in \cite{higgs1,higgs2,higgs3,higgs4,higgs5}. A different means of sourcing Inflation through the Higgs field was discussed in \cite{sanil_varun} where it was shown that the $SM$ Higgs potential with a non-canonical kinetic term fits the CMB data very well  by accounting for the currently observed values of the scalar spectral index $n_{_{S}}$ and the tensor-to-scalar ratio $r$. We shall  proceed to study Higgs inflation first in the non-minimal framework in section \ref{sec:higgs_non_min} followed by the same in the  non-canonical framework in section  \ref{sec:higgs_non_can}.

\subsection{Initial conditions for Higgs Inflation in the non-minimal framework}
\label{sec:higgs_non_min}

Inflation sourced by the Standard Model ($SM$) Higgs boson was first discussed  in \cite{higgs1}.  
In this model the Higgs non-minimally couples to gravity 
 with a moderate value of the non-minimal coupling 
\footnote{The value of the dimensionless non-minimal coupling $\xi\sim 10^4$
though in itself quite large, is much smaller than the ratio 
$\left(\frac{m_p}{M_W}\right)^2\simeq 10^{34}$, where $M_W \sim 100 \,GeV$ is the Electroweak scale.} \cite{higgs2,higgs3}. The model does not
 require an additional degree of freedom beyond the $SM$ and
 fits the observational data quite well \cite{CMB}. Reheating after inflation in this model has been studied 
in detail \cite{higgs3,higgs4,higgs4a} and quantum corrections to the potential at very high energies have been shown to 
be small \cite{higgs5}. In this section we assess the generality of Higgs inflation (in the Einstein frame)
 and determine the range of initial conditions which gives rise to adequate inflation (with $N_e \geq 60$)
 for a given value of the initial energy scale.

\subsection*{Action for Higgs Inflation}
 The action for a scalar field $\phi$ which couples non-minimally to 
gravity (\ie in the Jordan frame) is given by 
\cite{higgs1,higgs2,kaiser} 
\begin{equation}
S_{J}=\int d^{4}x\sqrt{-g} \bigg [ f(\phi)R-\frac{1}{2} g^{\mu\nu}\partial_{\mu}\phi\partial_{\nu}\phi-U(\phi) \bigg ]
\label{eqn:action_higgs}
\end{equation} 
where $R$ is the Ricci scalar and $g_{\mu\nu}$ is the  metric in the Jordan frame. 
The potential for the $SM$ Higgs field is given by            
\begin{equation}
U(\phi)=\frac{\lambda}{4}\Big ( \phi^{2}-\sigma^2 \Big )^{2}
\label{eqn:higgs_pot1}
\end{equation}
where $\sigma$ is  the vacuum expectation value of the Higgs field 
\begin{equation}
\sigma=246 \,GeV =1.1\times 10^{-16} m_{p}
\label{eqn:higgs_vev}
\end{equation}  
and the Higgs  coupling constant has the value 
$\lambda=0.1$.
Furthermore      
\begin{equation}
f(\phi)=\frac{1}{2}(m^{2}+\xi \phi^{2})
\label{eqn:fun_nm1}
\end{equation}
where $m$ is a mass parameter given by \cite{kaiser}
$$m^{2}=m_{p}^{2}-\xi\sigma^2$$
$\xi$ being the non-minimal coupling constant whose value
\begin{equation}
\xi=1.62\times10^{4}
\label{eqn:coup_nm}
\end{equation}
agrees with observations \cite{CMB} (see Appendix \ref{App:AppendixA}).
For the above values\footnote{Note that the observed vacuum expectation value of the Higgs field $\sigma=1.1\times 10^{-16} m_{p}$ is much smaller than the energy scale of inflation and hence we neglect it in our subsequent calculations.} of $\sigma$ and $\xi$, one finds $m\simeq m_{p}$, so that 
\begin{equation}
f(\phi) \simeq \frac{1}{2}(m_{p}^{2}+\xi \phi^{2})=\frac{m_{p}^{2}}{2}\Big ( 1+\frac{\xi\phi^{2}}{m_{p}^{2}} \Big )~.
\label{eqn:fun_nm2}
\end{equation}
 We now transfer to the Einstein frame by means of the following  conformal transformation of the metric \cite{kaiser}
\begin{equation}
g_{{\mu}{\nu}}\longrightarrow \hat{g}_{{\mu}{\nu}}=\Omega^{2} g_{{\mu}{\nu}}
\label{eqn:confo_trans}
\end{equation}
where the conformal factor is given by
\begin{equation}
\Omega^{2}=\frac{2}{m_{p}^{2}} f(\phi)=1+\frac{\xi\phi^{2}}{m_p^{2}}~.
\label{eqn:confo_fact}
\end{equation}
 After the field redefinition $\phi\longrightarrow \chi$  the action in the {\em Einstein frame}
 is given by \cite{kaiser}
\begin{equation}
S_{E}=\int d^{4}x\sqrt{-\hat{g}} \bigg [ \frac{m_{p}^{2}}{2}\hat{R}-\frac{1}{2} \hat{g}^{\mu\nu}\partial_{\mu}\chi\partial_{\nu}\chi-V(\chi) \bigg ]
\label{eqn:action_einstein}
\end{equation}
where 
\begin{equation}
V(\chi)=\frac{U[\phi(\chi)]}{\Omega^{4}}
\label{eqn:pot1}
\end{equation}
and 
\begin{equation}
\frac{\partial\chi}{\partial\phi}=\pm\frac{1}{\Omega^{2}}\sqrt{\Omega^{2}+\frac{6\xi^{2}\phi^{2}}{m_{p}^{2}}}~.
\label{eqn:pot2}
\end{equation}
Eq. (\ref{eqn:action_einstein}) describes General Relativity (GR) in the presence of a 
minimally coupled
scalar field $\chi$ with the potential $V(\chi)$.
(The full derivation of the action in the Einstein frame is  given in appendix \ref{App:AppendixB}.) 

\subsection*{Limiting cases of the potential in the Einstein Frame}

From equations (\ref{eqn:confo_fact}) and (\ref{eqn:pot2}) one finds  the following 
asymptotic forms for the potential (\ref{eqn:pot1})  
(for details see  appendix \ref{App:AppendixC} and \cite{higgs1,higgs3})
\begin{enumerate}
\item For $\phi\ll \sqrt{\frac{2}{3}}\frac{m_p}{\xi}$ one finds
\begin{equation}
\chi=\pm\phi,~~~~
V(\chi)\simeq \frac{\lambda}{4}\chi^4, ~~~~ |\chi| \ll \sqrt{\frac{2}{3}}\frac{m_p}{\xi}.
\label{eqn:pot_E1}
\end{equation}
\item For $\sqrt{\frac{2}{3}}\frac{m_p}{\xi} \ll \phi \ll \frac{m_p}{\sqrt{\xi}}$, 
\begin{equation}
\chi=\pm\sqrt{\frac{3}{2}}\frac{\xi\phi^2}{m_p}, ~~~~
V(\chi)\simeq \left (\frac{\lambda m_p^2}{6\xi^2}\right )\chi^2, ~~~~ \sqrt{\frac{2}{3}}\frac{m_p}{\xi}\ll |\chi| \ll \sqrt{\frac{3}{2}}m_p.
\label{eqn:pot_E2}
\end{equation}
\item For $\phi\gg \frac{m_p}{\sqrt{\xi}}$
\begin{equation}
\chi=\pm\sqrt{6}m_p\log{\left(\frac{\sqrt{\xi}\phi}{m_p}\right)}, ~~~
%V(\chi)\simeq \frac{\lambda m_p^4}{4\xi^2}\left(1+\exp\Big[-\sqrt{\frac{2}{3}}\frac{|\chi|}{m_p}\Big]\right)^{-2}, ~~~~ \chi ..... 
V(\chi)\simeq \frac{\frac{\lambda m_p^4}{4\xi^2}}
{\left(1+\exp\Big[-\sqrt{\frac{2}{3}}\frac{|\chi|}{m_p}\Big]\right)^2}, ~~~~ |\chi| \gg \sqrt{\frac{3}{2}}m_p
\label{eqn:pot_E3}
\end{equation}
\end{enumerate}
%The Einstein frame field $\chi(\phi)$ in these 
%three asymptotic regions is shown
 %in fig. \ref{fig:chi_phi}.
%\begin{figure}[H]
%\begin{center}
%\includegraphics[width=0.7\textwidth]{chivsphi.eps}
%\caption{This figure shows the dependence of the Einstein frame field $\chi$ on the Jordan frame field $\phi$ in three asymptotic regions. The green curve represents the linear region with $\chi\sim \phi$, the blue curve represents the quadratic reheating region where $\chi\sim \phi^2$ while the red curve represents the large-field behaviour of $\chi$ (relevant for inflation) where $\chi\sim \log{\phi}$; see appendix \ref{App:AppendixB} for details.
%}
%\label{fig:chi_phi}
%\end{center}
%\end{figure}
A good analytical approximation to the potential which can accommodate both 
(\ref{eqn:pot_E2}) and (\ref{eqn:pot_E3}) is 
%for large enough values of $\chi$ is given by 
\begin{equation}
V(\chi)\simeq V_0\left(1-\exp\Big[-\sqrt{\frac{2}{3}}\frac{|\chi|}{m_p}\Big]\right)^2,
~~~~ |\chi| \gg\sqrt{\frac{2}{3}}\frac{m_p}{\xi}
\label{eqn:higgs_pot_final}
\end{equation}
where $V_0$ is given by (Appendix \ref{App:AppendixA})
\begin{equation}
V_0=\frac{\lambda m_p^4}{4\xi^2}=9.6\times10^{-11}~m_p^4~.
\label{eqn:v_0}
\end{equation}

\subsection*{Generality Analysis of Higgs inflation in the Einstein Frame}

\begin{figure}[htb]
\begin{center}
\includegraphics[width=0.55\textwidth]{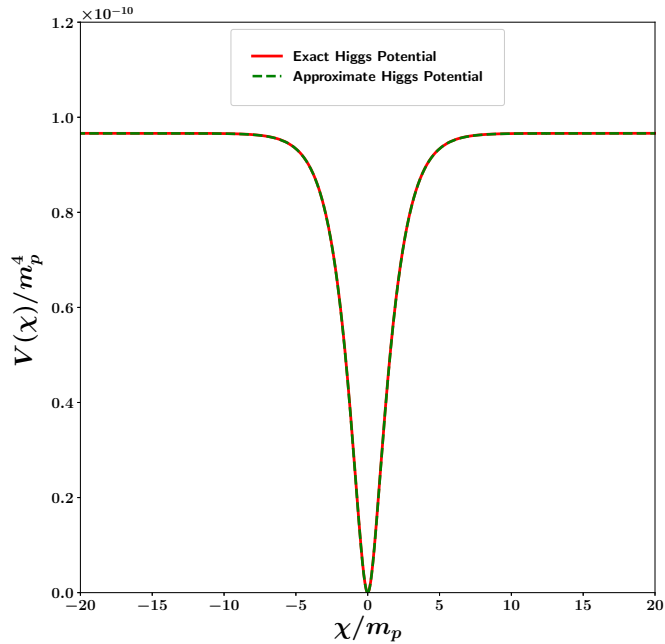}
\captionsetup{
	justification=raggedright,
	singlelinecheck=false
}
\caption{This figure shows the potential for Higgs inflation (in the Einstein frame)
 in units of $m_{p}^{4}$. The (solid) red curve shows the numerically 
determined value of the potential
from (\ref{eqn:pot1}) \& (\ref{eqn:pot2}), while the (dashed)
 green curve shows the approximate potential 
$V(\chi)=V_0\left (1-\exp{ \lbrack -\sqrt{\frac{2}{3}}
\frac{|\chi|}{m_{p}} \rbrack} \right )^2$.
Clearly the approximate form matches the exact one very well.
}
\label{fig:potential}
\end{center}
\end{figure}

\begin{figure}[htb]
\begin{center}
\includegraphics[width=0.55\textwidth]{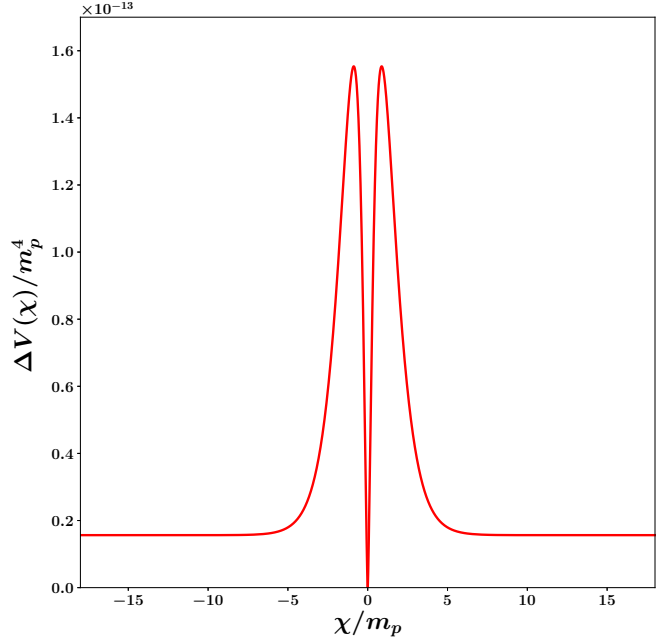}
\captionsetup{
	justification=raggedright,
	singlelinecheck=false
}
\caption{This figure shows the absolute value of the difference between the numerically determined Higgs potential (\ref{eqn:pot1}) \& (\ref{eqn:pot2}), 
and the approximate form 
(\ref{eqn:higgs_pot_final}).
We see that the maximum difference is near $\chi \sim m_p$ and its fractional
 value is only $0.16\%$.}
\label{fig:potential_diff}
\end{center}
\end{figure}

As we have seen, Higgs inflation in
 the Einstein frame can be described by
a minimally coupled canonical scalar field $\chi$ with a suitable potential  $V(\chi)$. 
We have analysed two different limits of the potential $V(\chi)$ which
 is asymptotically flat and has plateau like arms for $|\chi|\gg 1$. 
One notes that when $|\chi|\to 0$, $V(\chi)$ has a tiny kink with amplitude
 $\frac{\lambda}{4}\sigma^{4}\sim 10^{-66} m_{p}^{4}$. This kink is much smaller than the
 maximum height of the potential and can be neglected for all practical purposes.
(This is simply
 a reflection of the fact that the inflation energy scale is  much larger
than the  electro-weak scale.)
We have numerically evaluated the potential defined in (\ref{eqn:pot1}) \& (\ref{eqn:pot2})
 and compared it with the approximate form given in equation (\ref{eqn:higgs_pot_final}); see figure \ref{fig:potential}.  
The difference between the two potentials is shown in figure \ref{fig:potential_diff}. 
One finds that the maximum fractional difference between the two potentials is only
 $0.16\%$ which justifies the use of (\ref{eqn:higgs_pot_final}) for 
further analysis.

During Higgs inflation, the slow-roll parameter is given by 
\begin{equation}
\epsilon=\frac{m_{p}^{2}}{2}\left(\frac{1}{V}\frac{dV}{d\chi}\right)^{2} = \frac{4}{3}\frac{1}{\left(\exp\left(\sqrt{\frac{2}{3}}\frac{|\chi|}{m_{p}}\right)-1\right)^2}~,
\label{eqn:higgs_slowroll}
\end{equation}
since slow-roll ends when $\epsilon\simeq 1$, one finds
$$|\chi|\simeq 0.94 ~m_{p}\sim m_{p}~.$$

\begin{figure}[htb]
\begin{center}
\includegraphics[width=0.6\textwidth]{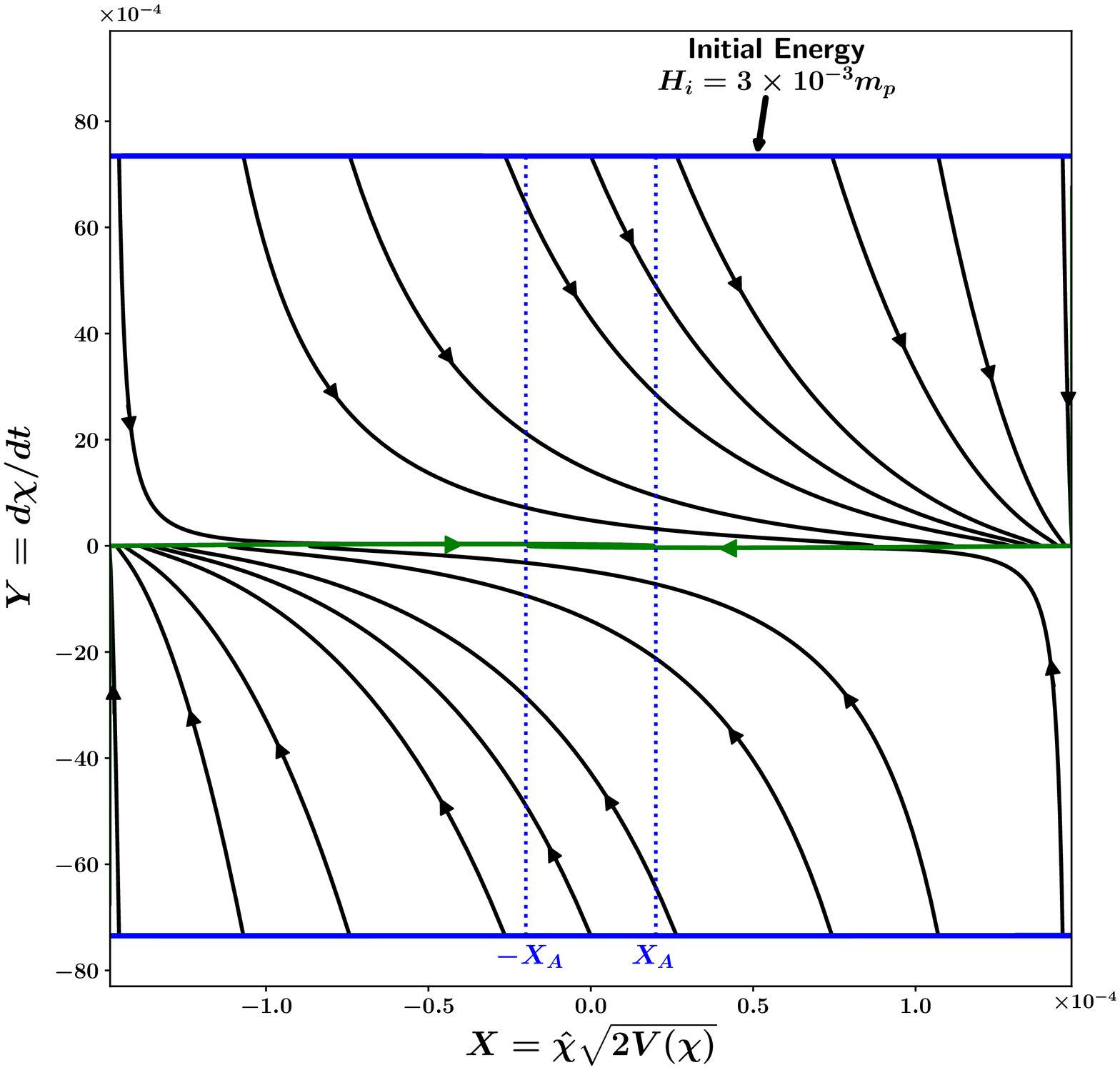}
\captionsetup{
	justification=raggedright,
	singlelinecheck=false
}
\caption{This figure shows the phase-space of Higgs inflation in the Einstein frame.
$Y=d\chi/dt$ is plotted against $X=\hat{\chi}\sqrt{2V(\chi)}$ for the initial energy scale 
$H_{i}=3\times 10^{-3}m_{p}$. 
($\hat{\chi}=\frac{\chi}{|\chi|}$ is the sign of field $\chi$.)
We see that commencing from a fixed initial energy (shown by the blue boundary lines), most 
solutions rapidly converge towards the two inflationary separatrices (horizontal green lines)
 corresponding to slow-roll inflation. 
We therefore find that inflation for the Higgs potential is remarkably general
and can commence from a very wide class of initial conditions.
%The vertical band corresponds to initial values $X_i$ and $Y_i$ lying close to 
%the origin. 
%(It is interesting that trajectories
%with opposite signs of $X_i$ and $Y_i$ (which commence in the first and fourth quadrants) are slanted and
%therefore able to meet the inflationary separatrices resulting in adequate inflation.)
Note that trajectories lying close to the origin, \ie within the vertical band 
marked by $( -X_A, X_A )$, are {\em strongly curved}. This property allows them 
 to converge to the 
inflationary separatrices giving rise to adequate inflation with $N_{e}\geq 60$.
It is interesting to contrast this behaviour 
 with that of chaotic inflation, shown in figure \ref{fig:chaotic},
 for which there is a small region with inadequate inflation
near the center. Because of this property, the Higgs scenario
 displays adequate inflation over a slightly larger range of initial conditions when compared with chaotic inflation.}
\label{fig:phase-space-higgs}
\end{center}
\end{figure}

\begin{figure}[htb]
\begin{center}
\includegraphics[width=0.48\textwidth]{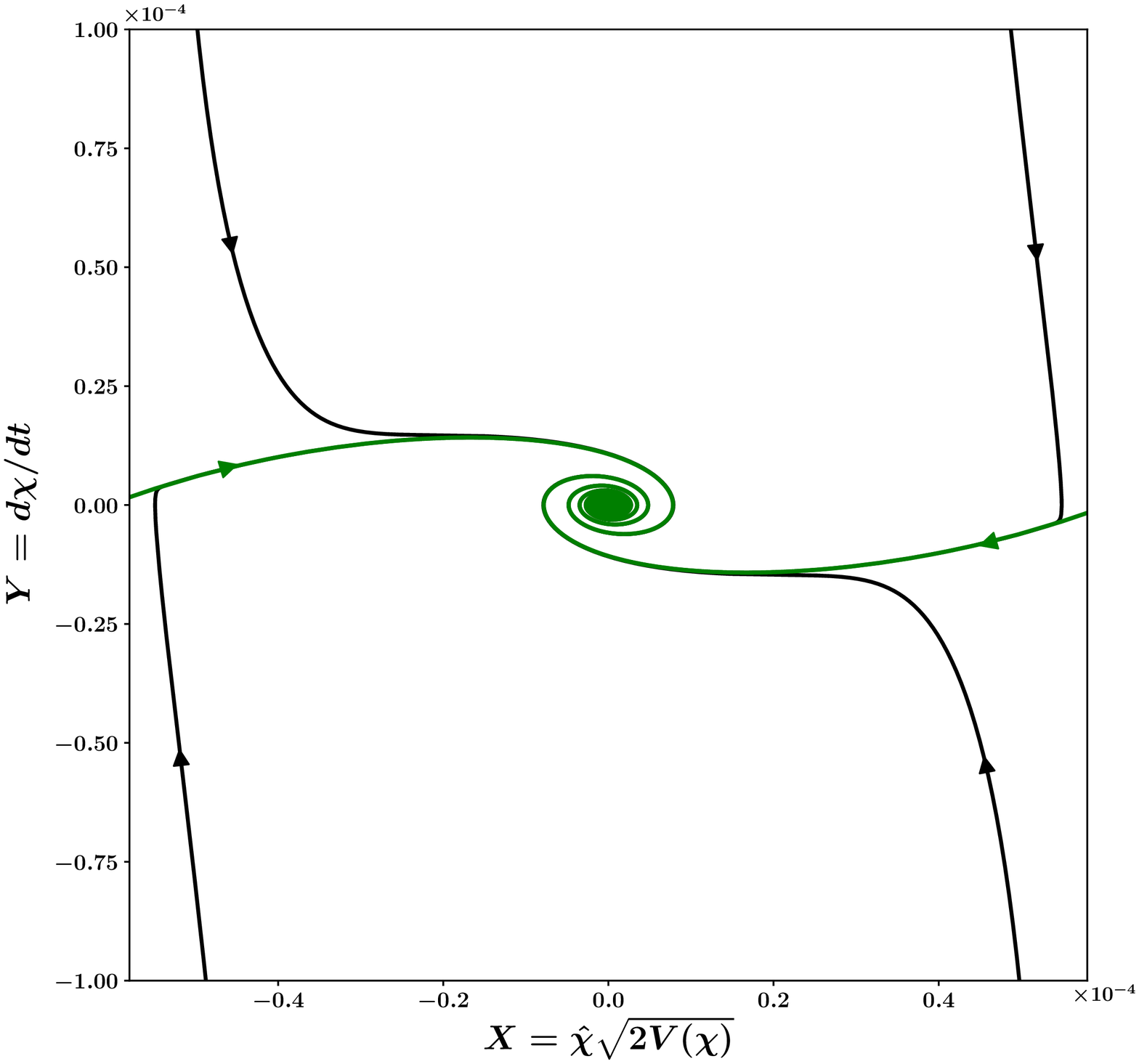}
\captionsetup{
	justification=raggedright,
	singlelinecheck=false
}
\caption{A zoomed-in view of the central region in figure \ref{fig:phase-space-higgs}.
% phase-space for Higgs inflation is shown for the 
%initial energy scale
%$H_{i}=3\times 10^{-3}m_{p}$. 
We see that most trajectories (associated with different initial conditions) initially
 converge towards the horizontal slow-roll inflationary separatrics (green lines)
 before 
spiralling in towards the center. (The spiral reflects
 oscillations of the inflaton about the minimum of its potential.)}
\label{fig:phase-space-higgs1}
\end{center}
\end{figure}

We study the generality of Higgs inflation in the Einstein frame by plotting the phase-space diagram for the 
potential (\ref{eqn:higgs_pot_final}) and determining the region of initial conditions which lead to
 adequate inflation (\ie $N_{e}\geq 60$). Our results are shown in figure \ref{fig:phase-space-higgs}
 and a zoomed-in view is presented
 in figure \ref{fig:phase-space-higgs1}.

We see that the phase-space diagram for Higgs inflation has very interesting properties. 
The asymptotically flat arms result in robust inflation as expected. 
However it is also possible to obtain adequate inflation if the inflaton commences from $\chi\simeq 0$.
This is because the scalar field is able to climb up the flat wings of $V(\chi)$.
This property is illustrated in figure \ref{fig:phase-space-higgs} by lines originating 
in the central region, 
%(lines starting from 
%the second and fourth quadrants),
which are slanted and hence can converge to the slow-roll inflationary separatrics resulting in
adequate inflation. This feature is not shared by chaotic inflation where one cannot obtain adequate
 inflation by starting from the origin
(provided the initial energy scale is not too large, \ie $H_i < m_{p}$.)

\begin{figure}[htb]
\begin{center}
\includegraphics[width=0.6\textwidth]{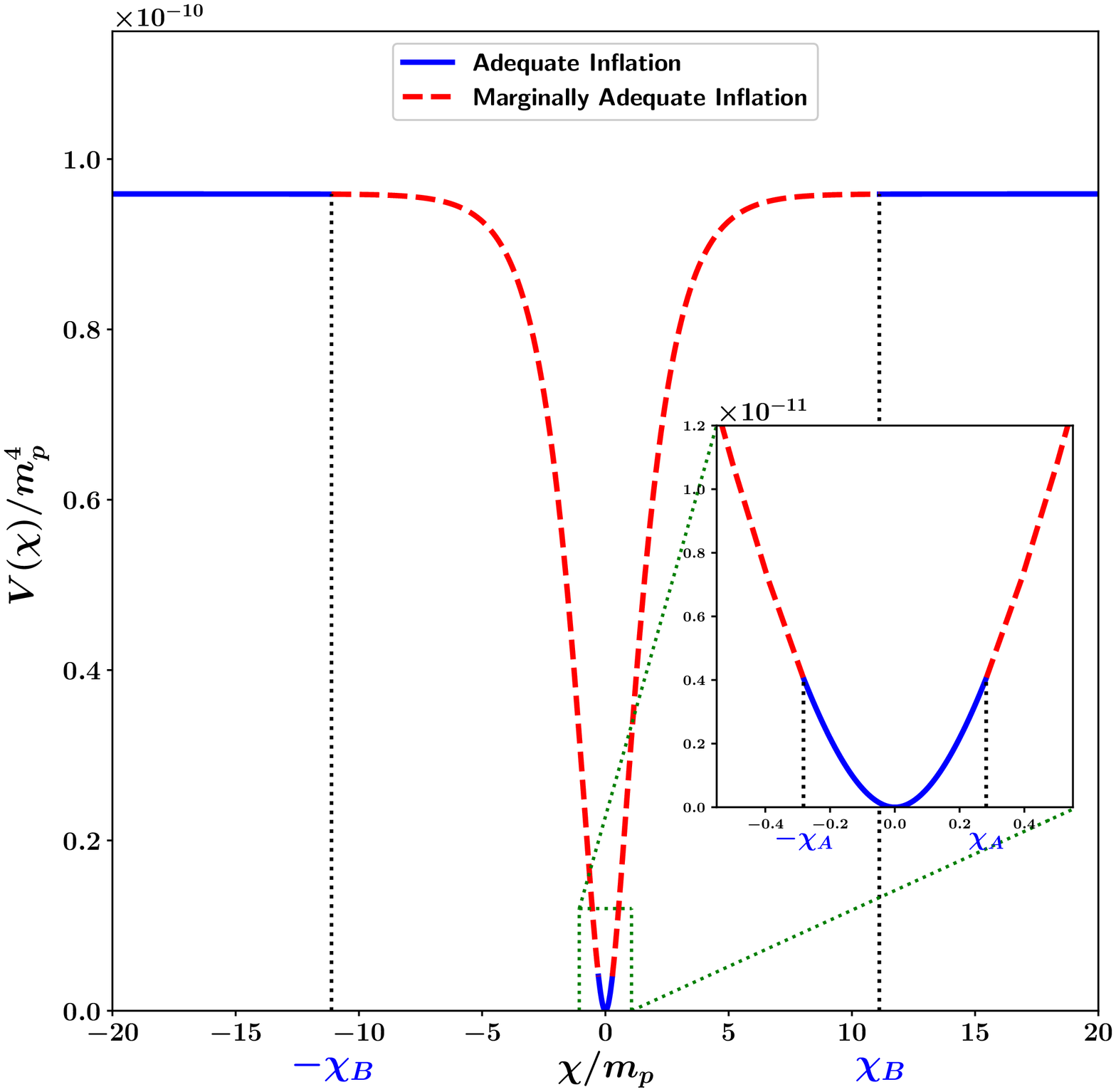}
\captionsetup{
	justification=raggedright,
	singlelinecheck=false
}
\caption{This figure shows initial field values, $\chi_i$,
 which either lead to adequate inflation (solid blue lines) or
 partially adequate inflation (dashed red lines). The region corresponding to 
$\chi_i\in [-\chi_{B},-\chi_{A}]\cup [\chi_{A},\chi_{B}]$ (dashed red)  
leads to partially adequate
inflation. Initial field values originating in this region result in inadequate inflation 
only when $\dot{\chi_i}$ is directed towards decreasing values of $V(\chi)$. The alternative case,
with $\dot{\chi_i}$ directed towards increasing $V(\chi)$, leads to adequate inflation
for the same subset $\chi_i\in [-\chi_{B},-\chi_{A}]\cup [\chi_{A},\chi_{B}]$.
This figure is shown for an initial energy scale $H_{i}=3\times 10^{-3} m_{p}$. The precise values of $\chi_{A}$ and $\chi_{B}$ depend on the initial energy scale $H_{i}$ 
as shown in table \ref{table:4}. Note that only a small portion of the full potential is shown in this figure.}
\label{fig:Higs_pot}
\end{center}
\end{figure}

This does not however imply that all possible initial conditions lead to adequate inflation in the Higgs scenario.
As shown in figure \ref{fig:Higs_pot} there is a small region 
of initial field values denoted by $|\chi_{A}|<|\chi_{i}|<|\chi_{B}|$ which does 
not lead to adequate inflation if $\chi_i$ and $\dot{\chi_i}$ have opposite signs 
(dashed red lines). 
By contrast,  the solid blue lines in the same figure   
show the region of $\chi_i$ that results in adequate inflation 
{\em independently of the direction} of the initial velocity $\dot{\chi_i}$.   
The dependence of  $\chi_{A}$ and $\chi_{B}$ on the initial energy scale is shown in table \ref{table:4} (also see figure  \ref{fig:Higs_pot}).
Note the surprising fact that the value of $\chi_{B}-\chi_{A}$ remains
{\em virtually unchanged} as $H_{i}$ increases.

\begin{table}[htb]
\begin{center}
 \begin{tabular}{||c|c|c|c|c|c|c||} 
 \hline
 $H_{i}$ (in $m_{p}$) & $\chi_{A}$ (in $m_{p}$) & $\chi_{B}$ (in $m_{p}$) & $\chi_{B}-\chi_{A}$ (in $m_{p}$)\\ [1ex] 
 \hline\hline
  $3\times 10^{-3}$ & $0.28$ & $11.11$ & $10.83$ \\ [1.2ex] 
 \hline
 $3\times 10^{-2}$ &  $2.16$ & $12.99$ & $10.83$ \\ [1.2ex] 
 \hline
 $3\times 10^{-1}$ &  $4.04$ & $14.87$ & $10.83$ \\ [1.2ex] 
 \hline

\end{tabular}
\captionsetup{
	justification=raggedright,
	singlelinecheck=false
}
\caption{Dependence of $\chi_{A}$ and $\chi_{B}$ on the initial energy scale $H_{i}$ for Higgs inflation (also see figure  \ref{fig:Higs_pot}).}
\label{table:4}
\end{center} 
\end{table}

The results of figures \ref{fig:phase-space-higgs}, \ref{fig:phase-space-higgs1} and \ref{fig:Higs_pot} lead us to 
conclude that 
there is a region lying close to the origin of $V(\chi)$, namely
 $\chi_i\in (-\chi_{A},\chi_{A})$, where one gets adequate 
inflation regardless of the direction of $\dot{\chi}_i$. 
One might note that this feature is absent in the power law family of 
potentials described in the previous section 
(compare figure \ref{fig:Higs_pot} 
with figures \ref{fig:chaotic2}, \ref{fig:mono3}, \ref{fig:monodromy3}). 
values that lead to partially adequate inflation, $|\chi_i|\in [|\chi_{A}|,|\chi_{B}|]$ 
We therefore conclude that a wide range of initial conditions 
can generate adequate inflation in the Higgs case\footnote{See \cite{salvio1,salvio2} for an analysis of classical and quantum initial conditions for Higgs inflation.}, which does not support
 some of the conclusions drawn in \cite{stein13}. 

Finally we would like to draw attention to the fact that the phase-space
analysis performed here for Higgs inflation is likely to carry over to the T-model
$\alpha$-attractor potential \cite{T-model}, since the two potentials
are qualitatively very similar.

\subsection{Initial conditions for Higgs Inflation in the non-canonical framework}
\label{sec:higgs_non_can}

The class of initial conditions leading to sufficient inflation widens considerably
if we choose to work with scalar fields possessing a non-canonical kinetic term.

The Lagrangian for this class of models is  \cite{Mukhanov-2006}
\begin{equation}
{\cal L}(\phi , F) = -F \left(\frac{F}{M^4}\right)^{\alpha-1} - V(\phi),
\label{Lagrangian_nc}
\end{equation}
where $F = \frac{1}{2}\pa_{\mu}\phi\; \pa^{\mu}\phi$, $M$ has the dimensions of mass and $\alpha$ is a dimensionless parameter. 
The associated energy density and pressure in a FRW universe are given by \cite{Mukhanov-2006,sanil_varun}
\begin{eqnarray}
\rho_{\phi} = -(2\alpha-1)F\left(\frac{F}{M^4}\right)^{\alpha-1} + V(\phi)~,\label{eqn:Energy_nc}\\
p_{\phi} = -F\left(\frac{F}{M^4}\right)^{\alpha-1} - V(\phi)~, ~~ F = -\frac{1}{2}{\dot\phi}^2~,
\label{eqn:Pressure_nc}
\end{eqnarray}
which reduce to the canonical form $\rho_{\phi} = -F + V, ~p_{\phi} = -F - V$
when $\alpha = 1$.
 The two Friedmann equations now acquire the form 
\begin{eqnarray}
H^{2} &=& \frac{8 \pi G}{3}\l[-(2\alpha - 1)F\left(\frac{F}{M^4}\right)^{\alpha-1} +\;
V(\phi)\r]~,\label{eqn: FR-eqn1 model_nc}\\
\frac{\ddot{a}}{a} &=& -\frac{8 \pi G}{3}\l[-(\alpha+1)F\left(\frac{F}{M^4}\right)^{\alpha-1} -\;  V(\phi)\r]~,
\label{eqn: FR-eqn2 model_nc}
\end{eqnarray}
and the equation of motion of the scalar field becomes
\begin{equation}
\ddot{\phi}+\frac{3}{2\alpha-1}H\dot{\phi}+\left(\frac{V'(\phi)}{\alpha(2\alpha-1)}\right)\left(\frac{2M^4}{\dot{\phi}^2}\right)^{\alpha-1} = 0,
\label{eqn:EOM_nc}
\end{equation}
which reduces to (\ref{eqn:motion}) when $\alpha = 1$.

Before discussing Higgs inflation in the non-canonical framework, we first examine the inflationary slow-roll parameter $\epsilon_{nc}$ which, for non-canonical inflation, is
given by \cite{sanil_varun}
\begin{equation}
\epsilon_{nc}=\left(\frac{1}{\alpha}\right)^{\frac{1}{2\alpha-1}}\left(\frac{3M^4}{V}\right)^{\frac{\alpha-1}{2\alpha-1}}\left(\epsilon_c\right)^{\frac{\alpha}{2\alpha-1}},
\label{eqn:slow-roll_nc}
\end{equation}
$\epsilon_c$ being the canonical slow-roll parameter (\ref{eqn:slow_roll1}).
Note that $\epsilon_{nc}<\epsilon_c$ for $3M^4\ll V$. This suggests
that for a fixed potential $V$,  the duration of inflation can be enhanced relative to
the canonical case
($\alpha = 1$), by a suitable choice of $M$.

\bigskip
\leftline{\bf The Higgs potential}
\bigskip

It is well known that the standard model Higgs boson, when coupled minimally to gravity,
cannot provide a working model of inflation due to the large value of the 
coupling constant, $\lambda\simeq 0.1$, in the potential 
%(\ref{eqn:higgs_pot1}). 
\begin{equation}
 V(\phi)=\frac{\lambda}{4}\left(\phi^2-\sigma^2\right)^2,
 \label{eqn:higgspot0}
 \end{equation}
where $\sigma$ is  the vacuum expectation value of the Higgs field
(\ref{eqn:higgs_vev}).
%\begin{equation}
%\sigma=246 \,GeV =1.1\times 10^{-16} m_{p}.
%\label{eqn:higgs_vev1}
%\end{equation}
%and the Higgs  coupling constant has the value
%$\lambda=0.1$.
Indeed $\lambda\simeq 0.1$ is many orders of magnitude larger than  the CMB constrained value $\lambda_c \simeq 1.43\times 10^{-13}$ in the canonical framework (see Appendix \ref{App:AppendixA}).
Additionally the potential (\ref{eqn:higgspot0}) gives too small a value for the inflationary scalar spectral index
$n_{_{S}}$ and too large a value for the tensor-to-scalar ratio $r$, to be in accord with observations.

However the situation
 changes when one examines the potential (\ref{eqn:higgspot0}) in the non-canonical framework.
%In this case
%Within the non-canonical framework the potential (\ref{eqn:higgspot0}) results in the following
The expression for the inflationary scalar spectral index now becomes \cite{sanil_varun}
\ber
n_{_{S}}= 1 -\l(\f{\gamma + 4}{N_e\gamma + 2}\r)~,
\label{eqn:n_s}
\eer
where 
\ber
\gamma \equiv \frac{2(3\alpha - 2)}{2\alpha - 1}~.
%\gamma \equiv \frac{2\alpha + n\,\l(\alpha - 1\r)}{2\alpha - 1}, ~~ \mu \equiv \frac{M}{\mpl}~.
\label{eqn: gamma}
\eer
Since $\gamma$ increases from $\gamma = 2$ for $\alpha = 1$ to
$\gamma = 3$ for $\alpha  \gg 1$, therefore
the scalar spectral index {\em increases}
from the canonical value $n_{_{S}} = 0.951$ ($\alpha = 1, N_e = 60$) to $n_{_{S}} = 0.962$, 
in non-canonical models (with $\alpha  \gg 1$).

Similarly one can show that the tensor-to-scalar ratio declines in non-canonical models. 
 For the Higgs potential one gets \cite{sanil_varun}
\ber
r =  \l(\f{1}{\sqrt{2\,\alpha - 1}}\r)\l(\f{32}{N_e\gamma + 2}\r)~,
\label{eqn: T-to-S phi-n pot}
\eer
which demonstrates that the value of $r$ decreases with an increase in
 the non-canonical parameter $\alpha$. 
Figure \ref{fig:inf_ns,r_a}  shows $n_{_{S}}$, $r$ plotted as functions of $\alpha$.
One finds that $n_{_{S}} \simeq 0.96$, 
 $r < 0.1$ for $\alpha \geq 3$, which agrees well with CMB observations.

\begin{figure}[htb]
\centering
\subfigure[]{
\includegraphics[width=0.49\textwidth]{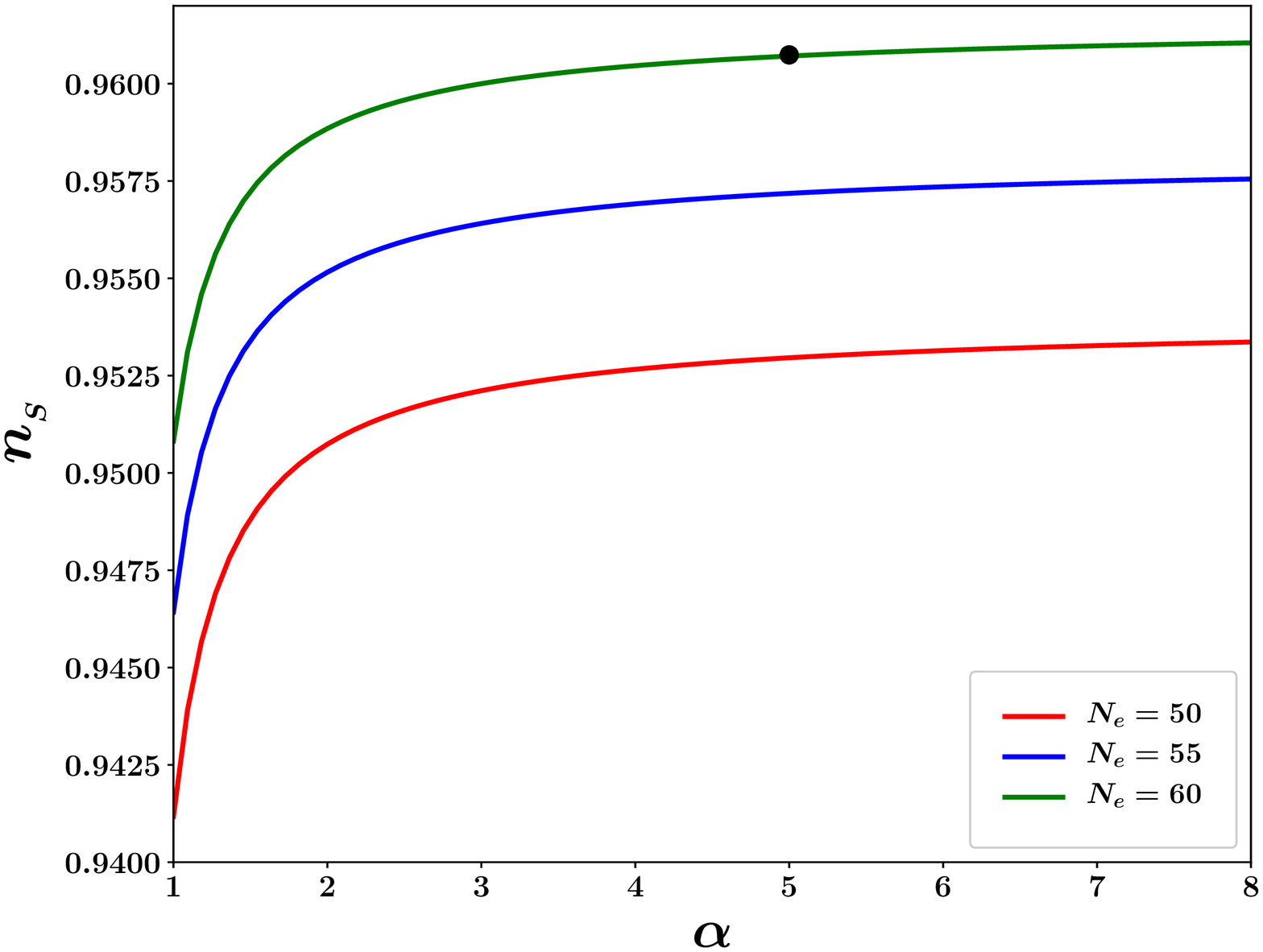}\label{fig:inf_ns_a}}
\subfigure[]{
\includegraphics[width=0.49\textwidth]{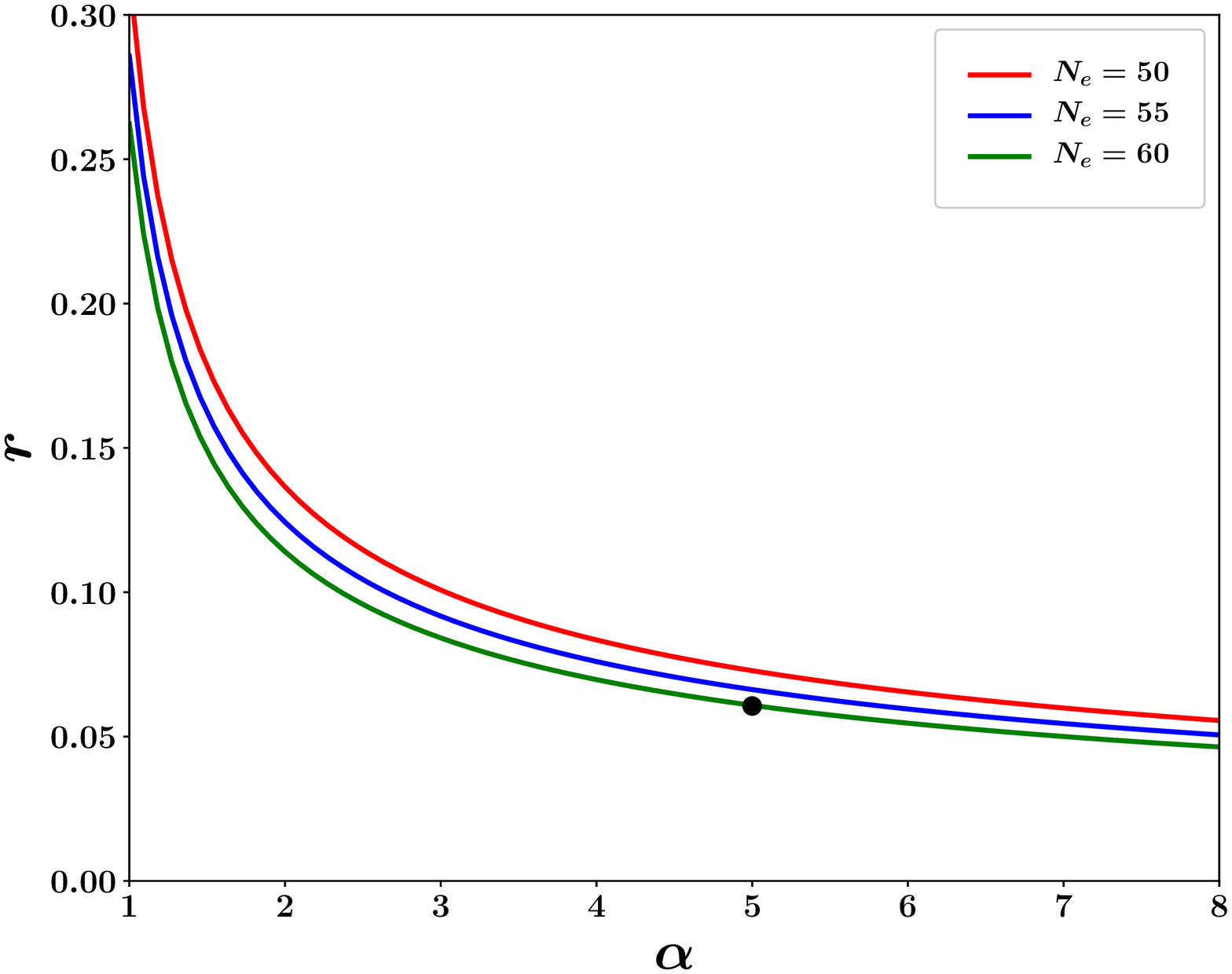}\label{fig:inf_r_a}}
\captionsetup{
	justification=raggedright,
	singlelinecheck=false
}
\caption{This figure shows:  
\textbf{(a)} the scalar spectral index $n_{_{S}}$, 
\textbf{(b)} the tensor to scalar ratio $r$,  as functions
 of the non-canonical parameter $\alpha$ and described respectively by
 equations (\ref{eqn:n_s}) and (\ref{eqn: T-to-S phi-n pot}). 
Three values of the  number of e-foldings, $N_e=50,~55$ and $60$, are chosen. 
One finds that larger values of $\alpha$ result in  
higher values of $n_{_{S}}$ and lower values of $r$. The black dot
in both figures indicates the value of $\alpha$,
 and the corresponding values of $n_{_{S}}$ and $r$, used in our subsequent analysis.} 
\label{fig:inf_ns,r_a}
\end{figure}

 The relation between the value of the Higgs self-coupling $\lambda\simeq 0.1$ in the non-canonical 
framework and the corresponding canonical value $\lambda_c$ is given by \cite{sanil_varun}
\ber
\lambda = 4\left[\frac{32\lambda_c(N_e+1)^3}{\sqrt{2\alpha-1}}\left(\frac{\alpha}{4}\Big(\frac{1}{6}\frac{m_p^4}{M^4}\Big)^{\alpha-1}\right)^{\frac{2}{3\alpha-2}}\left(\frac{1}{N_e\gamma+2}\right)^{\frac{\gamma+4}{\gamma}}\right]^{\frac{3\alpha-2}{\alpha}}~,
\label{eqn:lambda_vs_lambdac}
\eer
where consistency with CMB observations suggests $\lambda_c \sim 10^{-13}$.

\begin{figure}[htb]
\begin{center}
\includegraphics[width=0.55\textwidth]{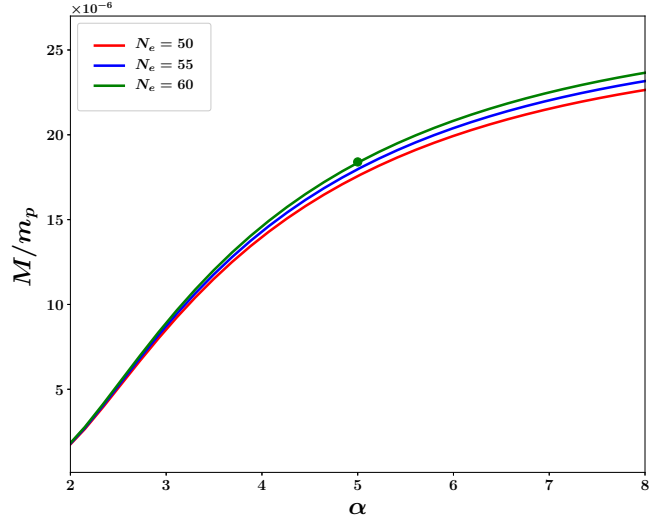}
\captionsetup{
	justification=raggedright,
	singlelinecheck=false
}
\caption{This figure illustrates the relation between the non-canonical parameters $M$ and $\alpha$, given by equation (\ref{eqn:lambda_vs_lambdac}), which results in the 
self-coupling value $\lambda=0.1$ in equation (\ref{eqn:higgspot0}).
 Results for three different e-folding values $N_e=50,~55~,60$ are shown. The green dot indicates the value of $M$ and $\alpha$ 
which is used in our subsequent analysis.}
\label{fig:inf_nc_Malpha}
\end{center}
\end{figure}

 Figure \ref{fig:inf_nc_Malpha}  describes the values of the non-canonical parameters $\alpha$ and $M$ that yield $\lambda\simeq 0.1$
 in (\ref{eqn:higgspot0}) 
-- the relation between $M$ and $\alpha$ being provided by
 equation (\ref{eqn:lambda_vs_lambdac}). 
In our subsequent analysis we choose $\alpha=5$
for simplicity. 
This is shown by the black color dot in figure \ref{fig:inf_ns_a} and \ref{fig:inf_r_a}.
 (The corresponding value of $M$ 
is shown by the green dot in figure \ref{fig:inf_nc_Malpha}.)

As in the case of canonical scalar fields (\ref{eqn:circle}), one
 can rewrite the Friedman equation for non-canonical scalars (\ref{eqn: FR-eqn1 model_nc}) as follows
\begin{equation}
R^{2}=Y_{nc}^{2}+X^{2}
\label{eqn:circle_nc}
\end{equation}
where
\begin{equation}
R=\sqrt{6}\frac{H}{m_{p}}, ~~
X=\hat{\phi}\frac{\sqrt{2V(\phi)}}{m_{p}^{2}},~~
Y_{nc}=\left[2(2\alpha-1)\left(-\frac{F}{m_p^4}\right)\left(\frac{F}{M^4}\right)^{\alpha-1}\right]^{1/2}.
\label{eqn:circle1_nc}
\end{equation}
 Therefore
commencing at some initial value of $R$ ($\equiv \sqrt{6}H/m_p$) one can set different
initial conditions by varying $X$ and $Y_{nc}$. Since $X$, $Y_{nc}$ satisfy the constraint
equation (\ref{eqn:circle_nc}) they lie on the boundary of a circle.

\begin{figure}[htb]
\begin{center}
\includegraphics[width=0.6\textwidth]{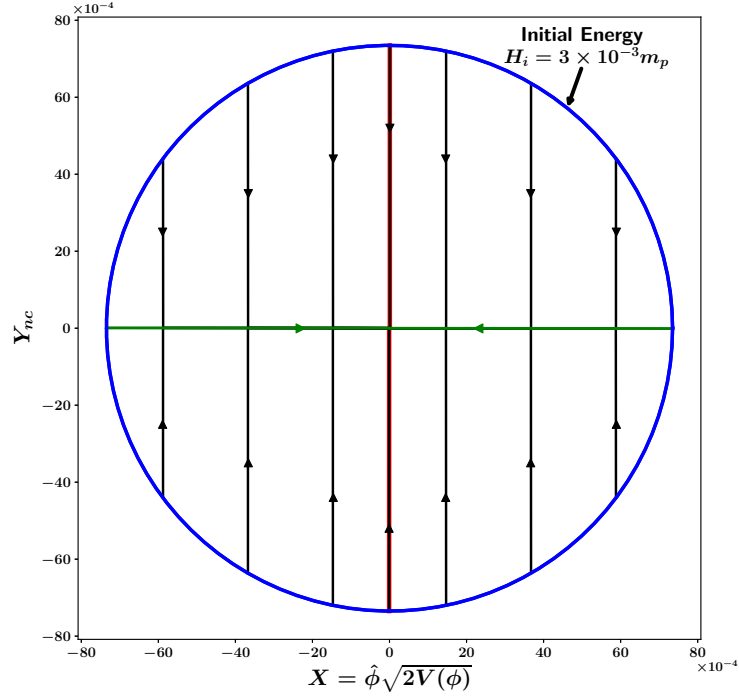}
\captionsetup{
	justification=raggedright,
	singlelinecheck=false
}
\caption{This figure shows
 the phase-space  of Higgs inflation in the non-canonical framework described by (\ref{eqn:higgspot0}).
$Y_{nc}$, given by (\ref{eqn:circle1_nc}) is plotted against $X$ ($=\hat{\phi}\sqrt{2V(\phi)}$) for 
different initial conditions all of which commence on the (blue) circle which represents the
 initial energy scale $H_{i}=3\times 10^{-3}m_{p}$. 
($\hat{\phi}=\frac{\phi}{|\phi|}$ is the sign of field $\phi$.)
One finds that commencing
from the circle, different inflationary trajectories rapidly
 converge to one of the two inflationary separatrices 
(green horizontal lines) before proceeding towards the center, which corresponds to the
minimum of the potential. The thin vertical central red band 
corresponds to the region in phase-space that {\em does not}
 lead to adequate inflation. Note that this band is {\em very small} which is indicative of the
robustness of Higgs inflation in the non-canonical framework.
The arc-length of the red band, when divided by the circumference of the circle with radius = $\sqrt{6}H_{i}/m_p$,
gives the fraction of initial conditions $\frac{2\Delta l_A}{l}$ which lead to inadequate inflation.
}	
\label{fig:higgsPS_nc}
\end{center}
\end{figure}

We probe the robustness of this model to initial conditions by plotting its phase-space diagram 
($Y_{nc}$ vs $X$) and determining the region of initial conditions which gives rise to 
adequate inflation ($N_e\geq 60$) for values of $M$ and $\alpha$ which satisfy CMB constraints (shown by the green dot in figure \ref{fig:inf_nc_Malpha}). The phase-space diagram corresponding to an initial energy scale $H_i=3\times 10^{-3}~m_p$ is shown in figure \ref{fig:higgsPS_nc}.

\begin{figure}[htb]
\centering
\subfigure[]{
\includegraphics[width=0.49\textwidth]{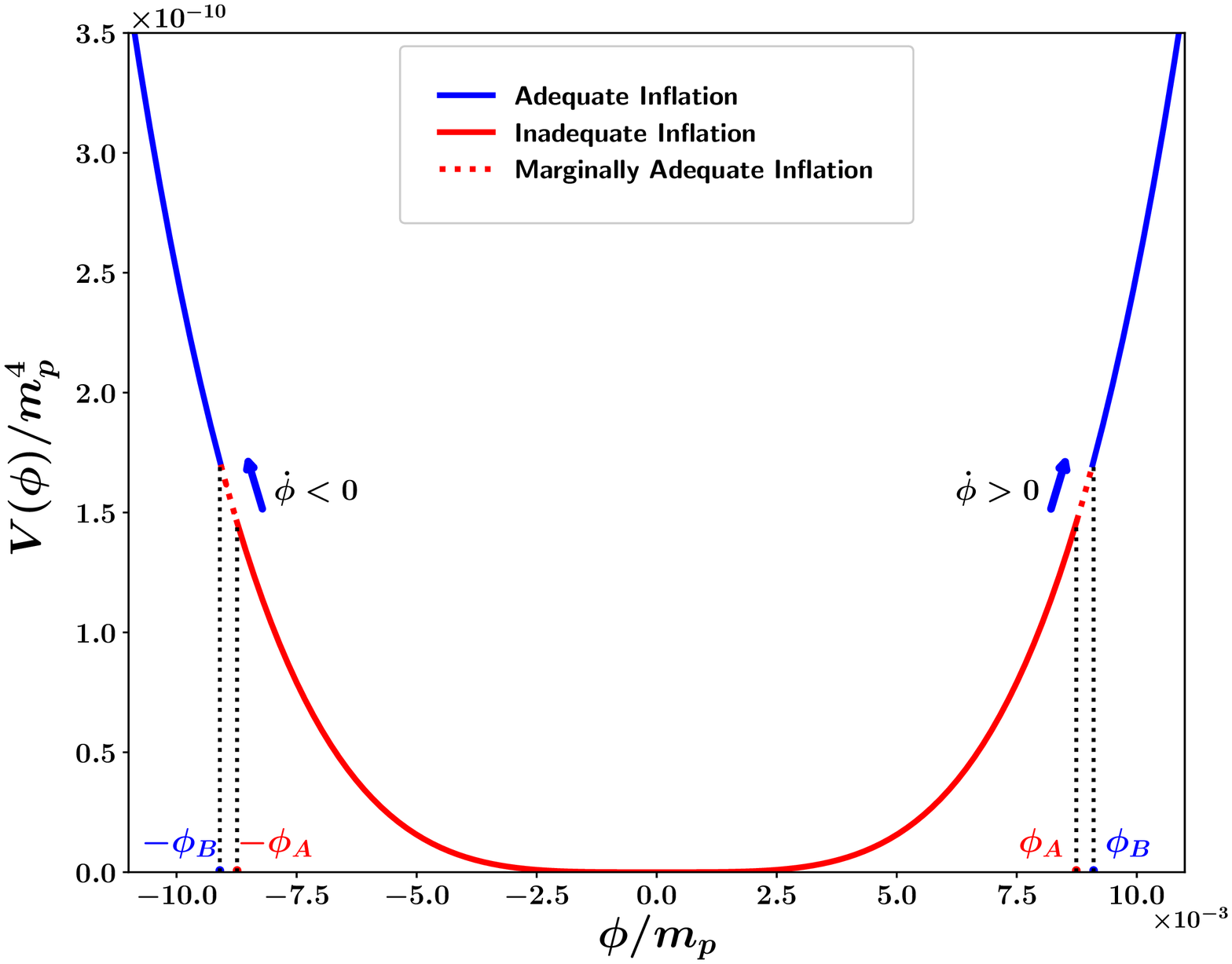}\label{fig:quartnc_A}}
\subfigure[]{
\includegraphics[width=0.49\textwidth]{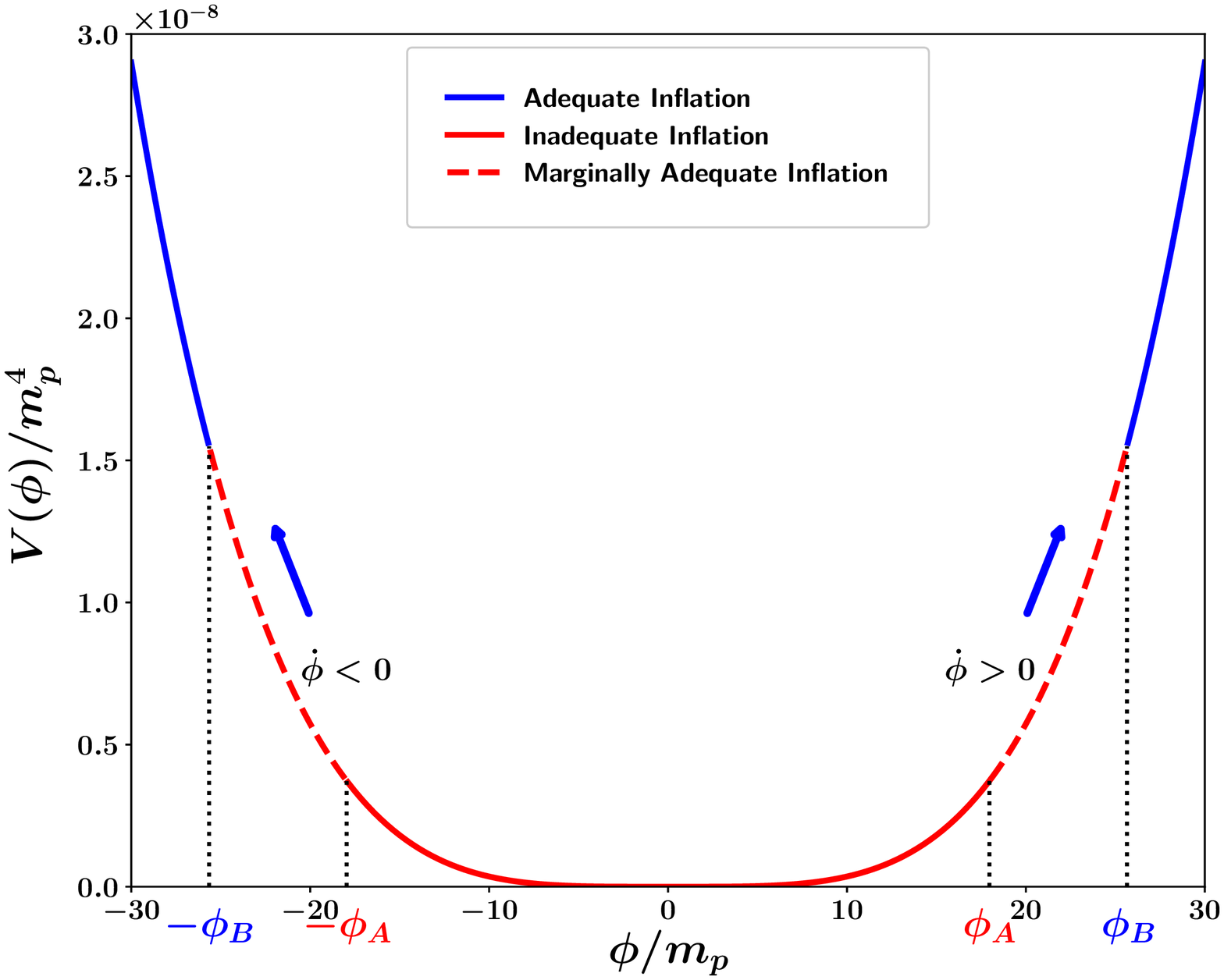}\label{fig:quartc_A}}
\captionsetup{
	justification=raggedright,
	singlelinecheck=false
}
\caption{ Initial field values, $\phi_i$,
which lead to adequate inflation with $N_e \geq 60$ (blue), marginally adequate (dashed red)  and
inadequate (red) inflation are schematically shown
 for the Higgs  inflation with the quartic potential (\ref{eqn:higgspot0}):  \textbf{(a)} in the  
\textbf{non-canonical} framework and \textbf{(b)} in the \textbf{canonical} framework.
The blue
lines represent regions of adequate inflation. The red lines come in two styles: dashed/solid
and correspond to the two possible initial directions of ${\dot\phi_i}$.
The solid red line represents initial values of $\phi$ for which inflation is never adequate
irrespective of the direction of ${\dot\phi_i}$.
In the region shown by the dashed line one gets
 adequate inflation only when ${\dot\phi_i}$ points in the direction of increasing $V(\phi)$. We note that  for the non-canonical case, the values of $\phi_A$ and $\phi_B$ are extremely small as shown in table \ref{table:5}.
(Only a small portion of the full potential is shown in this figure
which corresponds to the initial energy scale $H_{i}=3\times 10^{-3}m_{p}$.)
} 
\label{fig:quartA}
\end{figure}

\begin{figure}[htb]
\centering
\subfigure[]{
\includegraphics[width=0.49\textwidth]{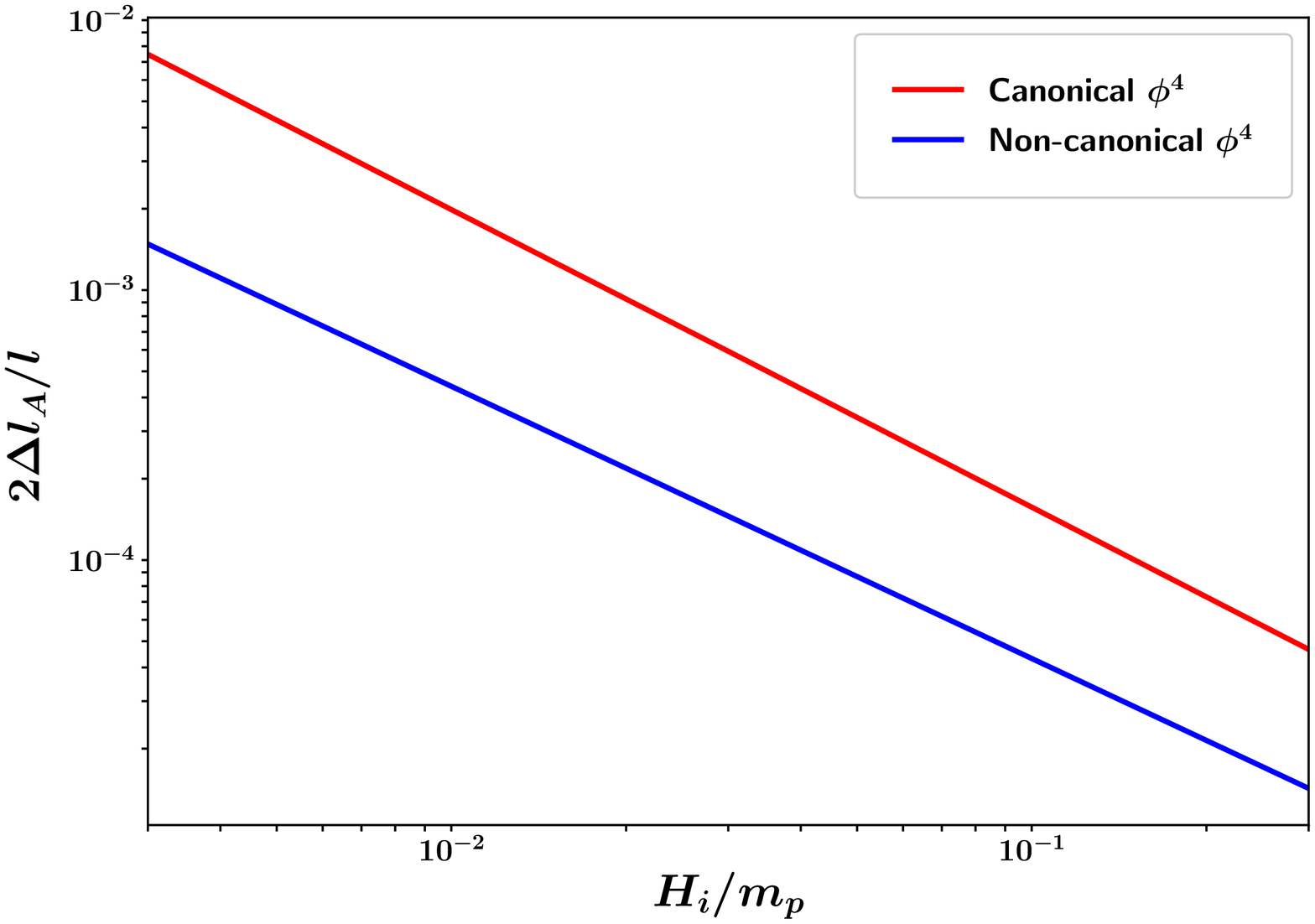}\label{fig:inf_cVSncA}}
\subfigure[]{
\includegraphics[width=0.49\textwidth]{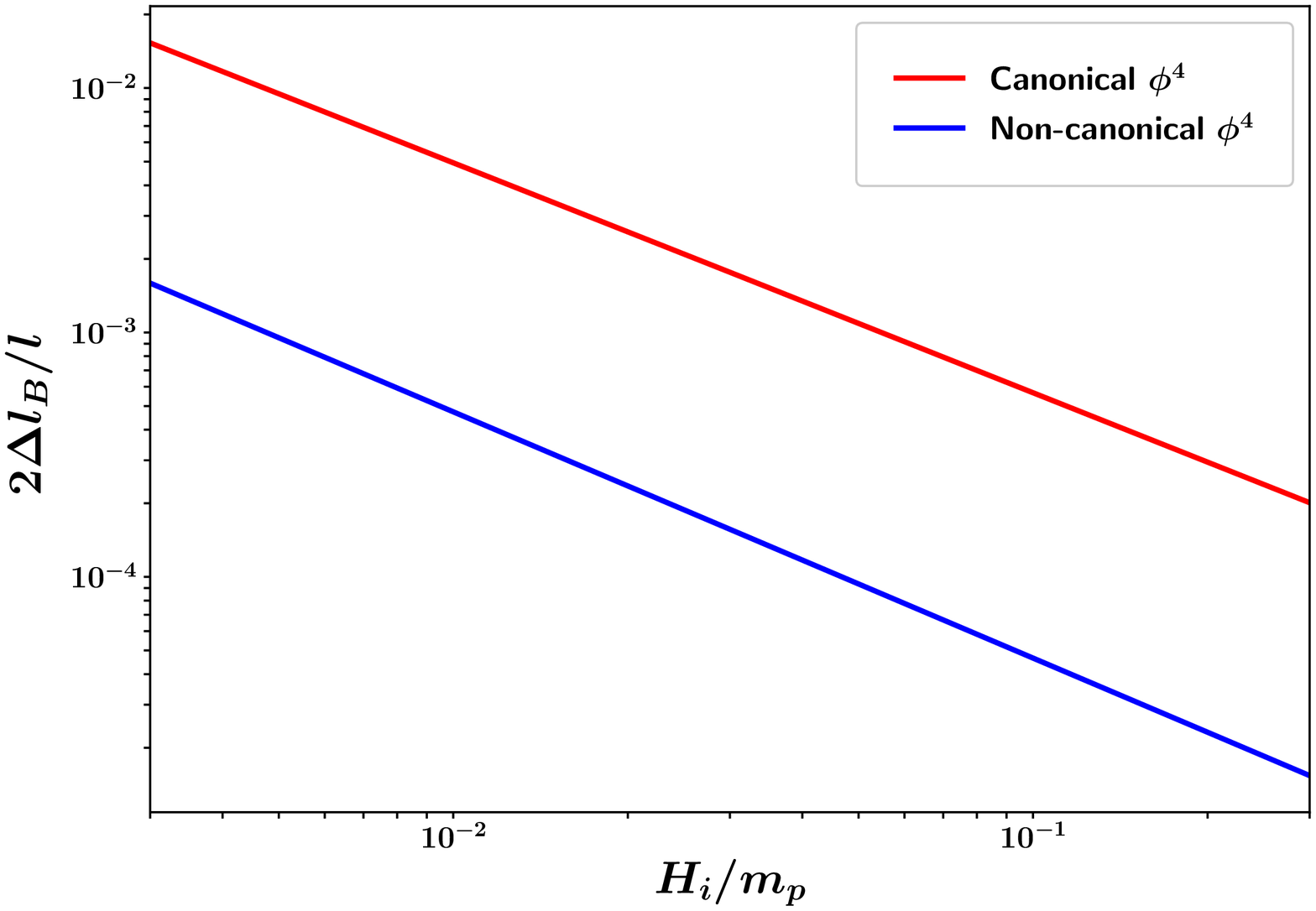}\label{fig:inf_cVSncB}}
\captionsetup{
	justification=raggedright,
	singlelinecheck=false
}
\caption{This figure compares the values of \textbf{(a)} $\frac{\Delta l_{A}}{l}$ and 
\textbf{(b)} 
$\frac{\Delta l_{B}}{l}$ 
for canonical and  non-canonical scalar fields with the
potential $V(\phi)\propto \phi^4$. 
$\frac{\Delta l_{A}}{l}$ and $\frac{\Delta l_{B}}{l}$ are shown as functions of the initial energy scale of inflation,
$H_{i}$. The red and blue curves correspond to  canonical and non-canonical quartic inflation respectively. 
The smaller amplitude of the blue curve in both panels indicates that non-canonical
inflation arises for a larger class of initial conditions than canonical inflation (red).
The decrease in $\frac{\Delta l_{A}}{l}$ and  $\frac{\Delta l_{B}}{l}$
with an increase in $H_{i}$ is indicative of the fact that the set of initial conditions which
give rise to adequate inflation (with $N_e \geq 60$) increases with the energy scale of inflation, $H_{i}$.} 
\label{fig:inf_cvsnc}
\end{figure}

The fraction of initial conditions which give rise to inadequate inflation, 
$\frac{2\Delta l_A}{l}$, and partially adequate inflation, $\frac{2\Delta l_B}{l}$,
are shown in table \ref{table:5}. 
(As earlier, a uniform distribution of $X$ and $Y_{nc}$ on the boundary of initial 
conditions has been assumed.)
From this table one finds that 
the values of $\phi_A$ and $\phi_B$ associated with an initial energy scale
$H_i$, are much smaller than their counterparts for 
canonical inflation (see figures \ref{fig:quartnc_A},
\ref{fig:quartc_A} and \ref{fig:inf_cvsnc}).
This is a consequence of the fact that for identical potentials,
the  slow-roll parameter in the non-canonical case is much smaller than its
canonical counterpart ($\epsilon_{nc}\ll\epsilon_c$), which permits inflation to commence
from {\em smaller values} of the inflaton field in the non-canonical case.
%(for fixed values of $M$, and $\alpha$). 
We also find that the fraction of non-inflationary initial conditions,
 $\frac{\Delta l_{A}}{l}$, decreases with an increase $H_i$, as expected.

\begin{table}[htb]
\begin{center}
 \begin{tabular}{||c|c|c|c|c|c|c||} 
 \hline
 $H_{i}$ (in $m_{p}$) & $\phi_{A}$ (in $m_{p}$) & $\phi_{B}$ (in $m_{p}$) &  $2\frac{\Delta l_{A}}{l}$   & $2\frac{\Delta l_{B}}{l}$\\ [1ex] 
 \hline\hline
  $3\times 10^{-3}$ & $8.74 \times 10^{-3}$ & $9.07 \times 10^{-3}$ & $1.48 \times 10^{-3}$ & $1.59 \times 10^{-3}$ \\ [1.2ex] 
 \hline
 $3\times 10^{-2}$ &  $8.66 \times 10^{-3}$ & $8.99 \times 10^{-3}$ & $1.45 \times 10^{-4}$ & $1.57 \times 10^{-4}$ \\ [1.2ex] 
 \hline
 $3\times 10^{-1}$ &  $8.58 \times 10^{-3}$ & $8.91 \times 10^{-3}$ & $1.43 \times 10^{-5}$ & $1.54 \times 10^{-5}$ \\ [1.2ex] 
 \hline
\end{tabular}
\captionsetup{
	justification=raggedright,
	singlelinecheck=false
}
\caption{Dependence of $\phi_A$, $\phi_B$, $\frac{\Delta l_{A}}{l}$  and $\frac{\Delta l_{B}}{l}$  on the initial energy scale $H_{i}$ for non-canonical Higgs inflation . Here $l = 2\pi R \equiv 2\pi\sqrt{6} H_i/m_p$.}
\label{table:5}
\end{center} 
\end{table} 
% From Table 7, it is clear that   that $\frac{\Delta l_{B}}{l}$ decreases with increase in $H_i$ as expected.  

%In figures \ref{fig:deginf_nc2VSnc4} and .......
%we compare the robustness of inflation in different models by plotting 
%$\frac{\Delta l_{B}}{l}$ as a function of the initial energy scale of inflation, $H_i$, for models
%with an $m^2\phi^2$ and $\lambda\phi^4$ potential in the canonical as well as non-canonical framework.
%Somewhat surprisingly we find that
%$\frac{\Delta l_{B}}{l}$ is least in Higgs inflation, which is indicative of the robustness of
%this model vis-a-vis different choices of intial conditions.

In figure \ref{fig:inf_cvsnc} we compare values of
$\frac{\Delta l_{A}}{l}$ and $\frac{\Delta l_{B}}{l}$ for canonical inflation with $V_c(\phi)=\frac{\lambda_c}{4}\phi^4 $ and non-canonical inflation\footnote{Note that the Higgs 
potential in equation(\ref{eqn:higgspot0}) can be rewritten as $V(\phi)\simeq \frac{\lambda}{4}\phi^4$, since $\sigma\ll m_p$.} with $V(\phi)=\frac{\lambda}{4}\phi^4$ where $\lambda$ and $\lambda_c$ are related by equation (\ref{eqn:lambda_vs_lambdac}).
We find that the values of $\frac{\Delta l_{A}}{l}$ and $\frac{\Delta l_{B}}{l}$ are 
significantly smaller for non-canonical inflation, which implies
that inflation arises from a larger class of initial conditions in the non-canonical framework.

\section{Starobinsky Inflation}
\label{sec:starobinsky}

\subsection{Action and Potential in the Einstein Frame}
  Starobinsky inflation \cite{star80} is based on the  action 
\begin{equation}
S=\int d^{4}x\sqrt{-g} \frac{m_{p}^{2}}{2}\bigg [R+\frac{1}{6m^{2}}R^{2}  \bigg ]~,
\label{eqn:star_action}
\end{equation}
where $m$ is a mass parameter. % which is the mass of inflaton field in the Einstein frame. 
The corresponding action in the Einstein frame is given by \cite{whitt84,maeda88,gorbunov13} 
\begin{equation}
S_{E}=\int d^{4}x\sqrt{-g} \bigg [ \frac{m_{p}^{2}}{2}\hat{R}-\frac{1}{2} \hat{g}^{\mu\nu}\partial_{\mu}\phi\partial_{\nu}\phi-V(\phi) \bigg ]
\label{eqn:star_action_einstein}
\end{equation}
where the inflaton potential is 
\begin{equation}
V(\phi)=\frac{3}{4}m^{2}m_{p}^{2}\Big(1-e^{-\sqrt{\frac{2}{3}}\frac{\phi}{m_{p}}}\Big)^{2}
\label{eqn:star_pot}
\end{equation}
\par
\begin{figure}[htb]
\begin{center}
\includegraphics[width=0.55\textwidth]{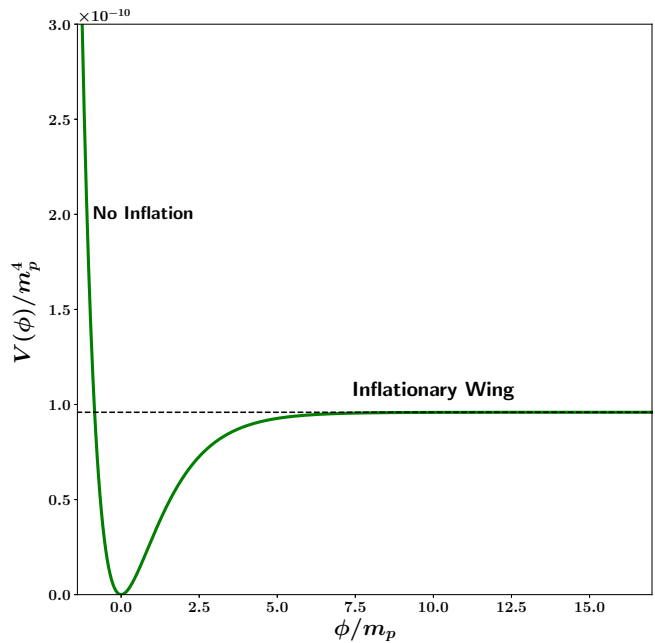}
\captionsetup{
	justification=raggedright,
	singlelinecheck=false
}
\caption{The effective potential in Starobinsky Inflation, (\ref{eqn:star_pot}), is plotted  in units of $m_{p}^{4}$. 
The potential is asymmetric about the origin and has a steep left wing and  plateau-like right wing. 
Inflation occurs along the flat plateau-like right wing,
the steep left wing being unable to sustain inflation.}
\label{fig:star_pot}
\end{center}
\end{figure}

\n
and $m=1.13\times 10^{-5}m_{p}$ is required 
from an analysis of scalar fluctuations
\cite{gorbunov13} (see Appendix \ref{App:AppendixA}). 
The potential (\ref{eqn:star_pot}) is shown in figure \ref{fig:star_pot}. 

\par
As shown in figure \ref{fig:star_pot}, the potential for Starobinsky inflation is asymmetric about 
the origin. One should note that the
flat right wing of the potential has the  same functional form as the Higgs inflation potential in the Einstein frame. However the left wing of $V(\phi)$ is very steep. The slow-roll parameter for this potential is given by
$$\epsilon=\frac{4}{3}
\left[\exp\left(\sqrt{\frac{2}{3}}\frac{\phi}{m_{p}}\right)-1\right]^{-2}.$$
 Inflation occurs for 
$\epsilon\leq 1$, which corresponds to
$\phi\geq 0.94~m_p$ and implies that
no inflation can happen on the steep left wing of the potential (for which $\phi < 0$). 

\subsection{Generality of Starobinsky Inflation}
\begin{figure}[htb]
\begin{center}
\includegraphics[width=0.6\textwidth]{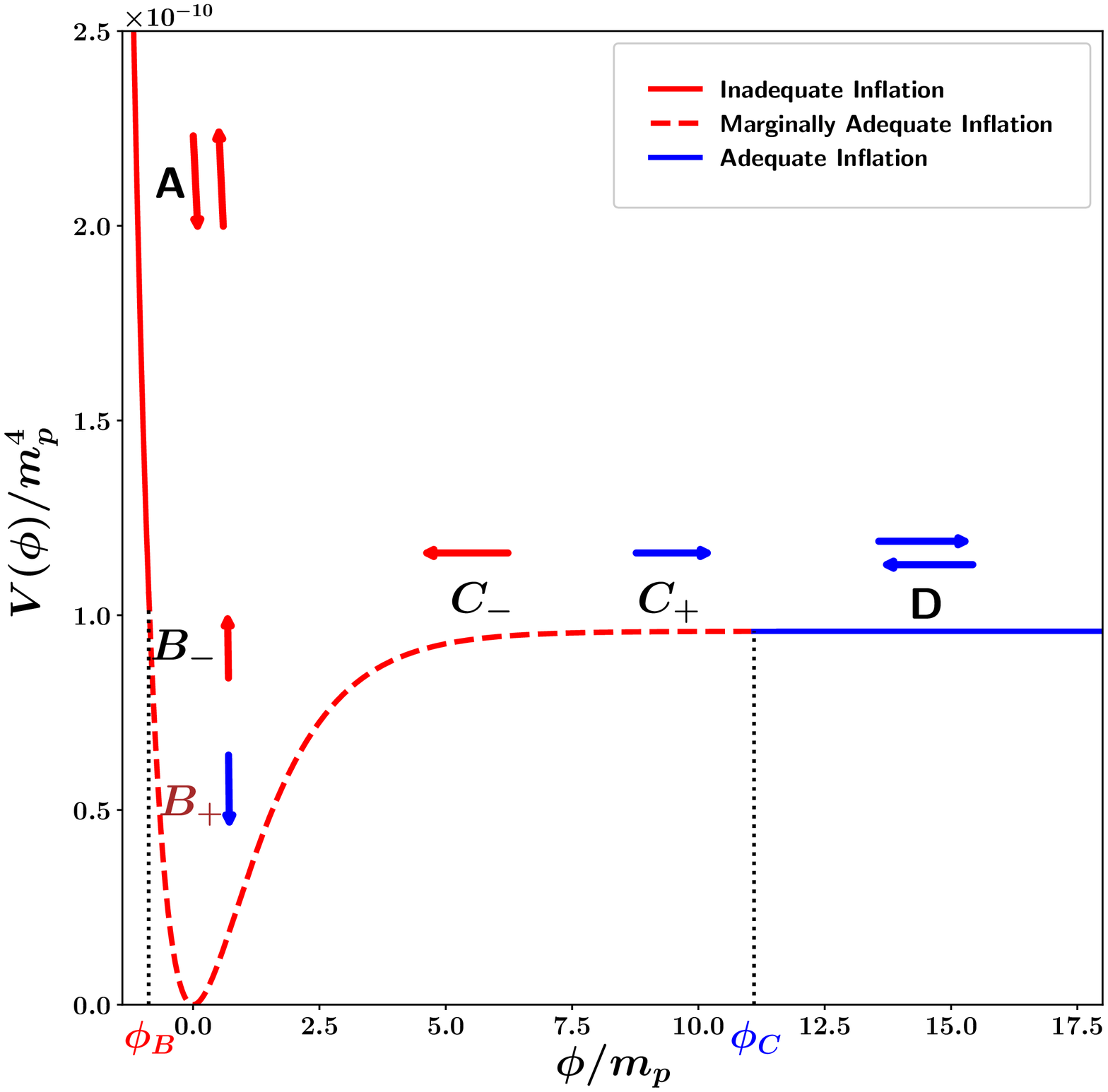}
\captionsetup{
	justification=raggedright,
	singlelinecheck=false
}
\caption{This figure schematically shows initial field values which lead to adequate  and inadequate 
Starobinsky inflation. The  initial energy scale is $H_{i}=3\times 10^{-3}m_{p}$. The solid blue line 
represents the region of adequate inflation while the solid red line displays the region of inadequate 
inflation. (Note that $\phi$ is unbounded on the right.)
 For initial field values lying in the interval
 $\phi_i\in[\phi_{B},\phi_{C}]$ (red dashed line),  one gets adequate inflation only
if the initial velocity $\dot{\phi_i}$ is positive. 
This figure shows that it is easy for inflation to commence from
 the flat right wing of the potential. Note that only a small portion of the full potential is shown in this figure.}
\label{fig:star_pot2}
\end{center}
\end{figure}

\begin{figure}[htb]
\begin{center}
\includegraphics[width=0.55\textwidth]{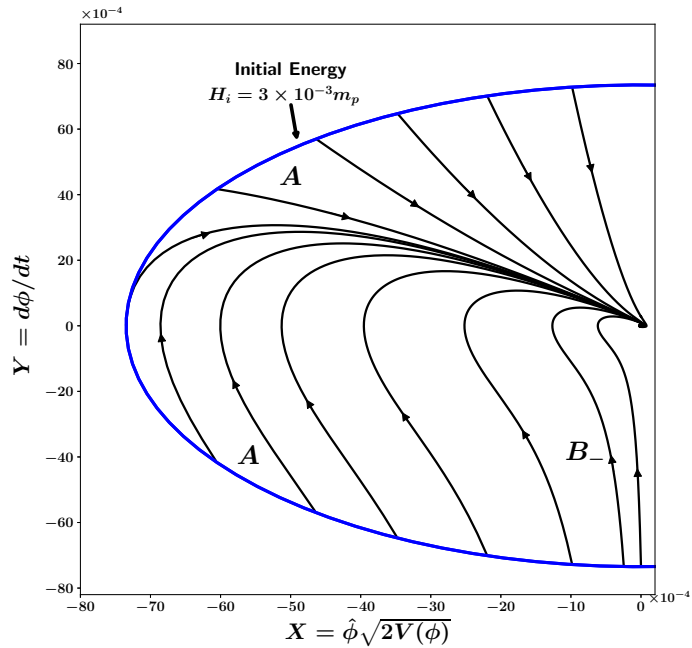}
\captionsetup{
	justification=raggedright,
	singlelinecheck=false
}
\caption{This figure illustrates the phase-space associated with the regions $A$ and $B_-$ on the steep left wing of the potential (\ref{eqn:star_pot}) illustrated
 in figure \ref{fig:star_pot2}. As earlier, $Y=\dot{\phi}$ is plotted against $X=\hat{\phi}\sqrt{2V(\phi)}$  for the fixed initial energy scale 
$H_{i}=3\times 10^{-3} m_{p}$ (blue line). 
($\hat{\phi}=\frac{\phi}{|\phi|}$ is the sign of field $\phi$.)
Note that the 
horizontal slow-roll inflationary separatrix
is absent which reflects the fact that commencing from region $A$ (and $B_-$) in 
figure \ref{fig:star_pot2}, one cannot get adequate inflation from the steep left wing of the Starobinsky potential.}
\label{fig:star_phase}
\end{center}
\end{figure}

\begin{figure}[htb]
\begin{center}
\includegraphics[width=0.6\textwidth]{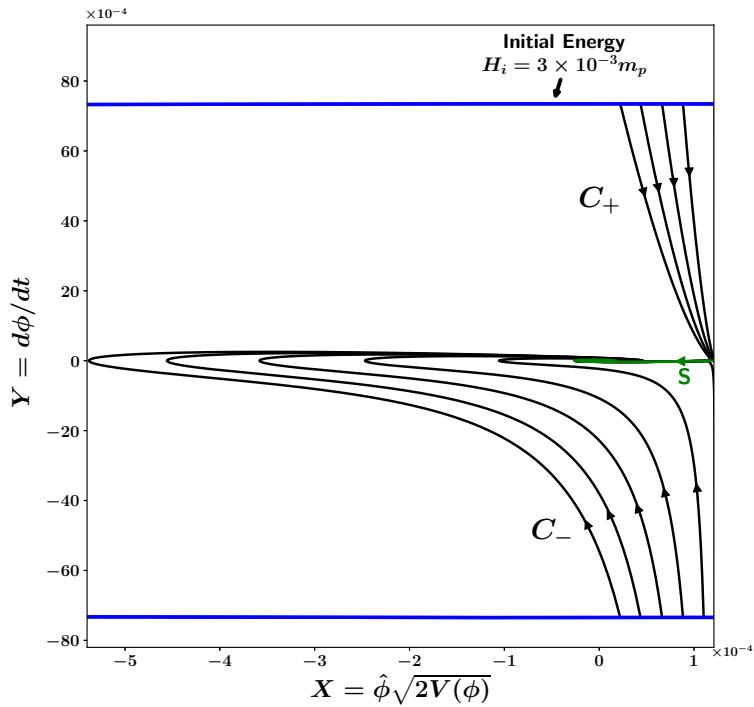}
\captionsetup{
	justification=raggedright,
	singlelinecheck=false
}
\caption{This figure  illustrates the phase-space associated with the flat
 right wing of the potential (\ref{eqn:star_pot}). 
$Y=\dot{\phi}$ is plotted against $X=\hat{\phi}\sqrt{2V(\phi)}$  for the fixed initial energy scale 
$H_{i}=3\times 10^{-3} m_{p}$ (denoted by blue lines at the boundary). 
($\hat{\phi}=\frac{\phi}{|\phi|}$ is the sign of field $\phi$.)
Note that trajectories commencing
 at the boundary with $\dot{\phi_i}>0$ (region $C_+$  in figure \ref{fig:star_pot2}) converge to the inflationary separatrix 
`S' before winding up in spirals around the center
(shown in detail in the next figure). By contrast,
 trajectories commencing on the right wing of $V(\phi)$ with $\dot{\phi_i}<0$ in the region $C_-$ in figure \ref{fig:star_pot2}
, do not lead to inflation. }
\label{fig:star_phase1}
\end{center}
\end{figure}

\begin{figure}[htb]
\begin{center}
\includegraphics[width=0.6\textwidth]{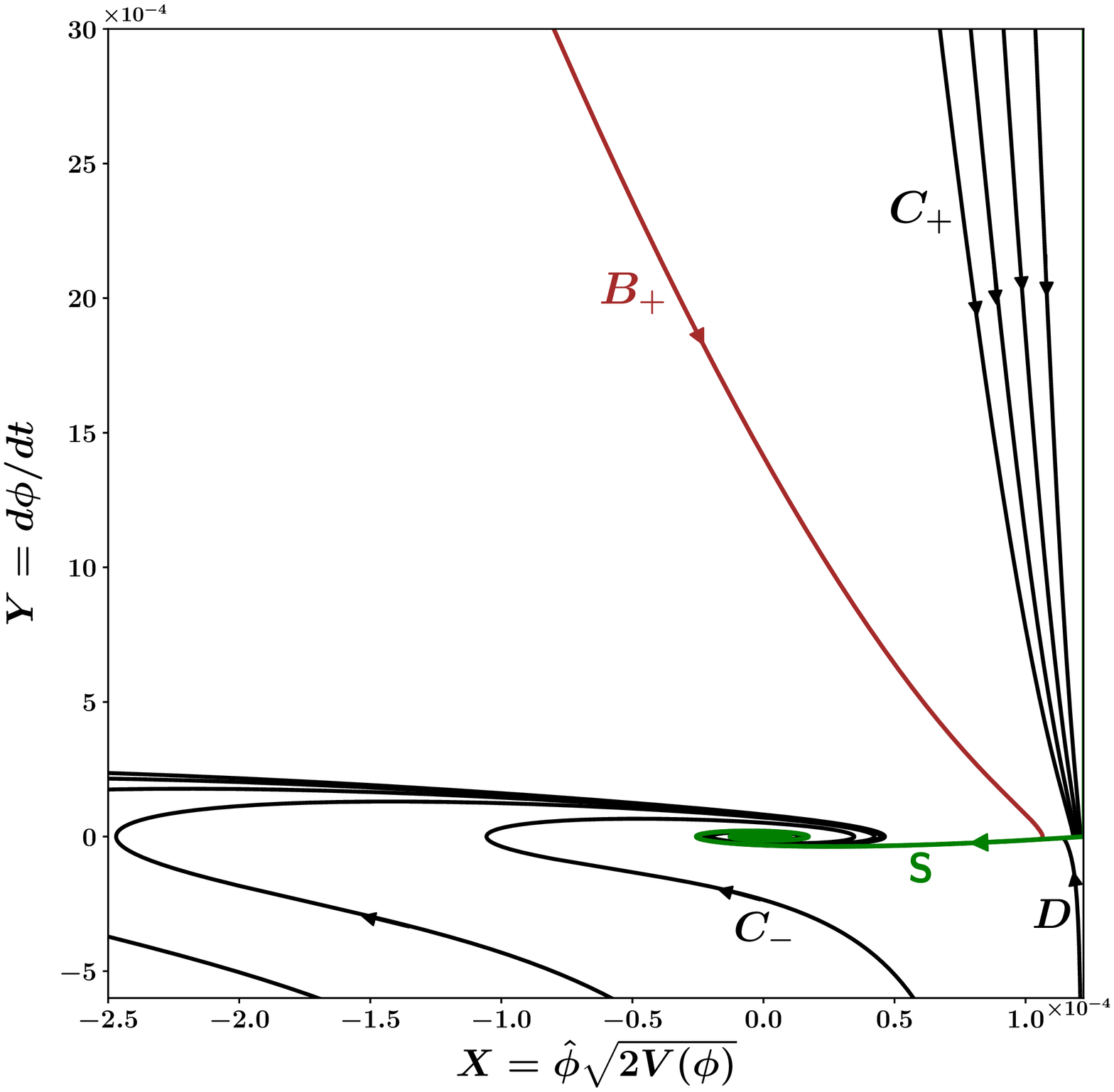}
\captionsetup{
	justification=raggedright,
	singlelinecheck=false
}
\caption{A zoomed-in view of the phase-space of Starobinsky inflation which highlights the
 existence of the slow-roll inflationary separatrix on the flat right wing (green line marked `S' in figure \ref{fig:star_pot2}).
 Most trajectories commencing on the right wing (from regions $C_+$ and $D$) converge to `S' before 
spiralling in towards the minimum of $V(\phi)$. 
(The spirals correspond to post-inflationary oscillations.) 
Such an inflationary separatrix does not exist for the steep left 
wing of the potential. However note
the brown trajectory which is able to meet the inflationary separatrix on 
the right wing even though it commences from region $B_+$ of $V(\phi)$ (but with $\dot{\phi_i}>0$) as shown in figure \ref{fig:star_pot2}.
The brown trajectory describes the motion of the field $\phi$ as it rolls
up the potential.
}
%negative $X$ (but with $\dot{\phi}>0$).}
\label{fig:star_phase2}
\end{center}
\end{figure}

The distinctive properties of the Starobinsky potential
discussed above, result in an interesting phase-space, which is
shown in figures \ref{fig:star_phase}, \ref{fig:star_phase1} and  \ref{fig:star_phase2} for an initial energy scale $H_{i}=3\times 10^{-3} m_{p}$. 
A deeper appreciation of this phase-space is obtained by dividing the potential in equation (\ref{eqn:star_pot}) into 4 regions $A,~B,~C$ and $D$ as shown in figure \ref{fig:star_pot2}. Note that adequate inflation is marked by blue arrows while inadequate inflation is marked 
by red arrows (this notation has been  consistently used throughout our paper).  
One gets adequate inflation in region $D$ independently of the direction of 
$\dot{\phi_i}$ (illustrated by blue arrows in region $D$). Similarly one gets 
inadequate inflation in region $A$ independently of the direction of $\dot{\phi_i}$ 
(red arrows). However one gets adequate inflation in region $B$ (called $B_+$) and 
$C$ (called $C_+$) provided $\dot{\phi_i}$ is positive (blue arrows) whereas
  negative $\dot{\phi_i}$ values in these regions ($B_-$ and $C_-$) lead to inadequate 
inflation (red arrows). With this basic picture in mind, we now proceed to discuss
 the nature of  the phase-space in figures \ref{fig:star_phase}, \ref{fig:star_phase1} and  \ref{fig:star_phase2}.
 
The asymmetry of the potential (\ref{eqn:star_pot}) is
 reflected in the asymmetry of the phase-space shown 
in figures \ref{fig:star_phase}, \ref{fig:star_phase1}, \ref{fig:star_phase2}. 
The phase-space associated with region $A$ on the 
steep left wing of $V(\phi)$ shows no slow-roll and consequently does not
 possess
 an inflationary separatrix; see figure \ref{fig:star_phase}.
 The flat right wing of $V(\phi)$, on the other hand,  has a slow-roll inflationary separatrix `S' (shown by the green line in figures \ref{fig:star_phase1} and \ref{fig:star_phase2}),  towards which most trajectories converge; see figures 
\ref{fig:star_phase1}, \ref{fig:star_phase2}. 
Some of the lines commencing from the left wing with $\dot{\phi}>0$ initially, represented by $B_+$ in figure \ref{fig:star_pot2}, 
(the brown line in figure \ref{fig:star_phase2}) 
are also able to meet the inflationary separatrix 
giving rise to adequate inflation. 
These interesting features of Starobinsky inflation  have been summarized in 
figure \ref{fig:star_pot2}. In this figure, 
the solid blue line corresponding to $\phi_i \geq \phi_{C}$ 
shows trajectories which lead to adequate inflation {\em regardless of the initial 
direction of} $\dot{\phi_i}$. By contrast, the red region corresponding to $\phi_i\leq\phi_{B}$ 
reflects inadequate inflation. 
The intermediate region $\phi_i\in[\phi_{B},\phi_{C}]$ leads to adequate inflation only 
when the 
initial velocity is positive \ie $\dot{\phi_i}>0$  (dashed line).  
Dependence of $\phi_{B}$ and $\phi_{C}$  on the initial energy scale $H_{i}$ 
is shown in table \ref{table:6}.

\par

\begin{table}[H]
\begin{center}
 \begin{tabular}{||c|c|c||} 
 \hline
 $H_{i}$ (in $m_{p}$)  & $\phi_{B}$ (in $m_{p}$) & $\phi_{C}$ (in $m_{p}$)\\ [1ex] 
 \hline\hline
 $3\times 10^{-3}$ & $-0.28$ & $11.11$ \\ [1.2ex] 
 \hline
 $3\times 10^{-2}$ & $-2.16$ & $12.99$ \\ [1.2ex] 
 \hline
 $3\times 10^{-1}$ & $-4.04$ & $14.87$ \\ [1.2ex] 
 \hline
\end{tabular}
\captionsetup{
	justification=raggedright,
	singlelinecheck=false
}
\caption{Dependence of $\phi_{B}$ and $\phi_{C}$ on the initial energy scale $H_{i}$ for Starobinsky Inflation.}
\label{table:6}
\end{center}
\end{table}
From table \ref{table:6} one observes that $\phi_{B}$ shifts to lower (more negative) 
values as the initial energy scale of inflation, $H_i$, is increased.
This is indicative of the fact that inflation can commence
 even from the steep left wing of $V(\phi)$ provided
 the scalar field has a sufficiently large positive velocity initially, 
which would enable the inflaton to climb up the flat right wing 
and result in inflation. \footnote{Pre-inflationary initial conditions for 
Starobinsky inflation have also been studied in \cite{brajesh} in the context of 
loop quantum gravity.}

It may be noted that our results do not support some of the claims made in
 \cite{stein13} that inflation in plateau-like potentials suffers from an {\em unlikeliness
problem} since only a small range of initial field values leads to adequate inflation.
The authors of \cite{stein13} made this claim on the basis of a flat Mexican hat
potential.
Our analysis, based on more realistic models including Higgs inflation and
Starobinsky inflation, has shown that, on the contrary,
 a fairly large range of initial field values
(and initial energy scales) can give rise to adequate inflation, as illustrated in figures \ref{fig:Higs_pot} and \ref{fig:star_pot2}.

Finally we would like to draw attention to the fact that the phase-space
analysis performed here for Starobinsky inflation
is likely to carry over to the E-model
$\alpha$-attractor potential \cite{E-model}, since the two potentials
are qualitatively very similar.

%\section{A comparison between power-law and asymptotically flat potentials}
%\label{sec:power_vs_flat}

\section{Discussion}
\label{sec:discussion}

In this paper we have addressed the issue of the robustness of inflation to different 
choices of initial conditions. 
We have widely varied the initial kinetic and potential terms $\frac{1}{2}{\dot \phi_i}^2$ and $V(\phi_i)$
for a given initial energy scale of inflation and determined the fraction of initial conditions which give
rise to adequate inflation ($N_e \geq 60$). Our analysis has primarily focussed on the following models:
(i) chaotic inflation and its extensions such as monodromy inflation, %which are commonly associated with convex potentials,
(ii) Higgs inflation,
(iii) Starobinsky inflation. 
For class (i) we have shown that inflation becomes more robust for lower values of the exponent $n$
in the inflaton potential $V \propto |\phi|^n$.
This is illustrated in figure \ref{fig:scaling_mono}.
Concerning (ii),
it is well known that Higgs inflation can arise from 
a non-minimal coupling of the Higgs field
to the Ricci scalar. 
 In this case the effective inflaton potential in the Einstein frame is
asymptotically flat and has plateau-like features for large absolute values of the 
inflaton field. 
This is also true in the Einstein frame representation of the Starobinsky potential, but in this case
one of the wings of $V(\phi)$ is flat while the other is steep (and cannot sustain inflation).
A remarkable feature which is shared by (non-minimally coupled) Higgs inflation 
and Starobinsky inflation, is that
one can get adequate inflation ($N_e \geq 60$) even if the inflaton commences to roll
from the {\em minimum of the potential} 
($\phi=0$) and not from its periphery. 
This remarkable property is typical of asymptotically flat potentials and is 
not shared by the power law potentials commonly associated with chaotic inflation.
This new insight forms one of the central results of our paper.\footnote{Our results for Higgs 
and Starobinsky inflation are likely to carry over to the ($\alpha$-attractor based)
T-model \cite{T-model} and E-model \cite{E-model} respectively, due to the great similarity
between the potentials of Higgs inflation and the T-model on the one hand, 
and Starobinsky inflation and the E-model on the other.}

We also show that inflation can be sourced by a Higgs-like field provided the 
Higgs has a non-canonical kinetic term.
In this case non-canonical inflation is more robust, and arises for a larger class of initial 
conditions, than
canonical inflation.

Using phase space analysis we have shown that the fraction of trajectories which inflate
{\em increases} with an increase in the value of the energy scale at which inflation commences. This observation
appears to be generic and applies to all of the models which have been studied in this paper.

One might note that
our analysis in this paper assumes a specific measure on the 
space of initial conditions. Namely
 we assume that $X=\hat{\phi}\sqrt{2V(\phi)}$ ($\hat{\phi}=\frac{\phi}{|\phi|}$ is the sign of field $\phi$) and $Y=\dot\phi$ are distributed uniformly at the 
boundary where initial conditions are set. 
Following this we determine the degree of inflation.
While this approach follows the seminal work of \cite{belinsky85}, it is also possible to construct
 alternative measures. For instance one could assume instead that $\phi$ and $\dot{\phi}$ were 
distributed uniformly at the initial boundary. In this case the boundary will no more be a circle,
as it was for chaotic inflation in figure \ref{fig:chaotic}.
Instead its shape will crucially depend upon the form of $V(\phi)$. 
However we feel that as long as the initial phase-space distribution is not
 sharply peaked near specific values of $\phi_i$, $\dot{\phi}_i$,
the broad results of our analysis will remain in place.
(In other words we suspect that
 inflation is likely to remain generic for a large class of potentials,
although we cannot prove this assertion.) 

For the sake of simplicity we have confined our analysis of inflationary
initial conditions to a spatially flat FRW universe.
The reader should note that by restricting ourselves to homogeneous and
isotropic cosmologies we do not address the larger problem
of inflation in an inhomogeneous and anisotropic setting.
Indeed, the issue as to whether inflation can successfully arise in a
universe which is either inhomogeneous or anisotropic (or both) is rather
complex and has been discussed in several papers including the recent review
\cite{branden16}.
In the case of a positive cosmological constant, it is well known that
classical fluctuations in an FRW Universe redshift and disappear
and the space-time approaches de Sitter space asymptotically \cite{gibbons}.
This result was extended to a `no-hair' theorem by a consideration of
more general space-times including the spatially homogeneous but anisotropic
Bianchi I-VIII family which was shown to rapidly isotropize and (locally)
approach de Sitter space in the future, provided all matter (with the
exception of the cosmological constant) satisfied the strong energy condition
\cite{wald_staro}. The no-hair theorem was subsequently extended to inflationary
cosmology in \cite{no-hair}. However
these studies primarily focussed on
anisotropic models and did not include the effects of
inhomogeneity for which even a semi-analytical treatment is difficult.
A recent discussion of this issue within a numerical setting suggests that,
for plateau-like potentials, inflationary expansion can arise even when the
scale of inhomogeneity exceeds the hubble length provided the mean spatial
curvature is not positive \cite{ekls15} (also see \cite{guth14}).
The exception to this rule is associated with scalar field variations
which exceed the inflationary plateau region and regions with large
positive spatial curvature.\footnote{The latter can prove
problematic for plateau-like potentials since, if the universe emerges
from an initial Planck scale era with a large positive value of the 
curvature,
then the latter would make the universe contract much before the energy
density of the inflaton dropped to that of the
inflationary plateau. A possible resolution of this problem is provided by 
potentials which, 
in addition to possessing a plateau-like region, 
also have monomial/exponential wings which
allow inflation to commence from Planck scale densities 
\cite{monomial_wing,monomial_wing1}.}
Overall it appears that the robustness of inflation 
(in relation to inhomogeneous initial
data) is related to the fact that while strongly inhomogeneous overdense
regions collapse to form black holes, underdense regions continue to expand
enabling inflation to eventually commence. It therefore appears that for
inhomogeneous models the inflationary slow-roll regime is a local but
not global attractor \cite{branden16}.

Finally it is important to note that since the simplest models of inflation
are not past-extendible \cite{borde_vilenkin}, the origin of the 
inflationary scenario 
remains an important open question.

%The reader is referred to \cite{belinsky87,belinsky88} for an early 
%analysis of chaotic inflation in closed/open FRW
%universe models and to the more recent discussion in
%\cite{guth14}. A discussion of inflation in homogeneous and anisotropic models
%can be found in \cite{no-hair}.

%While for asymptotically flat potentials, even these worst initial conditions yield adequate inflation.}\footnote{\textbf{Dear Sir, I also wanted to add a comment that we can not consider the most popular measure which assumes an equipartition of kinetic and potential energy for the asymptotically flat potentials as the potential energy term is bounded from above.} }

%\textbf{Hence we would like to conclude that for the inflaton potentials favored by  CMB observations (\cite{CMB}),  it is possible  to get adequate inflation starting from a large range of initial field values for  a flat and  uniform  initial patch in the early universe.}\footnote{\textbf{Sir, I wanted to add a small paragraph stating that a closed initial patch quickly contracts on itself while in an open initial patch, inflation proceeds to happen easily and universe is driven towards a spatially flat state. These points are clearly mentioned in section II of the paper by Guth, Kaiser and Namura, Phys.Lett. B733 (2014) 112-119 (arXiv:1312.7619v2 [astro-ph.CO]) as a response to steinhardt's paper \cite{stein13}. But I am not sure how and where to phrase it. }}
\section{Acknowledgements} 
A.V.T is supported by RSF Grant 16-12-10401 and by the Russian Government
Program of Competitive Growth of Kazan Federal University. A.V.T is thankful to IUCAA,  where this research work has been carried out, for the hospitality. S.S.M. thanks the Council of Scientific and Industrial Research (CSIR), India, for
financial support as senior research fellow. S.S.M would also like to thank Surya Narayan Sahoo, Remya Nair and Prasun Dutta  for  technical help in generating some of the figures. S.S.M would like to thank Sanil Unnikrishnan for useful discussions and comments on the non-canonical Higgs inflation section.

\appendix

%\section{A note on initial conditions}
%\label{sec:power_vs_flat}

\section{The values of $n_{_S}$ and $r$ for several inflationary models} \label{App:AppendixA}
\begin{figure}[htb]
\begin{center}
\includegraphics[width=0.65\textwidth]{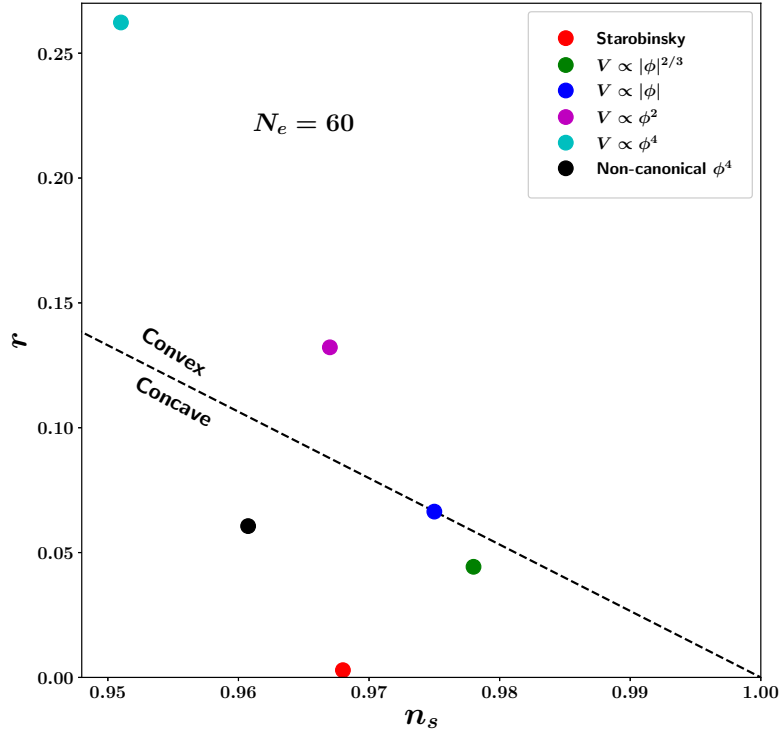}
\captionsetup{
	justification=raggedright,
	singlelinecheck=false
}
\caption{The values of tensor to scalar ratio $r$ and the corresponding values of scalar spectral index $n_{_{S}}$ are plotted in this figure for different inflationary potentials considered in this paper corresponding to $N_e=60$. Note that the values of $r$ and $n_{_{S}}$ for the Starobinsky inflation (\ref{eqn:star_pot}) and the Higgs inflation in the non-minimal framework (\ref{eqn:higgs_pot_final}) are same since  both the potentials  have the  same functional form as far as the flat inflationary wing is concerned. The value for the non-canonical
$\lambda\phi^4$ potential has been determined assuming $\alpha=5$ in 
(\ref{Lagrangian_nc}).}
\label{fig:rvsns}
\end{center}
\end{figure}

For single field slow-roll inflation, the amplitude of scalar fluctuations in given by \cite{baumann07}
\begin{equation}
\Delta_{_{S}}^2 = \frac{1}{24\pi^2}\frac{V(\phi_*)}{m_p^4}\frac{1}{\epsilon(\phi_*)}
\label{eqn:AppA1}
\end{equation}
where  $\phi_*$ is the value of $\phi$ at $N_e$ e-foldings before the end of inflation. 
CMB observations \cite{CMB} imply $\Delta_{_{S}}^2=2.2\times 10^{-9}$ so that
\begin{equation}
\frac{1}{24\pi^2}\frac{V(\phi_*)}{m_p^4}\frac{1}{\epsilon(\phi_*)}= 2.2\times 10^{-9}~.
\label{eqn:AppA2}
\end{equation}
Similarly, for single field slow-roll inflation, the scalar spectral index is given by \cite{baumann07}
\begin{equation}
n_{_{S}}=1+2\eta(\phi_*)-6\epsilon(\phi_*),
\label{eqn:AppA3}
\end{equation}
and the tensor to scalar ratio is given by \cite{baumann07}
\begin{equation}
r=16\epsilon(\phi_*).
\label{eqn:AppA4}
\end{equation}
Values of the CMB normalized parameters $n_{_{S}}$ and $r$ 
for some of the inflationary models discussed in this paper
 are listed in table \ref{table:7}, assuming $N_e = 60$. The corresponding
  $r$ vs $n_{_{S}}$ plot is shown in figure \ref{fig:rvsns}.  

\begin{table}[H]
\begin{center}
 \begin{tabular}{||c|c|c|c|c||} 
 \hline
 \bf{Model}  & \bf{$V(\phi)$} & \bf{Parameter} & \bf{$n_{_{S}}$}  & \bf{$r$} \\ [1ex] 
 \hline\hline
 
  Non-minimal Higgs & $V_0\left(1-e^{-\sqrt{\frac{2}{3}}\frac{|\phi|}{m_p}}\right)^2$ & $V_0=9.6\times 10^{-11}~m_p^4$ &$0.967$ & $0.003$\\ [1.2ex] 
 \hline
 Starobinsky & $\frac{3}{4}m^2 m_p^2\left(1-e^{-\sqrt{\frac{2}{3}}\frac{\phi}{m_p}}\right)^2$ & $m=1.13\times 10^{-5}~m_p$ &$0.967$ & $0.003$\\ [1.2ex] 
 \hline
 Fractional Monodromy & $V_0 \left|\frac{\phi}{m_p}\right|^{2/3}$ & $V_0=3.34\times 10^{-10}~m_p^4$ & $0.978$ & $0.044$ \\ [1.2ex] 
 \hline
 Linear Monodromy & $V_0 \left|\frac{\phi}{m_p}\right|$ & $V_0=1.97\times 10^{-10}~m_p^4$ & $0.975$ & $0.066$ \\ [1.2ex] 
 \hline
Quadratic Chaotic & $\frac{1}{2}m^2\phi^2$ & $m=5.97\times 10^{-6}~m_p$ & $0.967$ & $0.132$ \\ [1.2ex] 
 \hline
 Quartic Chaotic & $\frac{\lambda_c}{4}\phi^4$ & $\lambda_c=1.43\times 10^{-13}$ & $0.951$ & $0.262$ \\ [1.2ex] 
 \hline
\end{tabular}
\captionsetup{
	justification=raggedright,
	singlelinecheck=false
}
\caption{This table lists the CMB normalized value of parameter, scalar spectral index $n_{_{S}}$ and tensor to scalar ratio $r$ for different single field slow-roll 
inflationary models considered in this paper.}
\label{table:7}
\end{center}
\end{table}
 For Higgs inflation, substitution of the value $V_0=9.6\times 10^{-11}~m_p$ into equation (\ref{eqn:v_0}), 
gives $\xi=1.62\times 10^4$ for the non-minimal coupling parameter,
 which is in agreement with equation (\ref{eqn:coup_nm}).
\section{Jordan to Einstein frame transformation for Higgs inflation} \label{App:AppendixB}
A derivation of equations (\ref{eqn:pot1}) and (\ref{eqn:pot2}) is given below. 
Our derivation is similar to that given in \cite{kaiser}, however 
we calculate the field transformation $\phi\longrightarrow\chi$ explicitly.
We commence with the Jordan frame action (\ref{eqn:action_higgs}), 
namely
\begin{equation}
S_{J}=\int d^{4}x\sqrt{-g} \bigg [ f(\phi)R-\frac{1}{2} g^{\mu\nu}\partial_{\mu}\phi\partial_{\nu}\phi-U(\phi) \bigg ]
\label{eqn:action_higgs1}
\end{equation}
which is described by the metric $g_{{\mu}{\nu}}$.
The Einstein frame is described by $\hat{g}_{{\mu}{\nu}}$ where
\begin{equation}
%g_{{\mu}{\nu}}\longrightarrow 
\hat{g}_{{\mu}{\nu}}=\Omega^{2} g_{{\mu}{\nu}}
\label{A1}
\end{equation}
the conformal factor being given by
\begin{equation}
\Omega^{2}=\frac{2}{m_{p}^{2}} f(\phi)=1+\frac{\xi\phi^{2}}{m_p^{2}}~.
\label{A2}
\end{equation}
Furthermore $\sqrt{-g}$ transforms as
\begin{equation}
\sqrt{-g}\longrightarrow \sqrt{-\hat{g}}=\Omega^{4}\sqrt{-g}
\label{A3}
\end{equation}
and the Ricci scalar transforms as 
\begin{equation}
R\longrightarrow \hat{R}=\frac{1}{\Omega^{2}}\Big [ R-\frac{1}{\Omega}\Box\Omega \Big ]
\label{A4}
\end{equation}
where 
$$\Box\Omega=\frac{1}{\sqrt{-g}}\partial_{\mu}(\sqrt{-g}g^{{\mu}{\nu}}\partial_{\nu}\Omega).$$
As a result the action (\ref{eqn:action_higgs1}) transforms to
\begin{equation}
S=\int d^{4}x\sqrt{-\hat{g}} \bigg [ \frac{m_{p}^{2}}{2}\hat{R}-\frac{1}{2} \hat{g}^{\mu\nu}(\frac{1}{\Omega^{2}}\partial_{\mu}\phi\partial_{\nu}\phi+\frac{6m_p^{2}}{\Omega^{2}}\partial_{\mu}\Omega\partial_{\nu}\Omega)-\frac{U(\phi)}{\Omega^{4}}\bigg]. 
\label{A5}
\end{equation}
 Notice that the coupling of the scalar field to gravity has become minimal. 
However the kinetic term is non-canonical. In order to change this to the
 canonical form one redefines the field $\phi\longrightarrow\chi$ such that
%$$
%\int d^{4}x\sqrt{-\hat{g}} \bigg [ \frac{m_{p}^{2}}{2}\hat{R}-\frac{1}{2} \hat{g}^{\mu\nu}(\frac{1}{\Omega^{2}}\partial_{\mu}\phi\partial_{\nu}\phi+\frac{6m_p^{2}}{\Omega^{2}}\partial_{\mu}\Omega\partial_{\nu}\Omega)-\frac{U(\phi)}{\Omega^{4}}\bigg ]$$
%$$=\int d^{4}x\sqrt{-\hat{g}} \bigg [ \frac{m_{p}^{2}}{2}\hat{R}-\frac{1}{2} \hat{g}^{\mu\nu}\partial_{\mu}\chi\partial_{\nu}\chi-V(\chi) \bigg ]$$
\ber
\frac{1}{2} \hat{g}^{\mu\nu}(\frac{1}{\Omega^{2}}\partial_{\mu}\phi\partial_{\nu}\phi+\frac{6m_p^{2}}{\Omega^{2}}\partial_{\mu}\Omega\partial_{\nu}\Omega)+
\frac{U(\phi)}{\Omega^{4}} =
\frac{1}{2} \hat{g}^{\mu\nu}\partial_{\mu}\chi\partial_{\nu}\chi+V(\chi) 
\label{eqn:appendix}
\eer
where 
\ber
V(\chi)=\frac{U[\phi(\chi)]}{\Omega^{4}}.
\label{eq:pot_appendix}
\eer
Consequently the action in the Einstein frame becomes
$$S_{E}=\int d^{4}x\sqrt{-\hat{g}} \bigg [ \frac{m_{p}^{2}}{2}\hat{R}-\frac{1}{2} \hat{g}^{\mu\nu}\partial_{\mu}\chi\partial_{\nu}\chi-V(\chi) \bigg ].$$
Note that assuming a homogeneous and isotropic space-time one can
drop the spatial derivative terms in (\ref{eqn:appendix}) to get
$$\frac{1}{\Omega^{2}}\Big[\dot{\phi^{2}}+6m_{p}^{2}\dot{\Omega}^{2} \Big]=\dot{\chi}^{2}$$
$$\Rightarrow \frac{1}{\Omega^{2}}\Big[\dot{\phi^{2}}+6m_{p}^{2}(\frac{\partial\Omega}{\partial\phi})^{2} \dot{\phi}^{2}\Big]=(\frac{\partial\chi}{\partial\phi})^{2}\dot{\phi}^{2}$$
$$\Rightarrow (\frac{\partial\chi}{\partial\phi})^{2}=\frac{1}{\Omega^{4}}\Big[\Omega^{2}+\frac{6\xi^{2}\phi^{2}}{m_{p}^{2}} \Big]$$
$$\Rightarrow\frac{\partial\chi}{\partial\phi}=\pm\frac{1}{\Omega^{2}}\sqrt{\Omega^{2}+\frac{6\xi^{2}\phi^{2}}{m_{p}^{2}}}$$
which corresponds to (\ref{eqn:pot2}). Note that the '$\pm$' sign here leads to the symmetric potential in figure \ref{fig:Higs_pot}.

\section{Derivation of asymptotic forms of the Higgs potential in the Einstein frame} \label{App:AppendixC}

Equations (\ref{eqn:pot2}) and (\ref{eq:pot_appendix}) can be rewritten as 
\begin{equation}
\frac{\partial\chi}{\partial\phi}=\pm\frac{\sqrt{1+\frac{\xi\phi^2}{m_p^2}+\frac{6\xi^2\phi^2}{m_p^2}}}{1+\frac{\xi\phi^2}{m_p^2}}
\label{eqn:B1}
\end{equation}

\ber
V(\phi)=\frac{U[\phi(\chi)]}{\Omega^{4}} \simeq \frac{\frac{\lambda}{4}\phi^4}
{\left (1+\frac{\xi\phi^2}{m_p^2}\right )^2}~.
\label{eq:pot_app1}
\eer
Using these two equations we proceed to derive the 
following useful asymptotic formulae. \footnote{This analysis has been 
carried out assuming $\xi=1.62\times 10^4 \gg 1$.}
\begin{enumerate}
\item
For $\phi\ll \sqrt{\frac{2}{3}}\frac{m_p}{\xi}$ one finds $\frac{\partial\chi}{\partial\phi}\simeq \pm 1$, 
consequently (\ref{eq:pot_app1}) simplifies to
\begin{equation}
V(\chi)\simeq\frac{\lambda}{4}\chi^4.
\label{eqn:B2}
\end{equation}
\item
For $\phi\gg\sqrt{\frac{2}{3}}\frac{m_p}{\xi}$ one finds
 $\frac{\partial\chi}{\partial\phi}\simeq \pm \frac{\sqrt{6}\frac{\xi\phi}{m_p}}{\Omega^2}$ where $\Omega^2=1+\frac{\xi\phi^2}{m_p^2}$. Hence in this case 
\begin{equation}
\chi\simeq \pm\sqrt{\frac{3}{2}}m_p\log{\Omega^2(\phi)}.
\label{eqn:B3}
\end{equation}
%The constant of integration in the above expression has been set in order to satisfy the continuity of the function $\chi(\phi)$. We can prove it by noticing that when  $\phi\longrightarrow \frac{2m_p}{\sqrt{6}\xi}$, expression (\ref{eqn:B1}) yields $\chi\longrightarrow \pm\frac{2m_p}{\sqrt{6}\xi}$ and expression (\ref{eqn:B3}) also yields $\chi\longrightarrow \pm\frac{2m_p}{\sqrt{6}\xi}$.

For $\sqrt{\frac{2}{3 \xi^2}} \ll\frac{\phi}{m_p}\ll \frac{1}{\sqrt{\xi}}$,
%$\frac{\sqrt{\xi}\phi}{m_p} \ll 1 \ll \frac{\sqrt{6}\xi\phi}{m_p}$, 
expression (\ref{eqn:B3}) reduces to
\begin{equation}
\chi\simeq \pm\sqrt{\frac{3}{2}}\frac{\xi\phi^2}{m_p},
\label{eqn:B4}
\end{equation}
 consequently the potential in (\ref{eq:pot_app1}) acquires the form 
\begin{equation}
V(\chi)\simeq \left (\frac{\lambda m_p^2}{6\xi^2}\right )\chi^2 ~.
\label{eqn:B5}
\end{equation}
%which corresponds to the reheating region of the potential; see \cite{higgs3}. 
Finally for $\frac{\sqrt{\xi}\phi}{m_p} \gg 1$ one finds, from (\ref{eqn:B3}) 
\begin{equation}
\phi\simeq\frac{m_p}{\sqrt{\xi}}\exp\left(\frac{\pm\chi}{\sqrt{6}m_p}\right)
\label{eqn:B6}
\end{equation}
where the '$+$' sign is taken for $\chi>0$ and the '$-$' sign is taken for $\chi<0$, since the 
above solution is valid only in the limit when
 $|\frac{\sqrt{\xi}\phi}{m_{p}}|\gg 1$. Consequently we can rewrite our solution as 
\begin{equation}
\phi\simeq\frac{m_p}{\sqrt{\xi}}\exp\left(\frac{|\chi|}{\sqrt{6}m_p}\right)~,
\label{eqn:B7}
\end{equation}
and the potential in (\ref{eq:pot_app1}) is given by 
\begin{equation}
V(\chi)\simeq \frac{\lambda m_p^4}{4\xi^2}\left(1+\exp\Big[-\sqrt{\frac{2}{3}}\frac{|\chi|}{m_p}\Big]\right)^{-2}.
\label{eqn:B8}
\end{equation}
To summarize, the relation between $\chi$ and $\phi$ in the three asymptotic regions 
is given by
\[
\frac{\chi}{m_p}=\left\{
\begin{array}{ll}
\pm\frac{\phi}{m_p}, \quad \frac{\phi}{m_p}\ll \sqrt{\frac{2}{3 \xi^2}},\\
\pm\sqrt{\frac{3}{2}}\xi\left(\frac{\phi}{m_p}\right)^2, \quad \sqrt{\frac{2}{3 \xi^2}} \ll\frac{\phi}{m_p}\ll \frac{1}{\sqrt{\xi}},\\
\pm\sqrt{6}\log{\left(\frac{\sqrt{\xi}\phi}{m_p}\right)}, \quad \frac{\phi}{m_p}\gg \frac{1}{\sqrt{\xi}}
\end{array}
\right.
\]

\end{enumerate}

\end{document}